\pdfoutput=1
\documentclass[fullpage]{article}
\usepackage{graphicx}
\usepackage{xcolor}
\usepackage{tikz}
\usepackage{varwidth}
\usepackage{caption}
\usepackage{subcaption}
\usepackage{hyperref}
\usetikzlibrary{arrows}
\usetikzlibrary{shapes}
\usepackage{amsmath}

\usepackage{afterpage}
\usepackage{authblk}

\tikzset{
    *|/.style={
        to path={
            (perpendicular cs: horizontal line through={(\tikztostart)},
                                 vertical line through={(\tikztotarget)})
            -- (\tikztotarget) \tikztonodes
        }
    }
}

\definecolor{columbiablue}{rgb}{0.61, 0.87, 1.0}
\definecolor{classicrose}{rgb}{0.98, 0.8, 0.91}
\definecolor{lightgray}{rgb}{0.83, 0.83, 0.83} 
\definecolor{inchworm}{rgb}{0.7, 0.93, 0.36}

\begin{document}

\title{Backward Deep BSDE Methods and Applications to Nonlinear Problems}
\author{Yajie Yu} 
\author{Bernhard Hientzsch}
\author{Narayan Ganesan}
\affil{Corporate Model Risk, Wells Fargo}
\date{}

\maketitle
\begin{abstract}
In this paper, we present a backward deep BSDE method applied to Forward
Backward Stochastic Differential Equations (FBSDE) with given terminal condition
at maturity that time-steps the BSDE  backwards. We present an
application of this method to a nonlinear pricing problem - the differential
rates problem. To time-step the BSDE backward, one needs to solve a nonlinear
problem. For the differential rates problem, we derive an exact solution of this
time-step problem and a Taylor-based approximation. Previously backward deep BSDE
methods only treated zero or linear generators. While a Taylor approach for nonlinear generators was previously
mentioned, it had not been implemented or applied, while we apply our method to nonlinear generators and derive
details and present results. Likewise, previously backward deep BSDE methods were presented for fixed initial risk factor values
$X_0$ only, while we present a version with random $X_0$ and a version that
learns portfolio values at intermediate times as well. The method is able to solve nonlinear FBSDE problems in high 
dimensions.
\end{abstract}

\section{Introduction}
As proposed in E, Han and Jentzen \cite{weinan2017deep}, deep learning (DL) and
deep neural networks (DNN) method can be used to solve high dimensional
nonlinear PDEs by converting them to Forward Backward Stochastic Differential
Equations (FBSDE) and building neural networks to learn the control and initial
value of the corresponding stochastic control problem.
One example in that paper uses their proposed forward deep BSDE method to
price a combination of two call options with differential rates (different
borrowing and lending interest rates) as studied in Mercurio
\cite{mercurio2015bergman}.
Hientzsch \cite{hientzsch2019intro} also gives an overview of pricing different
instruments in quantitative finance via deep BSDE and FBSDE.
Ganesan, Yu and Hientzsch \cite{ganesan2019pricingbarriers} show how to price
Barrier options with deep BSDE and FBSDE.

Han, Jentzen and E \cite{han2018solving} propose time-stepping both forward and
backward SDE forward in time and transform the final value problem to a
stochastic control problem in which the objective function measures how well the
given final value has been approximated. We call their method ''forward deep BSDE''
method since it time-steps the BSDE forward. Wang et al \cite{wang2018deep} consider
a BSDE with zero drift term which can be trivially time-stepped backwards and
propose and demonstrate forward and backward methods with fixed $X_0$, describing the
first backward deep BSDE method. Liang,
Xu and Li \cite{liang2019deep} solve BSDEs with linear generators with both forward and backward
methods in their examples. They indicate that a nonlinear generator could be handled with 
a Taylor expansion approach, but do not work out nor implement the case of the nonlinear generator
in the backward method. In this paper, we will describe both the general approach as well as
the application to the differential rates setting for two variants of the backward method, 
demonstrating to the best of our knowledge the first application of the backward method to nonlinear problems. 

The main idea of the backward method is that
the BSDE is started at maturity with the given final value and then time-stepped
backwards until a given initial time ${t_0}$.
In the case of the dynamics of $X$ being started at ${t_0}$ at a fixed value
${X_0}$, with the right trading strategy, for the time-continuous case, all
realizations of $Y_{t_0}$(spot price of derivative at initial time) should have the same value at time ${t_0}$.
Thus, a measure of  the size of the range - variance in this particular case -
of $Y_{t_0}$ is picked as the objective function and the variance of the mini-batch is
chosen in mini-batch stochastic gradient descent. For the case of random
${X_0}$, we minimize the square distance from an also to-be-determined function
$\mathsf{yinit(X_0)}$ represented by a DNN. This function is also the
predictable adapted ${L^2}$-projection of the values obtained from the pathwise
roll-back.

The particular nonlinear pricing problem that we consider is the case of
differential rates together with Black-Scholes forward dynamics for European
option pricing problem involving, for example, a linear combination of two calls with
coefficients with opposite signs. Differential rates mean that positive cash
balances in the trading strategy accrue interest at a lower lending rate while
negative cash balances (debts/loans) accrue interest at a higher borrowing rate.
Standard self financing trading strategy arguments lead to a nonlinear BSDE.

For the differential rates problem, E, Han and Jentzen \cite{weinan2017deep}
present a nonlinear PDE which can be solved by appropriate nonlinear PDE solvers
in small dimensions (see, for instance, Forsyth and Labahn
\cite{forsyth2007numerical}).
For a more general setting, Mercurio \cite{mercurio2015bergman} presents PDEs
and proposes PDE solution or binomial tree methods. None of these methods works
in higher dimensions due to the curse of dimensionality.  All these methods
require problem specific implementation of nonlinear PDE or tree solver.

Standard Monte-Carlo approaches that simulate, discount, and average can not
handle nonlinear pricing or control problems that depend on the solution or its
gradient.

There are some other approaches to such nonlinear problems in high dimensions
such as Warin \cite{warin2018nesting} or Hur{\'e}, Pham and Warin
\cite{hure2019some}. However, they use nested Monte-Carlo or more
elaborate methods rather than a path-wise approach (and they do not treat the
differential rates problem as an example).

In this paper, we first introduce FBSDE for general nonlinear problems,  with
particular details for the differential rates problem, time-discretize them, and
then derive exact and Taylor approximations for the backward step. We then
quickly describe the forward and backward deep BSDE approaches that we consider
- both the batch-variance variant already described in the literature but also
the novel initial variable and network versions, the last one for random $X_0$,
together with the computational graphs for the implementations. Then we apply
these methods to the differential rates problem for the call combination case
from Han, Jentzen and E \cite{han2018solving} and for the straddle case from
Forsyth and Labahn \cite{forsyth2007numerical}. We compare the results for a
case with fixed $X_0$ and for a case with varying $X_0$ with the results from
Forsyth and Labahn \cite{forsyth2007numerical} and see that they agree well. We
visualize and discuss some of the results.
Finally, we conclude.

\section{FBSDE for Nonlinear Problems}
 
One type of nonlinear PDEs that we are interested in solving has the general
form:
\begin{equation}
u_{t}\left( t,x \right) + {\cal L}_t u\left( t,x \right) + f\left( t,x,u\left( t,x \right),\nabla
 u\left( t,x \right) \right) = 0, \label{eq:PDE}
\end{equation}
with
\begin{equation}
{\cal L}_t u\left( t,x \right) :=  
\frac{1}{2}\text{Tr}\left(\sigma_N\sigma_N^{T}\left( t,x \right)\left(
	\text{Hess}_{x}u \right)\left( t,x\right) \right) 
+ \mu\left( t,x \right)\nabla u\left( t,x \right), \label{eq:PDEOp}
\end{equation}
where $\text{Hess}_{x}u$ is the Hessian matrix, with terminal condition at maturity given as:
\begin{equation}
u(T, x) = g(x). \label{eq:PDEFV}
\end{equation}
A nonlinear Feyman-Kac theorem shows the solution of above PDE also satisfies
the following FBSDE system under appropriate assumptions:

The forward SDE (FSDE) for the underlying assets:
\begin{equation}
dX_{t} = \mu\left( t,X_{t} \right)dt + \sigma_N\left(t,X_{t}\right)dW_{t}, \label{XFSDE}
\end{equation}
\noindent and the backward SDE (BSDE) in terms of the coefficient of the
Brownian $Z_{t}$:
\begin{equation}
- dY_{t} = f_Z\left( t,X_{t},Y_{t},Z_{t} \right)dt\  - Z_{t}^{T}dW_{t}, \label{eq:YBSDE:Z}
\end{equation}
or in terms of values $\Pi_{t}$:\footnote{If $\Pi_{t}$ measures the hedging delta in the portfolio, it would
be $\sigma_{N}$ rather than $\sigma_{LN}$ in the stochastic term of the $Y$ BSDE, where $\sigma_N(t,X) = \sigma_{LN}(t,X) X$.}
\begin{equation}
- dY_{t} = f\left( t,X_{t},Y_{t},\Pi_{t} \right)dt\  - \Pi_{t}^{T}
\sigma_{LN}\left(t,X_{t}\right) dW_{t}, \label{eq:YBSDE:P}
\end{equation}
with terminal condition
\begin{equation}
Y_{T} = g(X_{T}), \label{eq:YBSDE:FV}
\end{equation}
where
\begin{equation}
Y_{t} = u(t, X_{t}),
\Pi_{t} = \nabla_X u(t, X_{t}),
Z_{t} = \sigma^T_{LN}\left(t, X_{t}\right) \Pi_{t} . \label{eq:BSDEvsPDE}
\end{equation}

In terms of pricing applications in finance, $g(X_T)$ is the final payoff of the
European option that one tries to replicate with a self-financing portfolio in
the underlying asset(s) $X_t$ and a remaining cash position. That portfolio will contain $\pi_i(t)$ worth of $X_i(t)$ ($\Pi_{t}$ being the
vector of $\pi_i(t)$).\footnote{If measured by value, or $\pi(t) X_i(t)$ worth of
$X_i(t)$ if measured by delta/size.} The portfolio (including cash position) is worth $Y_t$ at time $t$. 

Now $Y_t$ or equivalently $u(t,X_t)$ represent the needed wealth at $t$ to
exactly or approximately replicate the payoff when starting at $X_t$ at time
$t$.
This gives one of the possible ways to define price (pricing by replication):
$\mathsf{price}(t,X_t; X_T \mapsto g(X_T))$ as the solution of the FBSDE and/or
the nonlinear PDE.
Linear pricing satisfies (among other things)

\begin{equation}
\mathsf{price}(t,X_t; X_T \mapsto g(X_T)) = - \mathsf{price}(t,X_t; X_T \mapsto
- g(X_T)). \label{eq:longshortsymmetry}
\end{equation}

In nonlinear pricing in general (as for instance for differential rates, as we
will see in  examples later), these two prices are no longer necessarily the
same but will give an upper and a lower price.

\subsection{Time-discretizing Time-continuous FBSDE}

Using Euler-Maruyama method to discretize time direction forward for both $X_t$ and $Y_t$, we
have

\begin{equation}
X_{t_{i+1}} = X_{t_i} + \mu(t_i,X_{t_i}) \Delta t_i + \sigma(t_i,X_{t_i})
\Delta W^i \label{eq:XTimeStep}
\end{equation}
\noindent and
\begin{equation}
Y_{t_{i+1}} = Y_{t_i} - f\left( t_i,X_{t_i},Y_{t_i},\Pi_{t_i} \right) \Delta 
t_i + \Pi^T_{t_i} \sigma(t_i,X_{t_i}) \Delta W^i. \label{eq:YTimeStep}
\end{equation}

\subsection{Backward Time-stepping}
\subsubsection{Analytical Solution}
To backward step in time direction, we rewrite (\ref{eq:YTimeStep}) as:
\begin{equation} 
Y_{t_i} - f\left({t_i},X_{t_i},Y_{t_i},\Pi_{t_i} \right) \Delta {t_i} = 
Y_{t_{i+1}} - \Pi^T_{t_i}  \sigma(t_i,X_{t_i}) \Delta W^i \label{eq:timestepeq}
\end{equation}
and solve for $Y_{t_i}$.  

For a differential rates setup in a risk neutral measure, the $f$ generator
function in the BSDE is:
\begin{equation}
f(t,X_{t},Y_{t},\Pi_{t}) = -r^l(t) Y_{t} + (r^b(t)-r^l(t)) \left(\sum_{i=1}^n \pi_i(t) -
  Y_{t} \right)^{+} . \label{eq:drgenerator}
\end{equation}
\noindent This driver expresses that all assets $X_i(t)$ and positive cash
balances grow at a risk-neutral rate $r^l(t)$ unless the cash position $Y_t -
\sum_{i=1}^n \pi_i(t)$ is negative, and that negative cash balance will grow at a
rate $r^b(t)$ corresponding to the higher borrowing rate as compared to the lower
or equal lending rate.

There are two cases for equation (\ref{eq:drgenerator}):

1). If $\sum_{i=1}^n \pi_i(t)>  Y(t)$: 
\begin{equation}
f(t,X_{t},Y_{t},\Pi_{t}) = -r^l(t) Y_{t} + (r^b(t)-r^l(t)) \left(\sum_{i=1}^n \pi_i(t) -
  Y_{t}\right) . \label{eq:drgenerator1}
\end{equation}
Inserting this into equation (\ref{eq:timestepeq}) and solving, we obtain:
\begin{equation} 
Y_{t_i} =\frac{  Y_{t_{i+1}}  +  \left(r^b(t_i)-r^l(t_i) \right) \left(\sum_{j=1}^n \pi_j(t_i)\right)\Delta {t_i} 
- \Pi^T_{t_i}  \sigma(t_i,X_{t_i}) \Delta W^i }{1 + r^b(t_i)\Delta {t_i} } .
\label{eq:exstep1}
\end{equation}

2). If $\sum_{i=1}^n \pi_i(t) \leq  Y(t)$:
\begin{equation}
f(t,X_t,Y_t,\Pi_{t}) = -r^l(t) Y_t. \label{eq:drgenerator2}
\end{equation}
Inserting this into equation (\ref{eq:timestepeq}) and solving, we obtain:
\begin{equation} 
Y_{t_i} =\frac { Y_{t_{i+1}} - \Pi^T_{t_i}  \sigma(t_i,X_{t_i}) \Delta W^i }{1 +
r^l(t_i)\Delta {t_i} } .\label{eq:exstep2}
\end{equation}
However, we do not know ${Y_{t_i}}$ before solving the nonlinear equation
(\ref{eq:timestepeq}) for it. From (\ref{eq:exstep1}) and (\ref{eq:exstep2}) and the conditions involving ${Y_{t_i}}$, 
we obtain that $Y_{t_i} < \sum_{j=1}^n \pi_j(t_i) $ 
is equivalent to 
\begin{equation}
Y_{t_i+1} < \left(\sum_{j=1}^n
\pi_j(t_i)\right)\left\{\sigma(t_i,X_{t_i})\Delta W^i + (1+r^l(t_i))\Delta {t_i} \right\} \label{eq:excondition}
\end{equation} and the same for the 
relation with $\geq$. Thus, if (\ref{eq:excondition}) is satisfied, we use (\ref{eq:exstep1}), otherwise (\ref{eq:exstep2}).

\subsubsection{Taylor Expansion}
For first order Taylor expansion, we have:
\begin{eqnarray}
f\left(t_{i},X_{t_i},Y_{t_i},\Pi^T_{t_i}  \sigma(t_i,X_{t_i}) \right)
 & \approx &  
 f\left(t_{i},X_{t_i},Y_{t_i+1},\Pi^T_{t_i}  \sigma(t_i,X_{t_i})  \right) \nonumber \\ 
 & &  - \frac {\partial f}{\partial Y} \left( t_{i},X_{t_i},Y_{t_i+1},\Pi^T_{t_i}  \sigma(t_i,X_{t_i}) \right) 
\Delta Y^{\Delta t}_t . \label{eq:taylorgen}
\end{eqnarray} 
\noindent Inserting this into equation (\ref{eq:timestepeq}) and solving for
$Y_{t_i}$, we have the following:
\begin{equation}
Y_{t_i} = Y_{t_{i+1}}  + 
\frac{f\left( t_i,X_{t_i},Y_{t_{i+1}},\Pi^T_{t_i}  \sigma(t_i,X_{t_i})  \right) \Delta t_i -  \Pi^T_{t_i}  \sigma(t_i,X_{t_i})  \Delta W^i}
{1-\frac {\partial f}{\partial Y}\left(
t_{i},X_{t_i},Y_{t_{i+1}},\Pi^T_{t_i}  \sigma(t_i,X_{t_i})  \right) \Delta t} . 
\label{eq:taylortimestep}
\end{equation}
Note that ${f}$ and $\frac{\partial f}{\partial u}$ are evaluated at $Y_{t_{i+1}}$.

With the same setup for the differential rates problem, it is clear that there are only two possible forms for $f$:

1). If $\sum_{j=1}^n \pi_j(t_i)>  Y_{t_{i+1}}$: 
\begin{equation}
f(t_i,X_{t_i},Y_{t_{i+1}},\Pi_{t_i}) = -r^l(t_i) Y(t_i) + (r^b(t_i)-r^l(t_i)) \left(\sum_{j=1}^n \pi_j(t_i) -
  Y_{t_{i+1}}\right) \label{eq:drgenerator1a}
\end{equation}
and 
\begin{equation}
\frac{\partial f}{\partial Y} = -r^b(t_i). \label{eq:derdrgenerator1}
\end{equation}
Inserting this into equation (\ref{eq:taylortimestep}), we obtain:
\begin{equation} 
Y_{t_i} = \frac{Y_{t_{i+1}}  +  
\left(r^b(t_i)-r^l(t_i) \right)\left(\sum_{j=1}^n \pi_j(t_i) \right) \Delta {t_i} 
- \Pi^T_{t_i}  \sigma(t_i,X_{t_i}) \Delta W^i }{1 + r^b(t_i)\Delta {t_i}}
. \label{eq:taylorstep1}
\end{equation}

2). If $\sum_{j=1}^n \pi_j(t_i) \leq Y_{t_{i+1}}$: 
\begin{equation}
f(t_i,X_{t_i},Y_{t_{i+1}} ,\Pi_{t_i}) = -r^l(t_i) Y_{t_{i+1}} \label{eq:drgenerator2a}
\end{equation}
and
\begin{equation}
\frac{\partial f}{\partial Y} = -r^l(t_i) . \label{eq:derdrgenerator2}
\end{equation}
\noindent Inserting this into equation (\ref{eq:taylortimestep}), we have:
\begin{equation} 
Y_{t_i} = \frac{ Y_{t_{i+1}} - \Pi^T_{t_i}  \sigma(t_i,X_{t_i}) \Delta W^i }{1 +
r^l(t_i)\Delta {t_i}} . \label{eq:taylorstep2}
\end{equation}

Notice that (\ref{eq:exstep1}) and (\ref{eq:taylorstep1}) are the same
and that (\ref{eq:exstep2}) and (\ref{eq:taylorstep2}) are the same. The only difference
lies in the conditions when they are applied.

\section{Deep BSDE Approach}
\subsection{Forward Approach}

As introduced in E, Han and Jentzen \cite{weinan2017deep}, with forward
time-stepped equations (\ref{eq:XTimeStep}) and (\ref{eq:YTimeStep}), one 
minimizes the loss function 
\begin{equation}
E(||Y_{T} - g({X_N})||^2).
\end{equation} 

The initial portfolio value $Y_0$ is a parameter of the minimization problem as
are all the parameters of the DNN functions $\pi_i(t_i,X_{t_i})$ treated as
functions of $X_{t_i}$ (that give the stochastic vector process $\Pi_t$ as value
or size of the holdings of the risky underlying securities in the portfolio).
Since $X_0$ is fixed, instead of learning a function $\pi_0(X_0)$, one learns a
parameter $\pi_0$. Alternatively, one can learn a single function $\pi(t_i,X_{t_i})$
as function of $t_i$ and $X_{t_i}$ which means that all the parts of the
computational graph that represent the evaluation of $\pi(t,x)$ share the same
DNN parameters.\footnote{There are many introductions into DL, DNN, and common
forms of DNN. For a minimal one geared towards deep BSDE, see Hientzsch
\cite{hientzsch2019intro}.} The minimization problem is then solved with
standard deep learning approaches such as mini-batch stochastic gradient
methods, using approaches such as Adam, pre-scaling and/or batch-normalization,
etc.

For the case of random $X_0$, one also learns the initial value of $Y_0$ as a
function $\mathsf{Yinit}(X_0)$ of $X_0$, using the same loss function.
Han, Jentzen, and E \cite{han2018solving} mention this approach for random $X_0$
on page 8509. We are not aware of any publication presenting results or
implementations of the random $X_0$ approach except in our own work. 

\subsection{Backward Approach}

In the backward approach, one time-steps equations (\ref{eq:XTimeStep}) forward
but time-steps equations (\ref{eq:YTimeStep}) backward, starting from 
$Y_{T}=g(X_{T})$. As discussed in
the previous section, one can use an analytical solution of
(\ref{eq:timestepeq}) or some Taylor expansion approach. Using either approach, 
one will obtain an expression or implementation 
\begin{equation}
Y_{t_i} = \mathsf{ybackstep}(t_i,Y_{t_{i+1}},X_{t_i},\Pi_{t_i},\Delta W^i) .
\label{eq:ybstep}
\end{equation}
\noindent For fixed $X_0$, the loss function will be 
\begin{equation}
\mathsf{var}(Y_0).
\end{equation}
\noindent For the mini-batch stochastic gradient step, the loss function will be
the mini-batch variance
\begin{equation}
E(||Y_{0} - \bar{Y}_0||^2),
\end{equation} 
where $\bar{Y}_0$ will be the mean over the mini-batch. For MC (Monte Carlo) estimates for
$Y_0$, one can use the last mini-batch mean or one can compute the mean of $Y_0$
over a larger sample of paths (but fixing the trading strategy).

Instead of using the mini-batch mean in the loss function, one can learn
$\bar{Y}_0$ as a parameter/variable (resulting in the same loss function but
with different meaning of $\bar{Y}_0=\mathsf{Yinit}$).

Once $X_0$ is random, one can no longer use batch variance in a straightforward
way.
Instead (and inspired by the parameter version just discussed), one uses a loss
function
\begin{equation}
E(||Y_{0} - \mathsf{Yinit}(X_0)||^2),
\end{equation} 
where the $\mathsf{Yinit}(X_0)$ is a function represented by a DNN which is  
learned as part  of the DL approaches. 

Similarly, one can introduce additional terms
\begin{equation}
E(||Y_{t_i} - \mathsf{Ylearned}_i(X_{t_i})||^2)
\end{equation}
at some (or all) intermediate times $t_i$ to learn some approximations for the
solution function $\mathsf{Ylearned}_i(X_{t_i})=u(t_i,X_{t_i})$ (or one could learn 
the  trading strategy and intermediate solution functions in stages in a
roll-back fashion).

All the methods except the one using batch variance are novel, to the best of our knowledge. 

For these different backward approaches, the first time step translates into the
three different computational graphs shown in figures
\ref{fig:FirstStepBkwdMCmean}, \ref{fig:FirstStepBkwdLearnY0}, and
\ref{fig:FirstStepBkwdLearnY0Network}. The general time step for all three
methods is shown in figure \ref{fig:GeneralStepBkwd}, while the last time step
is shown in figure \ref{fig:laststep}. First, one would simulate $X$ forward,
starting at the first time step, proceeding through intermediate time steps, and
reaching the final time step. At the final time step, $Y_T$ is set to $g(X_T)$,
and backward steps $\mathbf{ybstep}$ are taken, proceeding through intermediate
steps, until one reaches back at the first time step. In the figures, gray boxes
are given implementations/operations that do not change, pink boxes (circles)
are networks (variables/parameters) to be trained, blue circles are random and
green circles are input parameters.

\begin{figure}[p]
\begin{center}
\begin{tikzpicture}

\node[draw,circle] (Yi) at (0,0) {$Y^B_0$};
\node[draw,circle,fill=inchworm] (Xi) at (0,5) {$X_0$};
\node[draw,fill=lightgray] (volstep) at (2,5) {$\mathbf{volstep}$};
\node[draw,circle] (sigmai) at (4,5) {$\Sigma_0$};
\node[draw,circle,fill=classicrose] (pi) at (4,3) {$\Pi_0$};
\node[draw,fill=lightgray] (mult) at (6,2) {$\mathbf{mult}$};
\node[draw,circle] (zi) at (8,1) {$Z_0$};
\node[draw,fill=lightgray] (ystep) at (10,0) {$\mathbf{ybstep}_0$};
\node[draw,fill=columbiablue,circle] (dWi) at (10,2.5) {$\Delta W_0$};
\node[draw,fill=lightgray] (xstep) at (10,5) {$\mathbf{xstep}_0$};
\node[draw,circle] (Yi1) at (12,0) {$Y_{1}$};
\node[draw,circle] (Xi1) at (12,5) {$X_{1}$};


\draw[->] (ystep) to[bend left] (Yi);
\draw[->] (Xi) to (volstep);
\draw[->] (Xi) to[bend right] (ystep);
\draw[->] (volstep) to (sigmai);
\draw[->] (sigmai) to (xstep);
\draw[->] (sigmai) to (mult);
\draw[->] (pi) to (mult);
\draw[->] (mult) to (zi);
\draw[->] (zi) to (ystep);
\draw[->] (dWi) to (xstep);
\draw[->] (dWi) to (ystep);
\draw[->] (xstep) to (Xi1);
\draw[->] (Yi1) to (ystep);

\end{tikzpicture}
\end{center}
\caption{First time step in backward method when using MC mean for $Y_0$ for a fixed $X_0$
\label{fig:FirstStepBkwdMCmean}}
\end{figure}
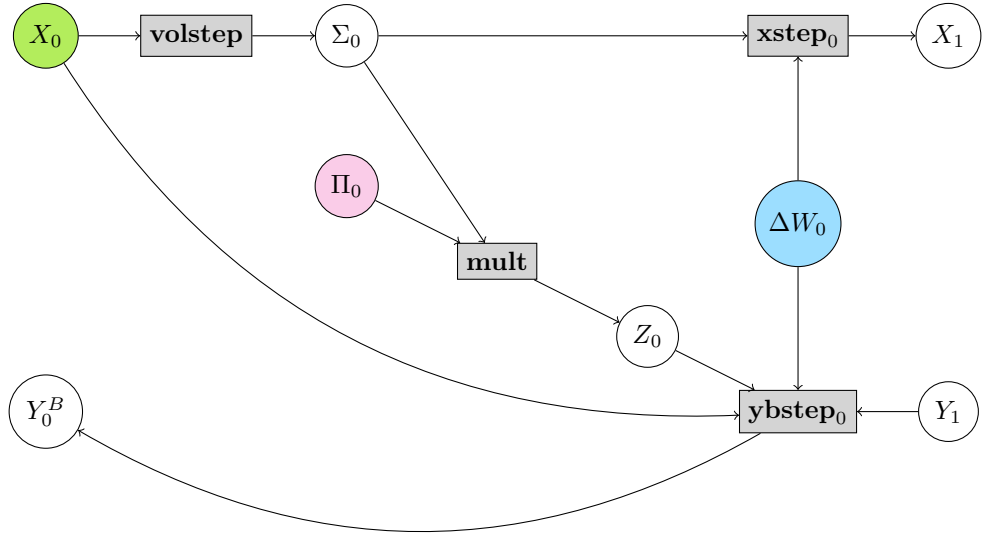

\begin{figure}[p]
\begin{center}
\begin{tikzpicture}

\node[draw,circle,fill=classicrose] (Yib) at (0,0) {$\bar{Y}_0=\Theta^y_0$};
\node[draw,circle] (Yi) at (2,0) {$Y^B_0$};
\node[draw,circle,fill=inchworm] (Xi) at (0,5) {$X_0$};
\node[draw,fill=lightgray] (volstep) at (2,5) {$\mathbf{volstep}$};
\node[draw,circle] (sigmai) at (4,5) {$\Sigma_0$};
\node[draw,circle,fill=classicrose] (pi) at (4,3) {$\Pi_0$};
\node[draw,fill=lightgray] (mult) at (6,2) {$\mathbf{mult}$};
\node[draw,circle] (zi) at (8,1) {$Z_0$};
\node[draw,fill=lightgray] (ystep) at (10,0) {$\mathbf{ybstep}_0$};
\node[draw,fill=columbiablue,circle] (dWi) at (10,2.5) {$\Delta W_0$};
\node[draw,fill=lightgray] (xstep) at (10,5) {$\mathbf{xstep}_0$};
\node[draw,circle] (Yi1) at (12,0) {$Y_{1}$};
\node[draw,circle] (Xi1) at (12,5) {$X_{1}$};


\draw[->] (ystep) to[bend left] (Yi);
\draw[->] (Xi) to (volstep);
\draw[->] (Xi) to[bend right] (ystep);
\draw[->] (volstep) to (sigmai);
\draw[->] (sigmai) to (xstep);
\draw[->] (sigmai) to (mult);
\draw[->] (pi) to (mult);
\draw[->] (mult) to (zi);
\draw[->] (zi) to (ystep);
\draw[->] (dWi) to (xstep);
\draw[->] (dWi) to (ystep);
\draw[->] (xstep) to (Xi1);
\draw[->] (Yi1) to (ystep);

\end{tikzpicture}
\end{center}
\caption{First time step in backward method when
learning $Y_0$ for a fixed $X_0$ \label{fig:FirstStepBkwdLearnY0}}
\end{figure}
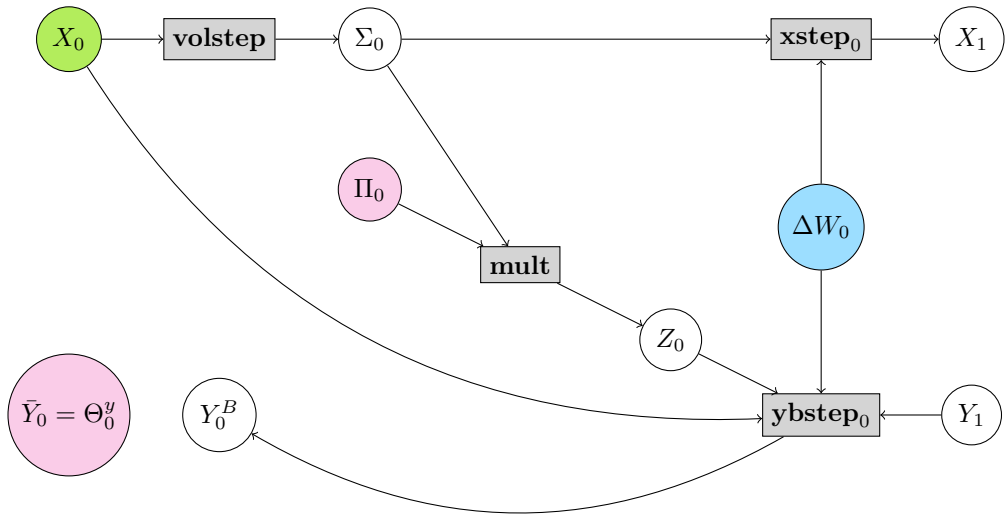

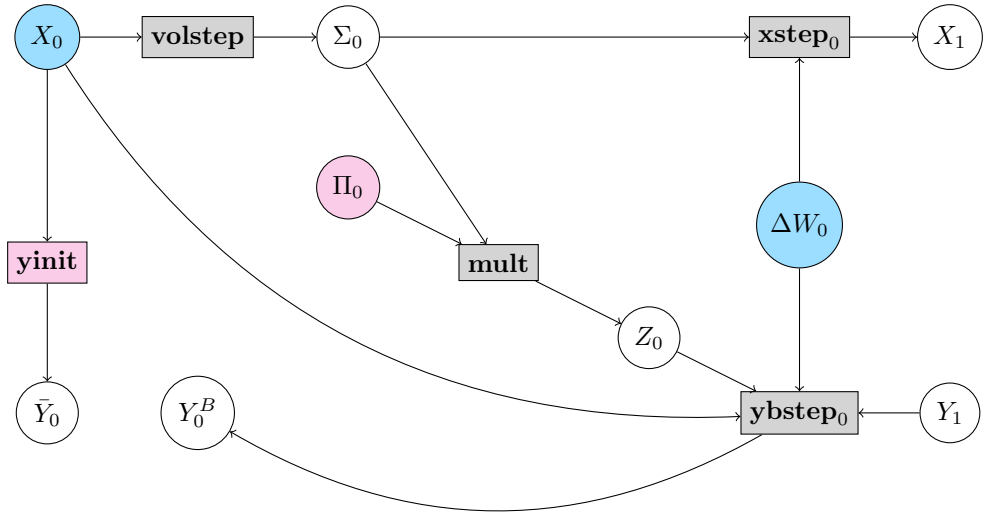
\begin{figure}[p]
\begin{center}
\begin{tikzpicture}

\node[draw,circle] (Yib) at (0,0) {$\bar{Y}_0$};
\node[draw,fill=classicrose] (yinit) at (0,2) {$\mathbf{yinit}$};
\node[draw,circle] (Yi) at (2,0) {$Y^B_0$};
\node[draw,circle,fill=columbiablue] (Xi) at (0,5) {$X_0$};
\node[draw,fill=lightgray] (volstep) at (2,5) {$\mathbf{volstep}$};
\node[draw,circle] (sigmai) at (4,5) {$\Sigma_0$};
\node[draw,circle,fill=classicrose] (pi) at (4,3) {$\Pi_0$};
\node[draw,fill=lightgray] (mult) at (6,2) {$\mathbf{mult}$};
\node[draw,circle] (zi) at (8,1) {$Z_0$};
\node[draw,fill=lightgray] (ystep) at (10,0) {$\mathbf{ybstep}_0$};
\node[draw,fill=columbiablue,circle] (dWi) at (10,2.5) {$\Delta W_0$};
\node[draw,fill=lightgray] (xstep) at (10,5) {$\mathbf{xstep}_0$};
\node[draw,circle] (Yi1) at (12,0) {$Y_{1}$};
\node[draw,circle] (Xi1) at (12,5) {$X_{1}$};


\draw[->] (ystep) to[bend left] (Yi);
\draw[->] (Xi) to (volstep);
\draw[->] (Xi) to[bend right] (ystep);
\draw[->] (volstep) to (sigmai);
\draw[->] (sigmai) to (xstep);
\draw[->] (sigmai) to (mult);
\draw[->] (pi) to (mult);
\draw[->] (mult) to (zi);
\draw[->] (zi) to (ystep);
\draw[->] (dWi) to (xstep);
\draw[->] (dWi) to (ystep);
\draw[->] (xstep) to (Xi1);
\draw[->] (Yi1) to (ystep);
\draw[->] (Xi) to (yinit);
\draw[->] (yinit) to (Yib);

\end{tikzpicture}
\end{center}
\caption{First time step in backward method when learning $Y_0$ as a network from
a random $X_0$}
\label{fig:FirstStepBkwdLearnY0Network}
\end{figure}

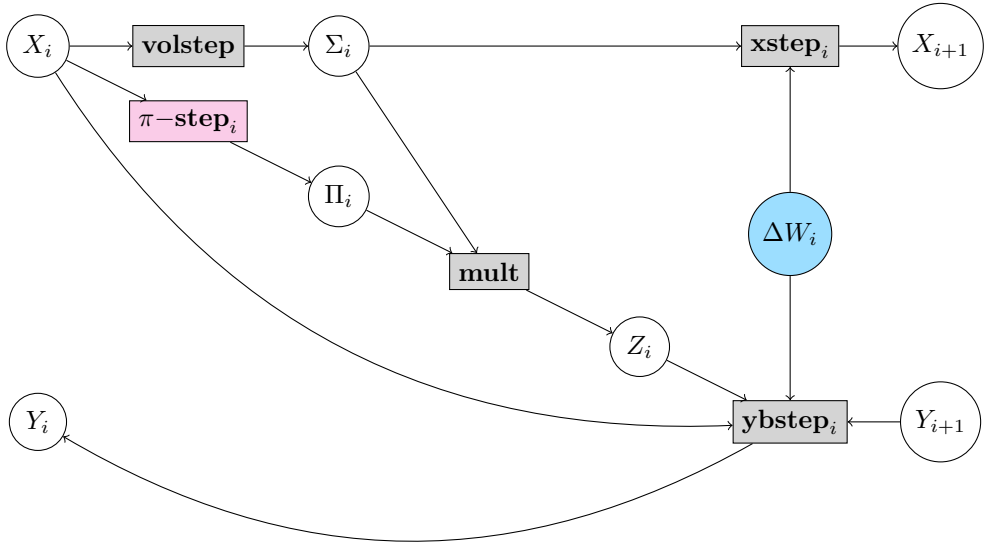
\begin{figure}[p]
\begin{center}
\begin{tikzpicture}

\node[draw,circle] (Yi) at (0,0) {$Y_i$};
\node[draw,circle] (Xi) at (0,5) {$X_i$};
\node[draw,fill=lightgray] (volstep) at (2,5) {$\mathbf{volstep}$};
\node[draw,fill=classicrose] (pstep) at (2,4) {$\pi\mathbf{-step}_i$};
\node[draw,circle] (sigmai) at (4,5) {$\Sigma_i$};
\node[draw,circle] (pi) at (4,3) {$\Pi_i$};
\node[draw,fill=lightgray] (mult) at (6,2) {$\mathbf{mult}$};
\node[draw,circle] (zi) at (8,1) {$Z_i$};
\node[draw,fill=lightgray] (ystep) at (10,0) {$\mathbf{ybstep}_i$};
\node[draw,fill=columbiablue,circle] (dWi) at (10,2.5) {$\Delta W_i$};
\node[draw,fill=lightgray] (xstep) at (10,5) {$\mathbf{xstep}_i$};
\node[draw,circle] (Yi1) at (12,0) {$Y_{i+1}$};
\node[draw,circle] (Xi1) at (12,5) {$X_{i+1}$};


\draw[->] (ystep) to[bend left] (Yi);
\draw[->] (Xi) to (volstep);
\draw[->] (Xi) to (pstep);
\draw[->] (Xi) to[bend right] (ystep);
\draw[->] (volstep) to (sigmai);
\draw[->] (pstep) to (pi);
\draw[->] (sigmai) to (xstep);
\draw[->] (sigmai) to (mult);
\draw[->] (pi) to (mult);
\draw[->] (mult) to (zi);
\draw[->] (zi) to (ystep);
\draw[->] (dWi) to (xstep);
\draw[->] (dWi) to (ystep);
\draw[->] (xstep) to (Xi1);
\draw[->] (Yi1) to (ystep);

\end{tikzpicture}
\end{center}
\caption{General step in backward method \label{fig:GeneralStepBkwd}}
\end{figure}

\begin{figure}
\begin{center}
\begin{tikzpicture}

\node[draw,circle] (Yi) at (0,0) {$Y_{N-1}$};
\node[draw,circle] (Xi) at (0,5) {$X_{N-1}$};
\node[draw,fill=lightgray] (volstep) at (2,5) {$\mathbf{volstep}$};
\node[draw,fill=classicrose] (pstep) at (2,4) {$\pi\mathbf{-step}_{N-1}$};
\node[draw,circle] (sigmai) at (4,5) {$\Sigma_{N-1}$};
\node[draw,circle] (pi) at (4,3) {$\Pi_{N-1}$};
\node[draw,fill=lightgray] (mult) at (6,2) {$\mathbf{mult}$};
\node[draw,circle] (zi) at (8,1) {$Z_{N-1}$};
\node[draw,fill=lightgray] (ystep) at (10,0) {$\mathbf{ybstep}_{N-1}$};
\node[draw,fill=columbiablue,circle] (dWi) at (10,2.5) {$\Delta W_{N-1}$};
\node[draw,fill=lightgray] (xstep) at (10,5) {$\mathbf{xstep}_{N-1}$};
\node[draw,circle] (Yi1) at (12,0) {$Y_{N}$};
\node[draw,fill=lightgray] (payoff) at (12,2.5) {$\mathbf{payoff}$};
\node[draw,circle] (Xi1) at (12,5) {$X_{N}$};


\draw[->] (ystep) to[bend left] (Yi);
\draw[->] (Xi) to (volstep);
\draw[->] (Xi) to (pstep);
\draw[->] (Xi) to[bend right] (ystep);
\draw[->] (volstep) to (sigmai);
\draw[->] (pstep) to (pi);
\draw[->] (sigmai) to (xstep);
\draw[->] (sigmai) to (mult);
\draw[->] (pi) to (mult);
\draw[->] (mult) to (zi);
\draw[->] (zi) to (ystep);
\draw[->] (dWi) to (xstep);
\draw[->] (dWi) to (ystep);
\draw[->] (xstep) to (Xi1);
\draw[->] (Yi1) to (ystep);
\draw[->] (Xi1) to (payoff);
\draw[->] (payoff) to (Yi1);

\end{tikzpicture}
\end{center}
\caption{Step for last time step in backward method \label{fig:laststep}} 
\end{figure}
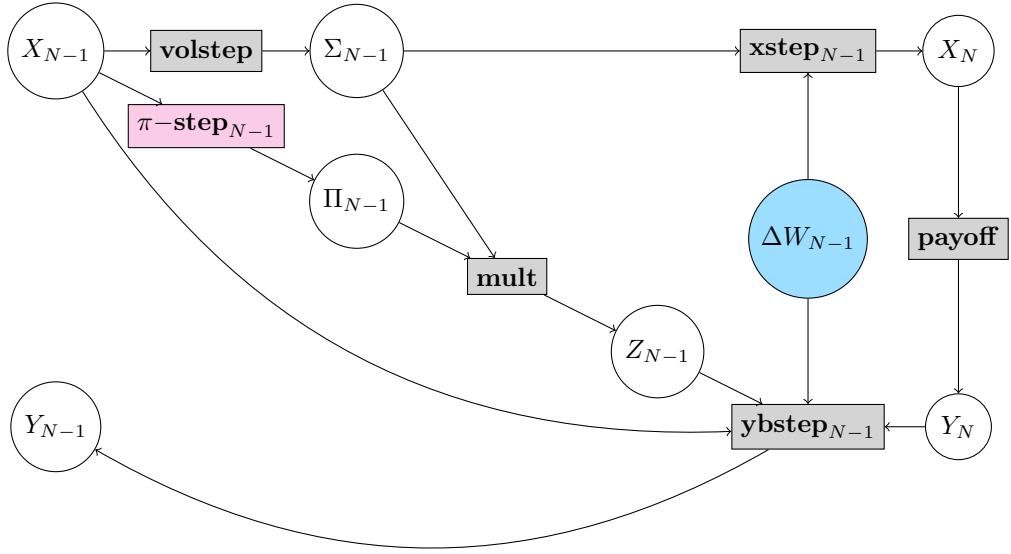

\section{Results}

We present results on two financial derivatives treated in the literature
so that we can compare our results more easily with those of others.
The two financial derivatives are a call combination (long a call on the maximum
across assets with strike 120.0 and short two calls on the maximum with strike
150.0, with maturity 0.5 years) as in E, Han and Jentzen \cite{weinan2017deep} and
a straddle on the maximum (both long and short a straddle with strike 100.0 with maturity 1 year)
as in Forsyth and Labahn \cite{forsyth2007numerical}.

For ease of visualization, testing, and presentation, we present results for the
one-dimensional case (which is also the case treated in Forsyth and Labahn
\cite{forsyth2007numerical}).

\subsection{Call Combination}

For the E, Han, and Jentzen example, we picked $\sigma=0.2$,
$\mu=0.06$, $r_l=0.04$, and $r_b=0.06$. We used 50 time steps.

For the call combination example, for the fixed $X_0$ case, we picked
$X_0=120.0$. For the random/varying $X_0$ case, we picked a uniform
distribution in $[70,170]$. We used a batch-size of 512, pre-scaling, two hidden
layers with $dim+10$ (dim is the dimension of PDE we are trying to solve) neurons and activitation functions ELU for the first two
layers and then identity on the output layer.

We first show results for the fixed $X_0$ case for the call combination. 

\begin{figure}[p]
\begin{subfigure}{.5\textwidth}
  \centering
  \includegraphics[width=.9\linewidth]{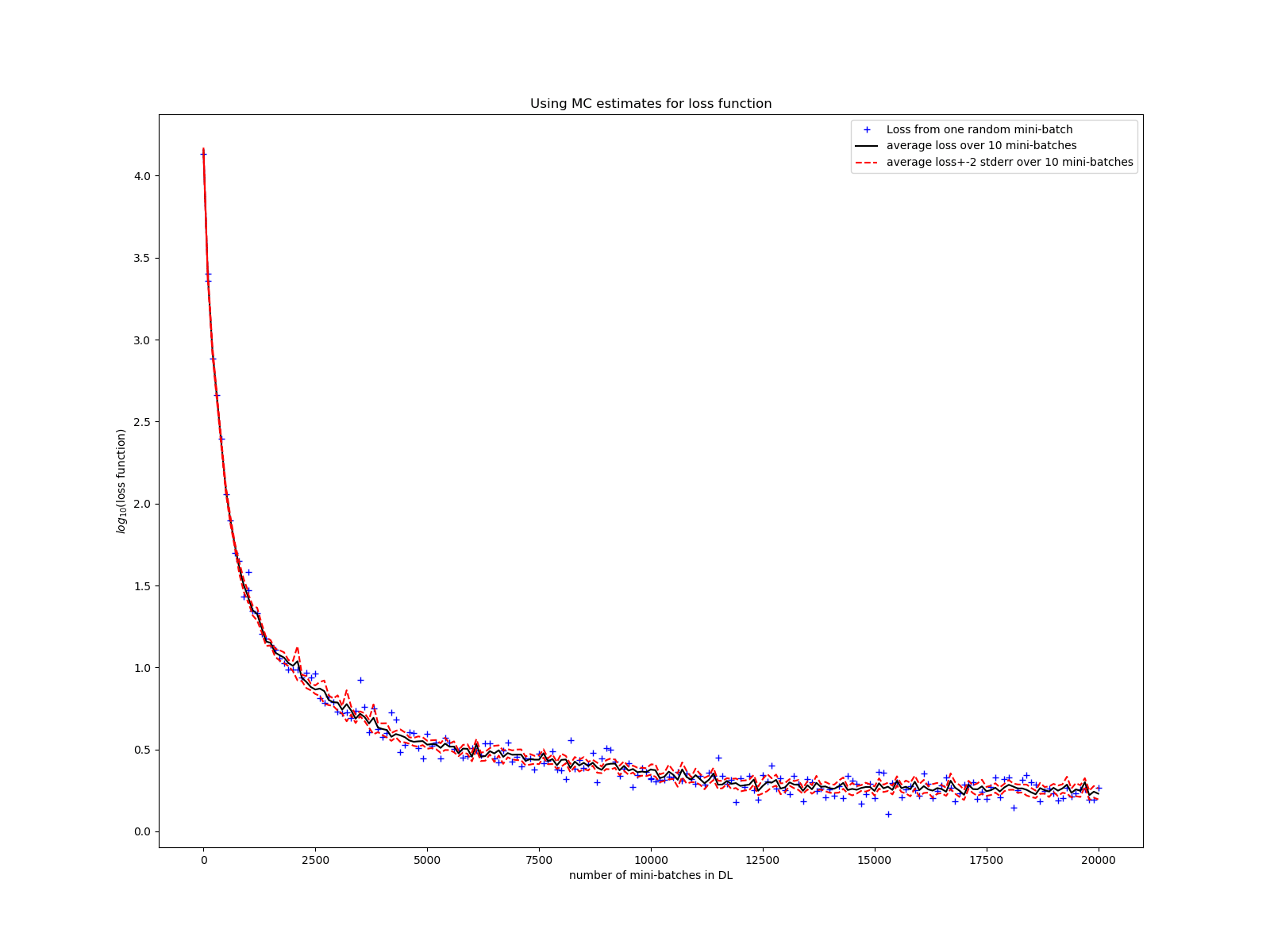}  
  \caption{Batch variance method - exact}
\end{subfigure}
\begin{subfigure}{.5\textwidth}
  \centering
  \includegraphics[width=.9\linewidth]{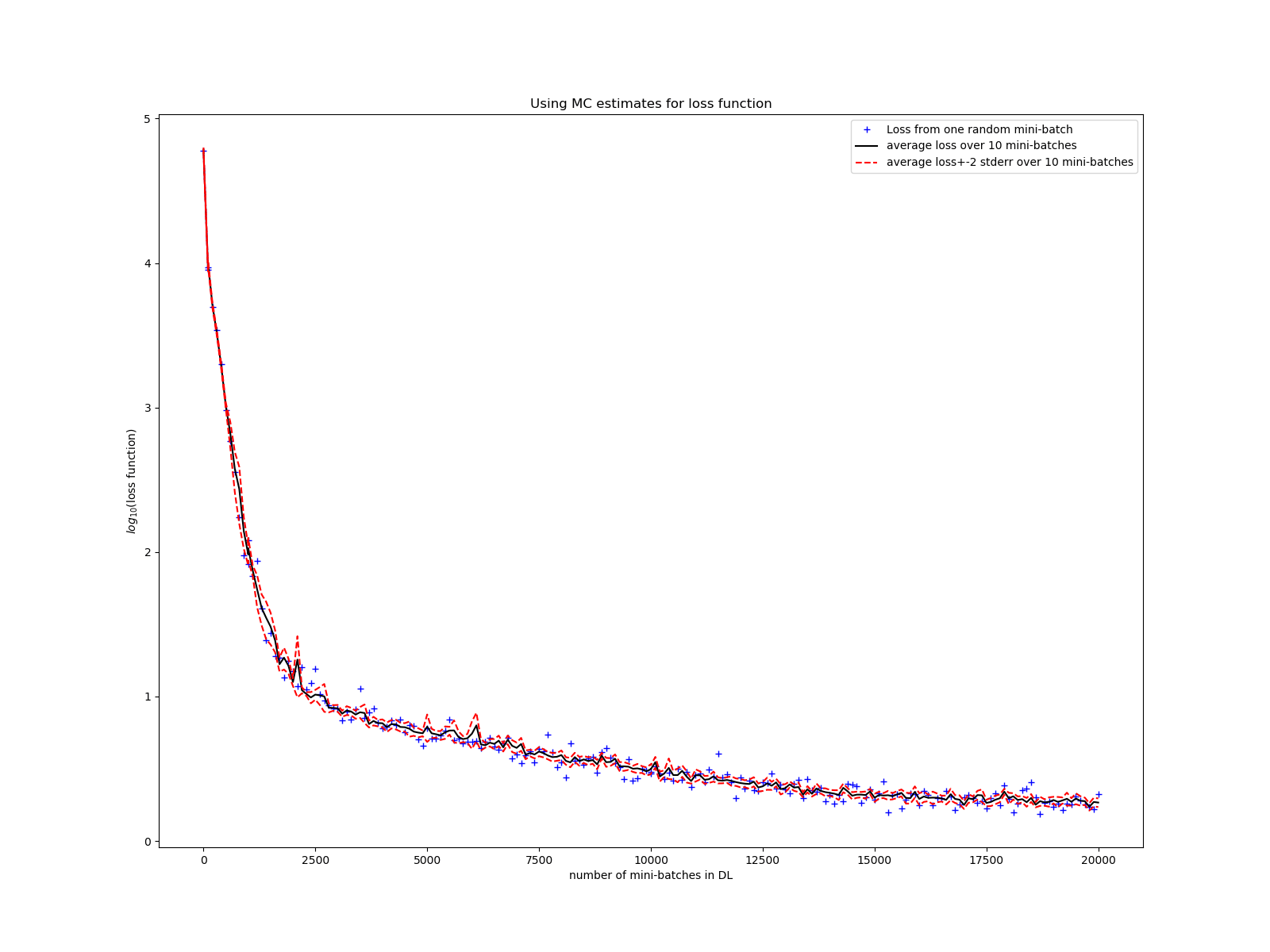}  
  \caption{Batch variance method - Taylor}
\end{subfigure}
\newline
\begin{subfigure}{.5\textwidth}
  \centering
  \includegraphics[width=.9\linewidth]{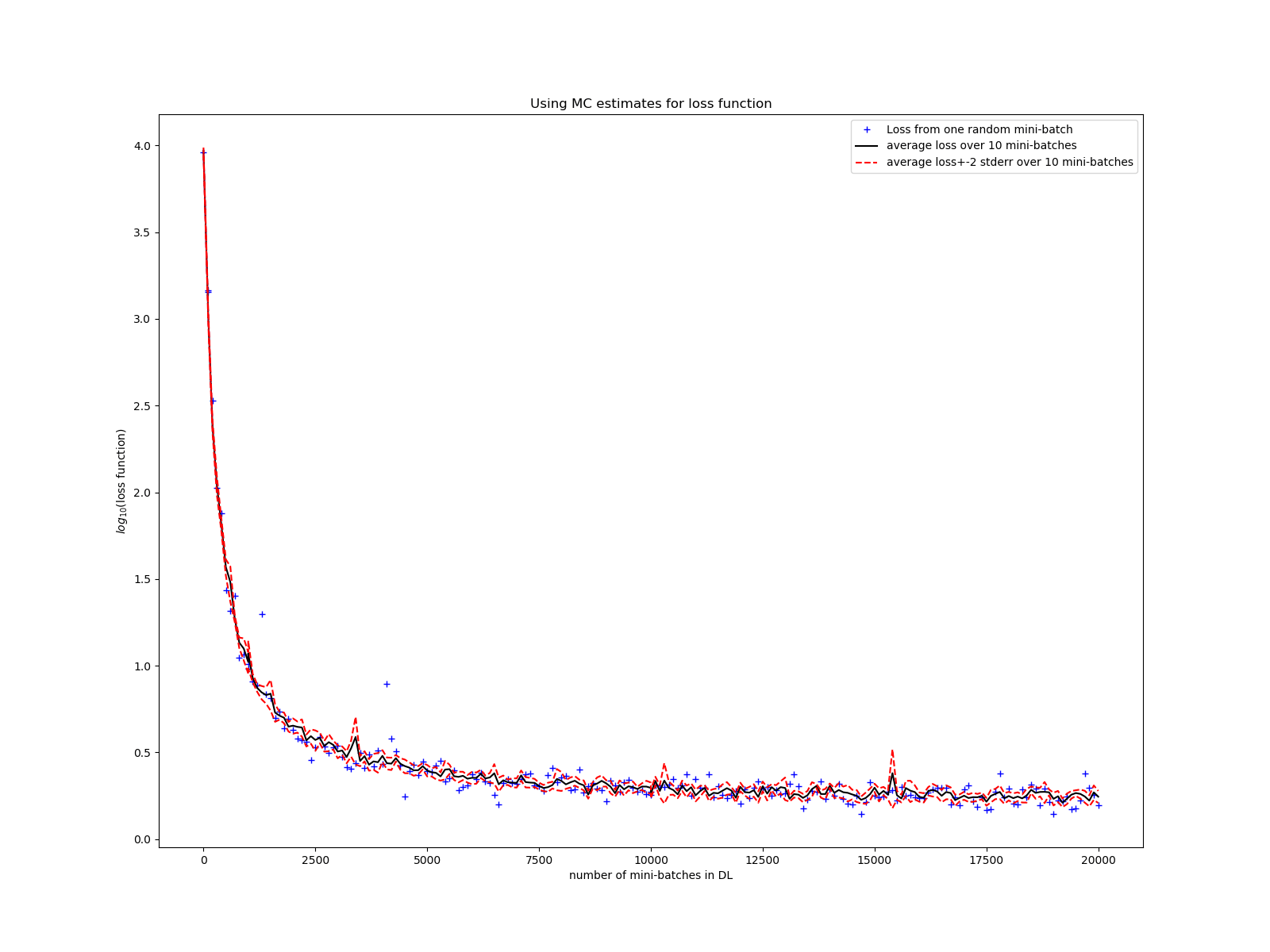}  
  \caption{Method with $Y_0$ as a parameter - exact}
\end{subfigure}
\begin{subfigure}{.5\textwidth}
  \centering
  \includegraphics[width=.9\linewidth]{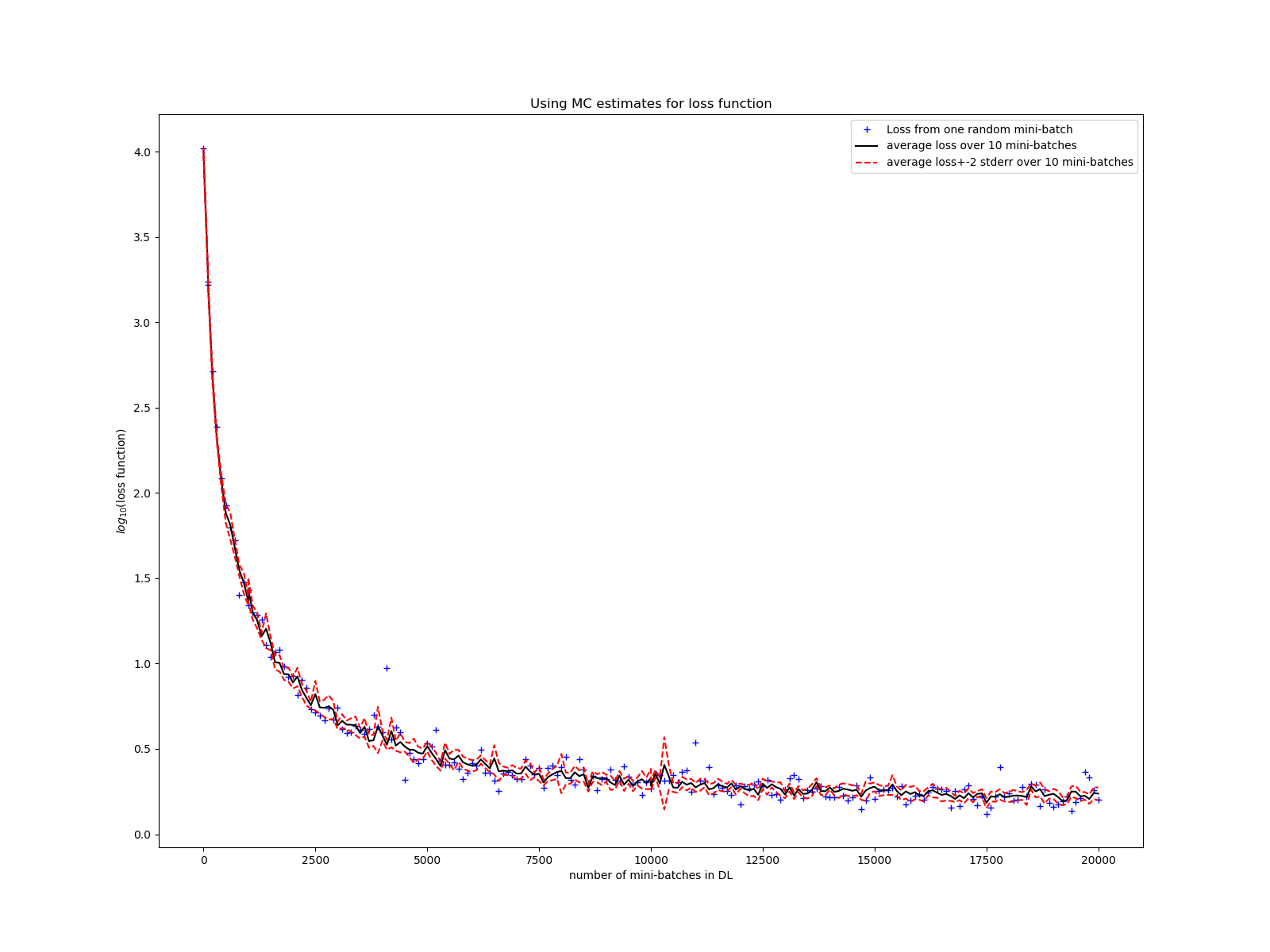}  
  \caption{Method with $Y_0$ as a parameter - Taylor}
\end{subfigure}
\caption{Loss function over 20000 mini-batches for different backward methods (fixed $X_0$) for the call combination example} \label{fig:CallComboFixedX0losses}
\end{figure}

Figure \ref{fig:CallComboFixedX0losses} shows how the loss function values
behave over the number of minibatches. 

\begin{figure}[p]
\begin{subfigure}{.5\textwidth}
  \centering
  \includegraphics[width=.9\linewidth]{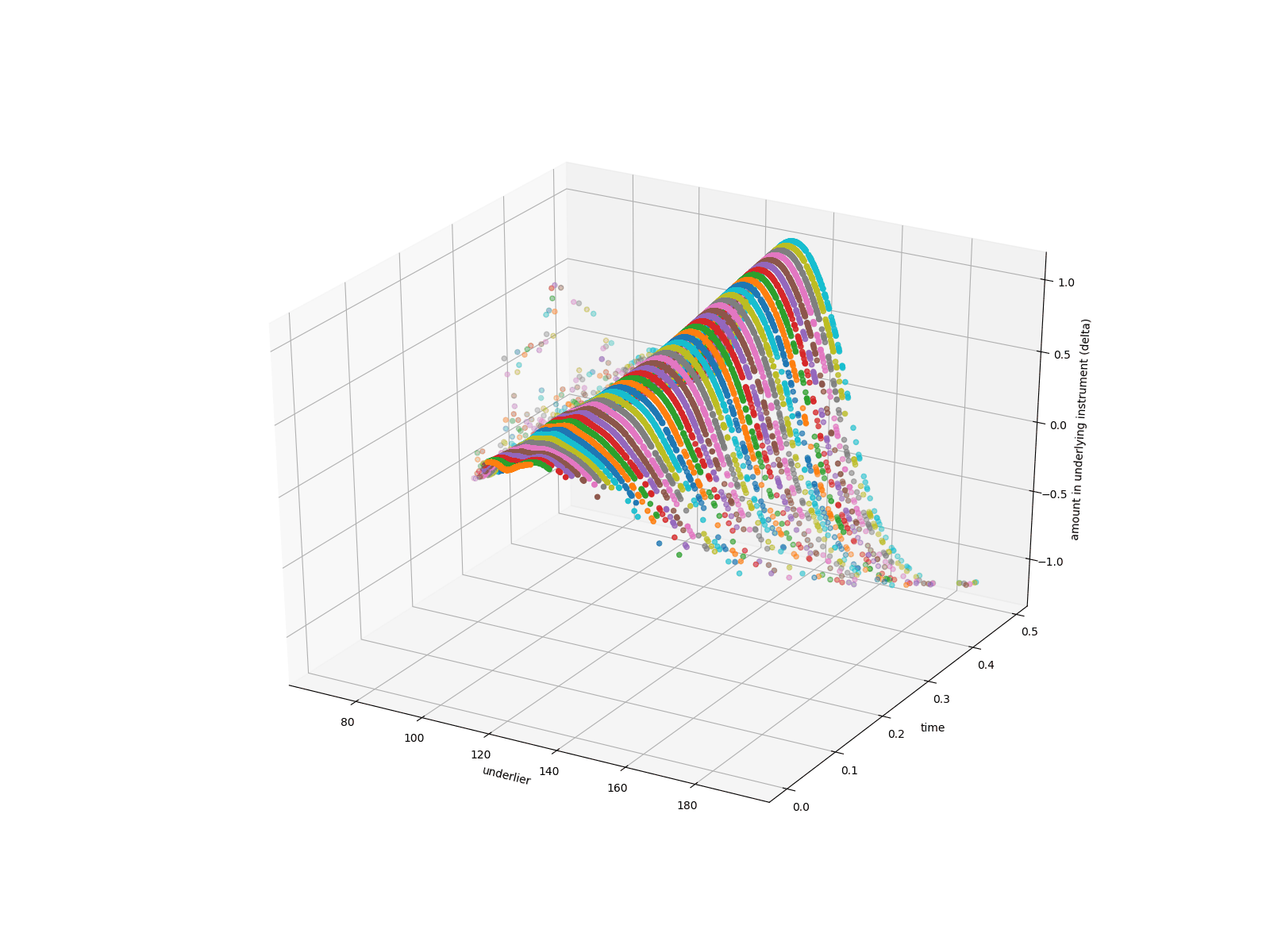}  
  \caption{Batch variance - exact}
\end{subfigure}
\begin{subfigure}{.5\textwidth}
  \centering
  \includegraphics[width=.9\linewidth]{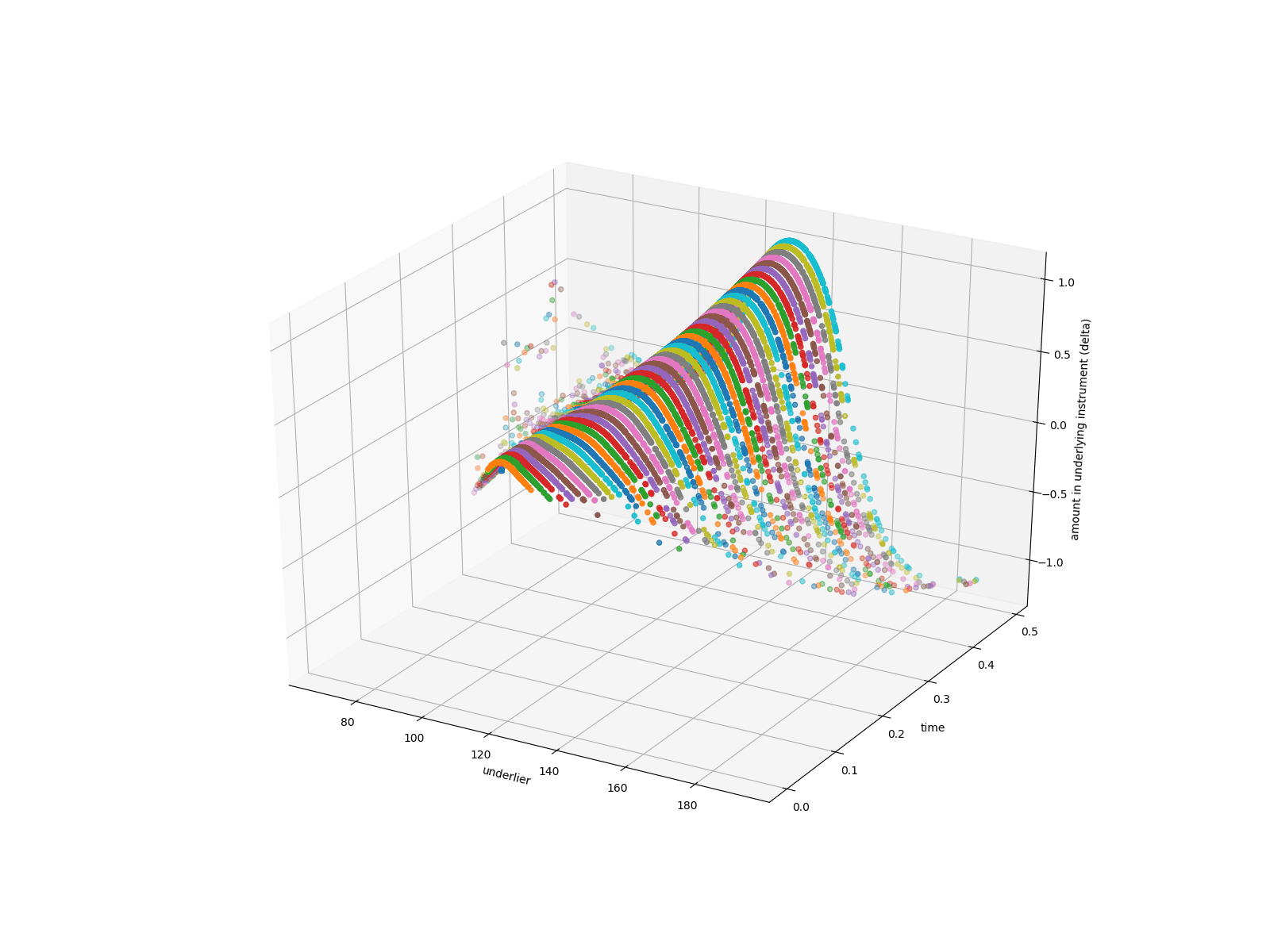}  
  \caption{Batch variance - Taylor}
\end{subfigure}
\newline
\begin{subfigure}{.5\textwidth}
  \centering
  \includegraphics[width=.9\linewidth]{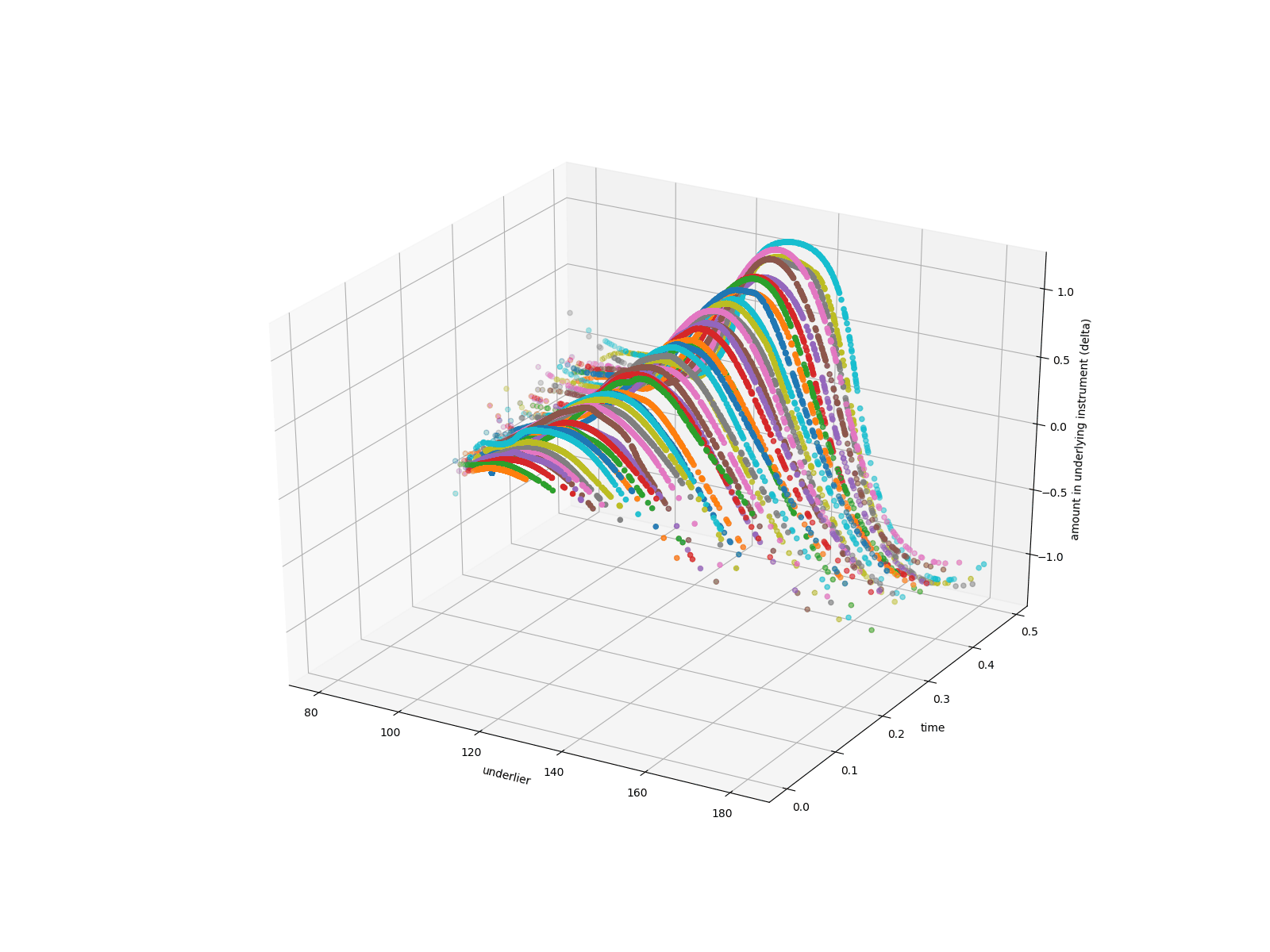}  
  \caption{Method with $Y_0$ as a parameter - exact}
\end{subfigure}
\begin{subfigure}{.5\textwidth}
  \centering
  \includegraphics[width=.9\linewidth]{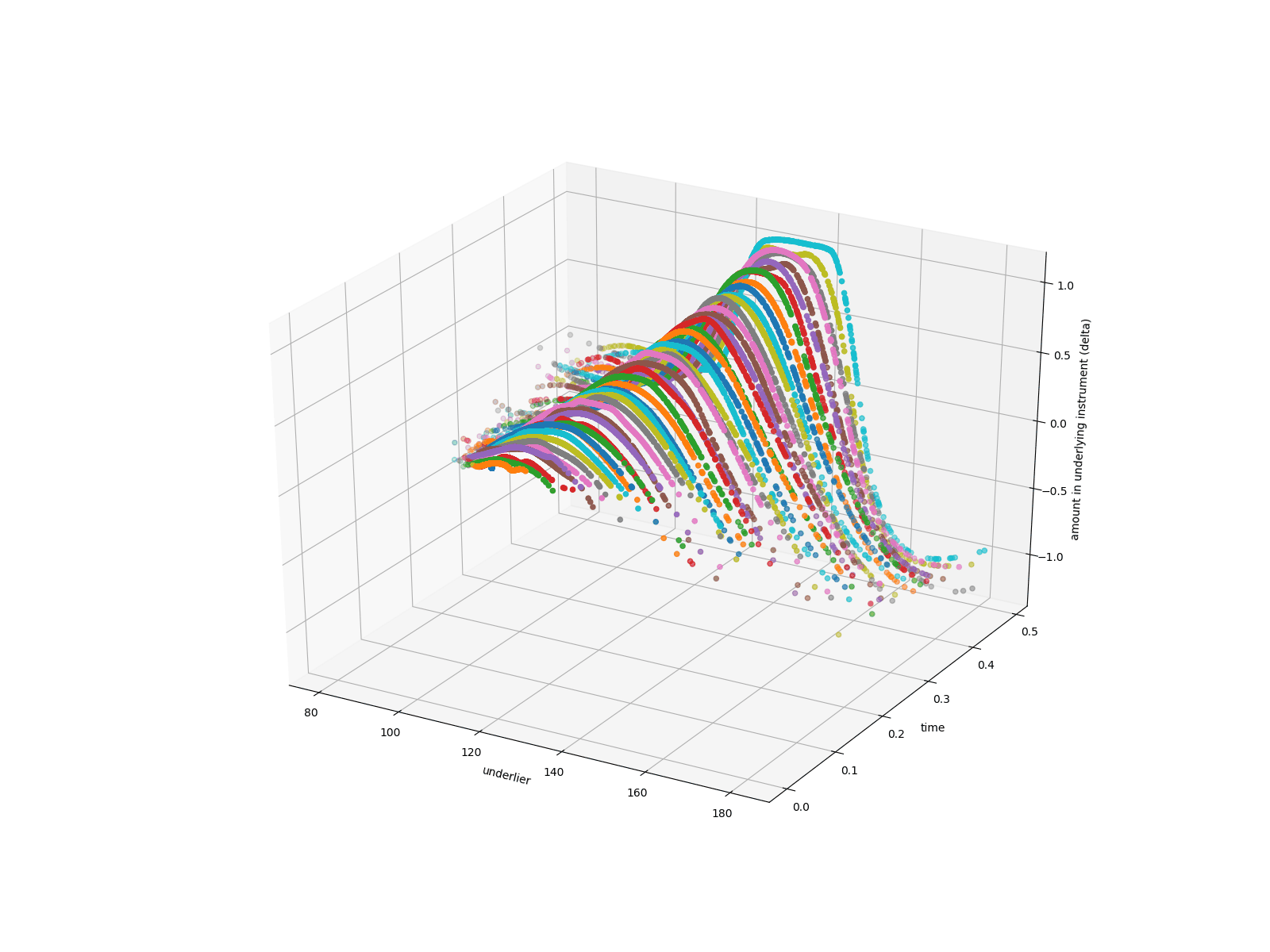}  
  \caption{Method with $Y_0$ as a parameter - Taylor}
\end{subfigure}
\caption{Strategy DNN (delta) outputs at 20000 mini-batches for different backward methods (fixed $X_0$) for the call combination example} \label{fig:CallComboFixedX0Strategy}
\end{figure}

\begin{figure}[p]
\begin{subfigure}{.5\textwidth}
  \centering
  \includegraphics[width=.9\linewidth]{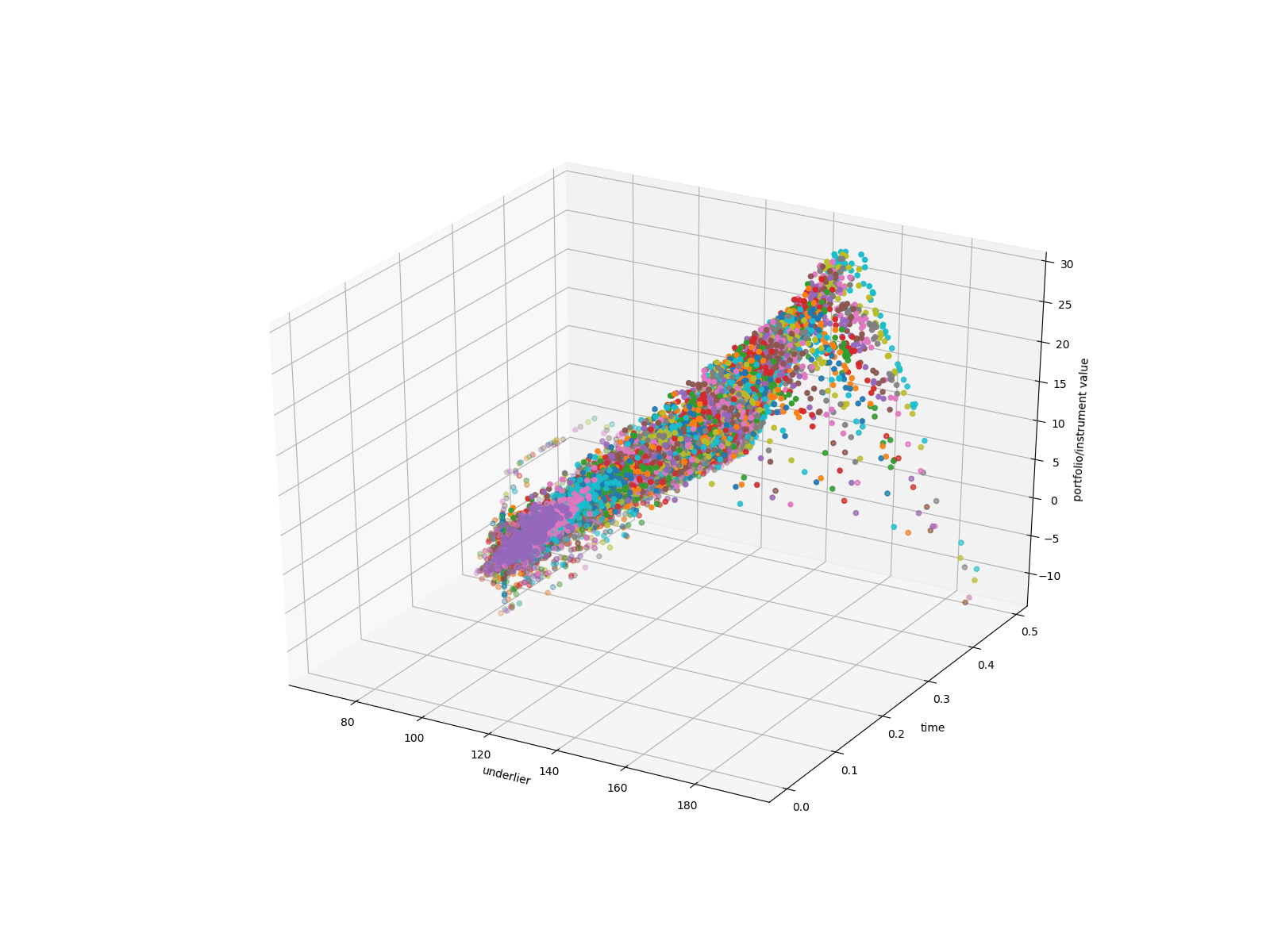}  
  \caption{Batch variance - exact}
\end{subfigure}
\begin{subfigure}{.5\textwidth}
  \centering
  \includegraphics[width=.9\linewidth]{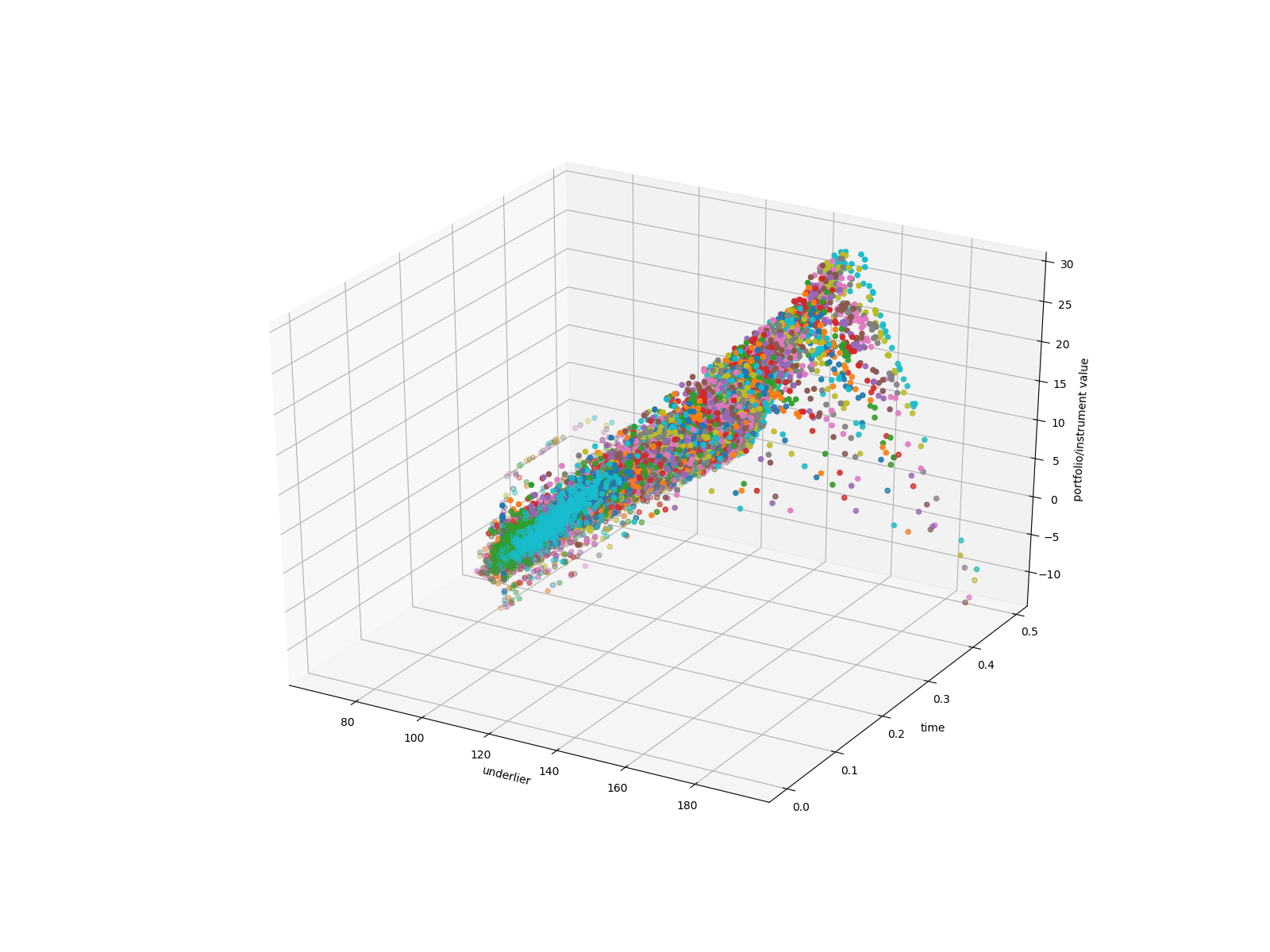}  
  \caption{Batch variance - Taylor}
\end{subfigure}
\newline
\begin{subfigure}{.5\textwidth}
  \centering
  \includegraphics[width=.9\linewidth]{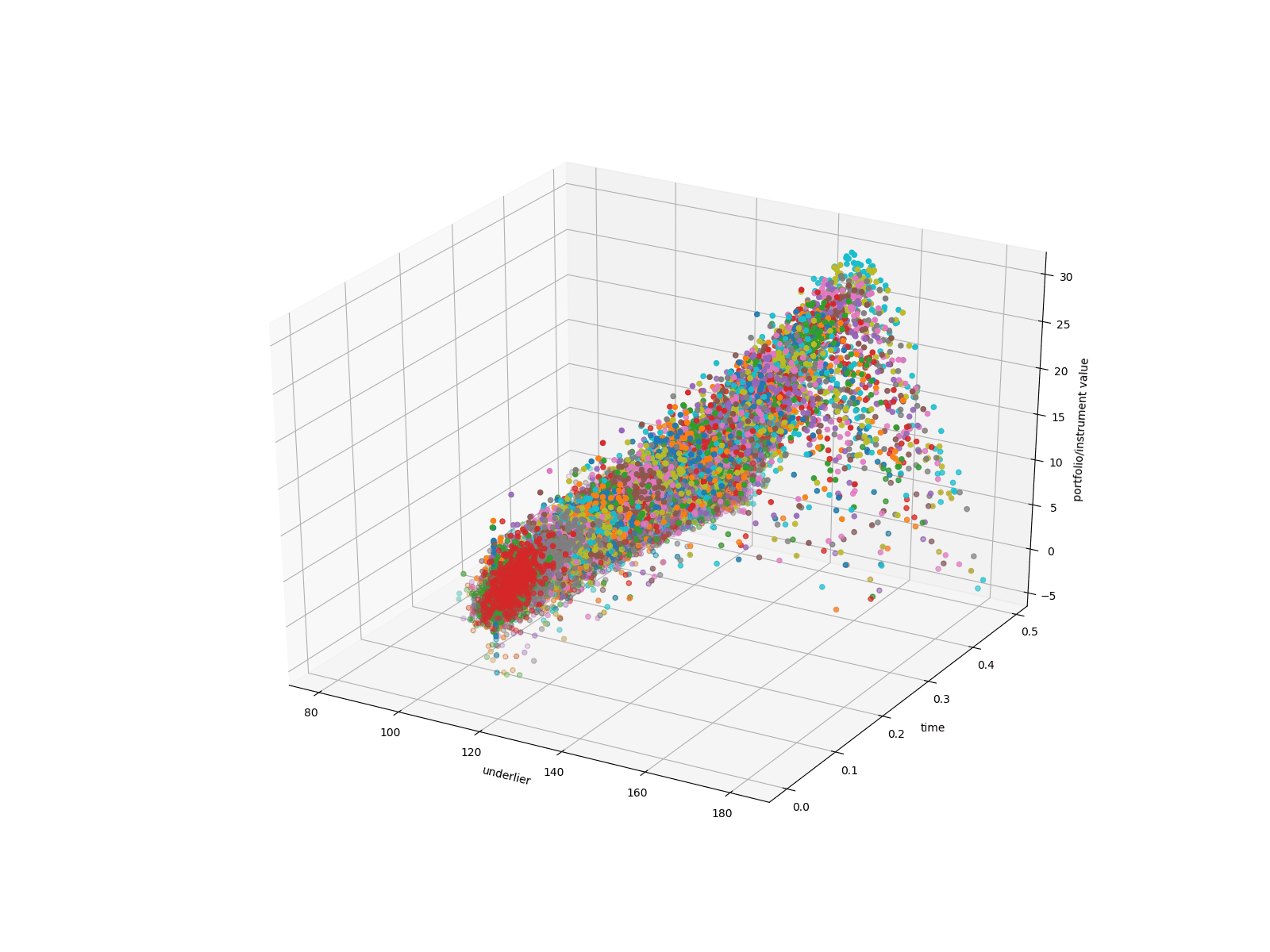}  
  \caption{Method with $Y_0$ as a parameter - exact}
\end{subfigure}
\begin{subfigure}{.5\textwidth}
  \centering
  \includegraphics[width=.9\linewidth]{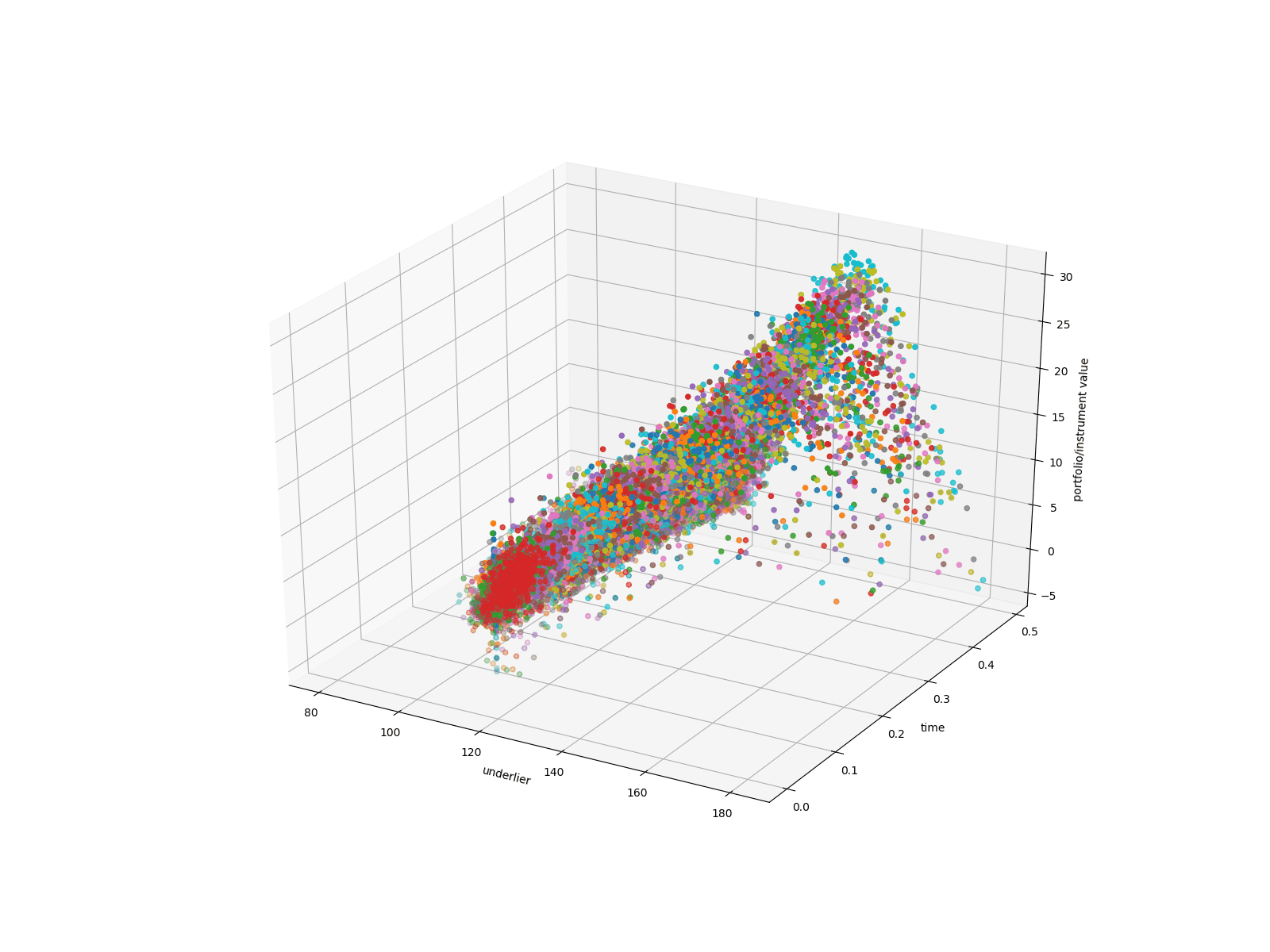}  
  \caption{Method with $Y_0$ as a parameter - Taylor}
\end{subfigure}
\caption{$Y$ path values at 20000 mini-batches for different backward methods (fixed $X_0$) for the call combination example} \label{fig:CallComboFixedX0YValues}
\end{figure}

Figures \ref{fig:CallComboFixedX0Strategy} and \ref{fig:CallComboFixedX0YValues}
show the trading strategy and the portfolio value $Y$.
The batch variance method uses shared parameters for the risky portfolio vector functions $\pi(t,x)$
while the initial parameter method uses separate DNNs with separate parameters for different times.

\begin{figure}[p]
\begin{subfigure}{.5\textwidth}
  \centering
  \includegraphics[width=.9\linewidth]{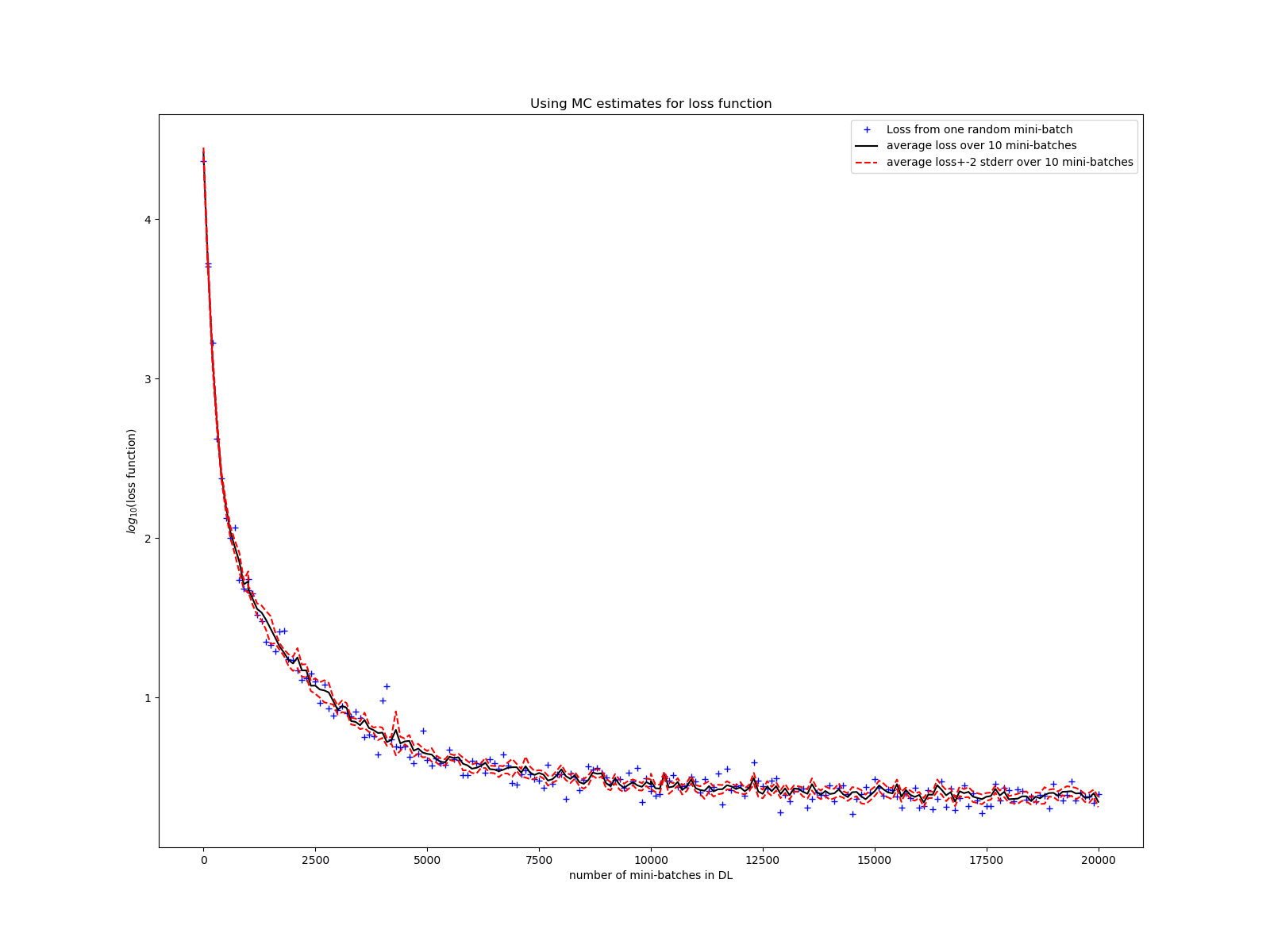}  
  \caption{Initial network - exact}
\end{subfigure}
\begin{subfigure}{.5\textwidth}
  \centering
  \includegraphics[width=.9\linewidth]{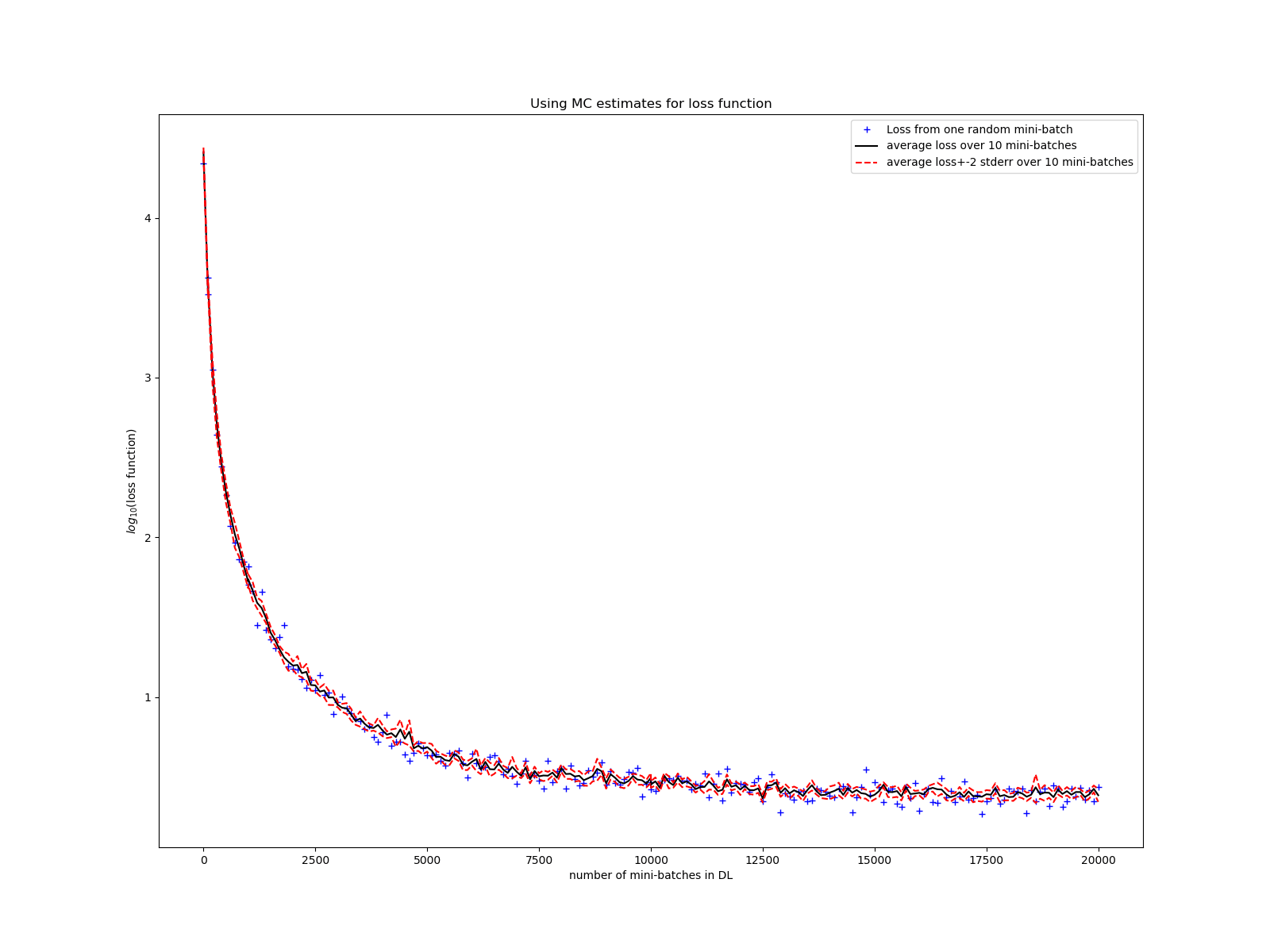}  
  \caption{Initial network - Taylor}
\end{subfigure}
\caption{Loss function over 20000 mini-batches for different backward methods (random $X_0$) for the call combination example} \label{fig:CallComboRandomX0losses}
\end{figure}

\begin{figure}[p]
\begin{subfigure}{.5\textwidth}
  \centering
  \includegraphics[width=.9\linewidth]{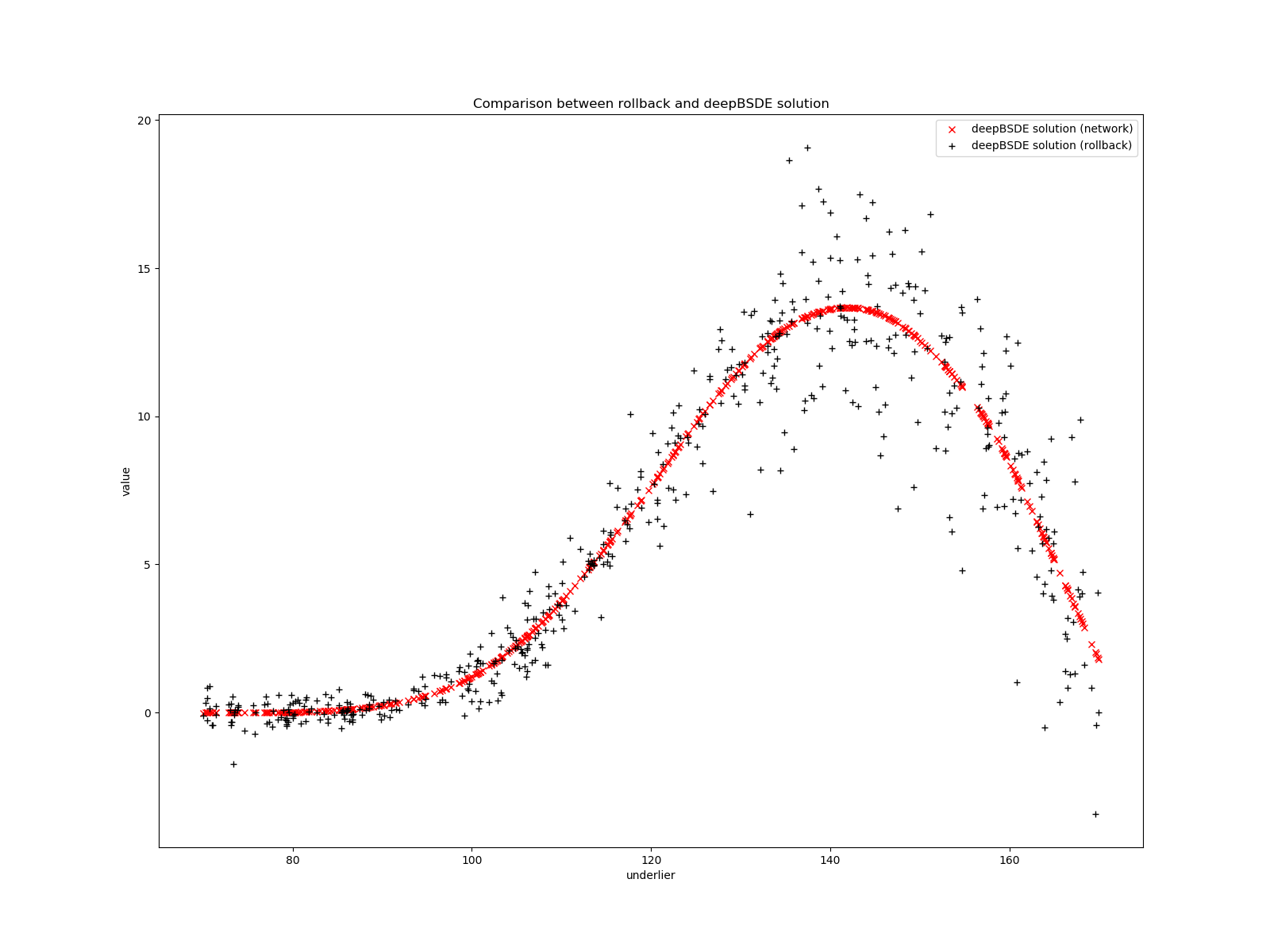}  
  \caption{$Y_0$ network vs roll-back - exact}
\end{subfigure}
\begin{subfigure}{.5\textwidth}
  \centering
  \includegraphics[width=.9\linewidth]{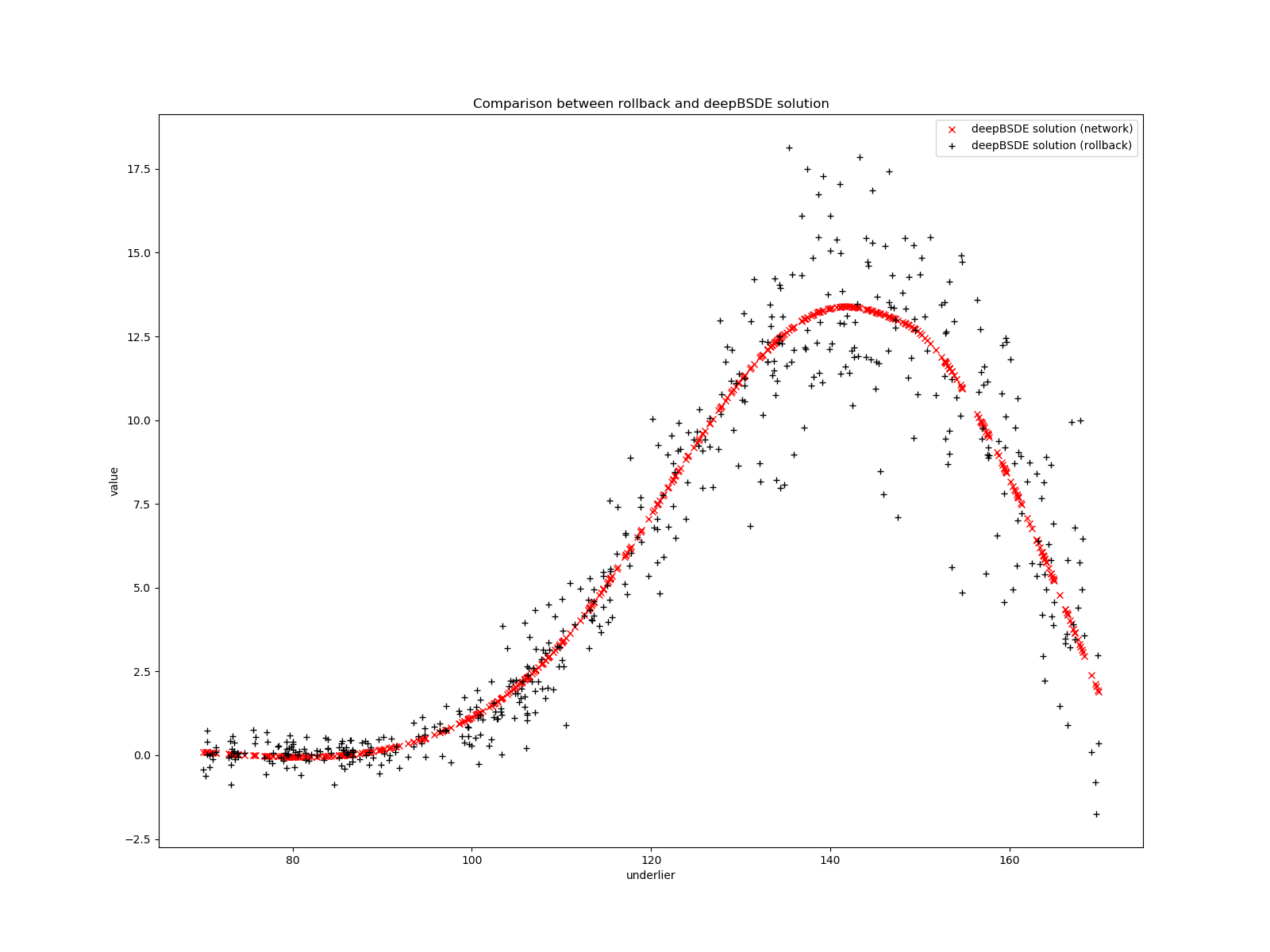}  
  \caption{$Y_0$ network vs roll-back - Taylor}
\end{subfigure}
\newline
\begin{subfigure}{.5\textwidth}
  \centering
  \includegraphics[width=.9\linewidth]{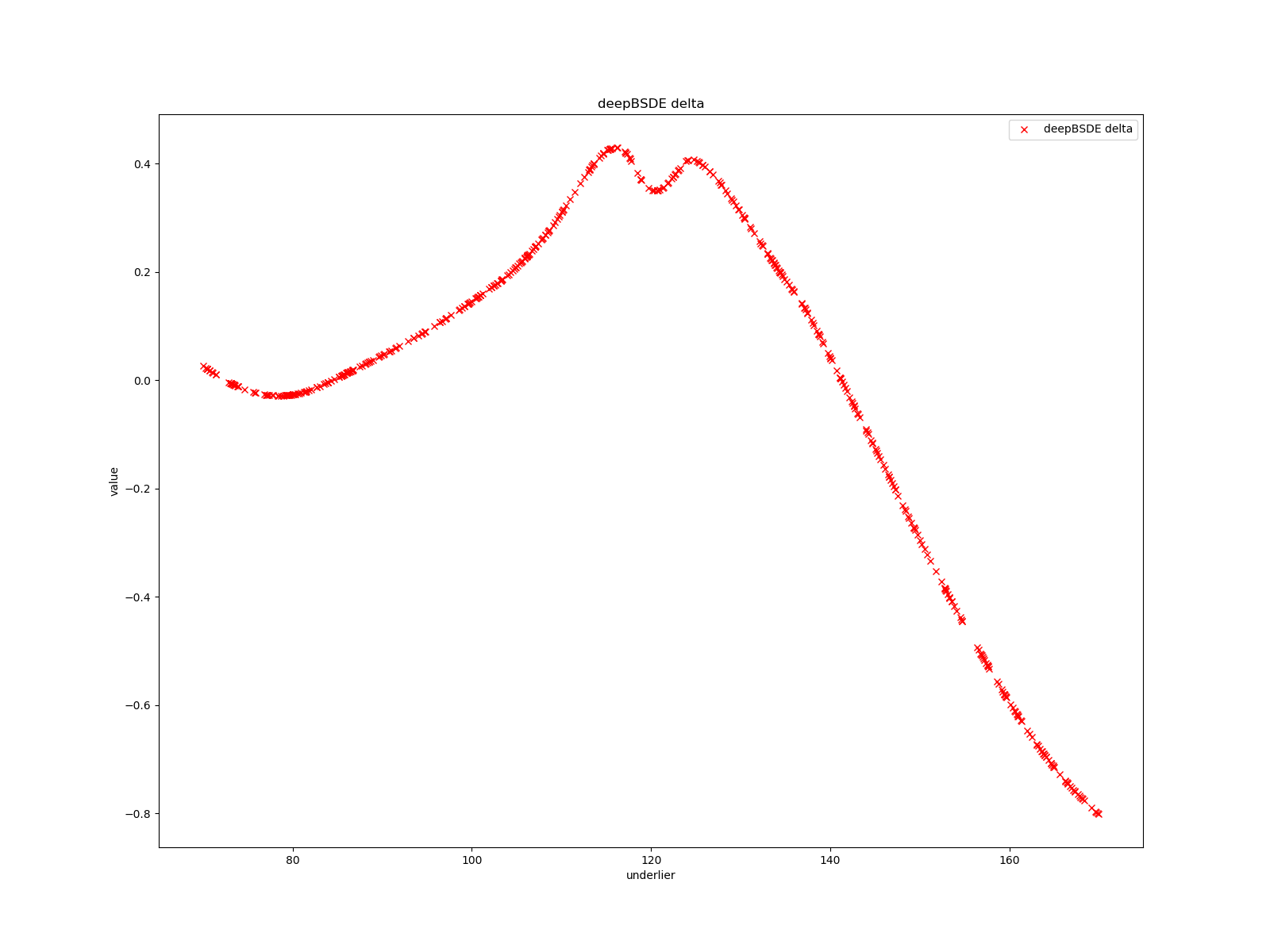}  
  \caption{$\pi_0$  network - exact}
\end{subfigure}
\begin{subfigure}{.5\textwidth}
  \centering
  \includegraphics[width=.9\linewidth]{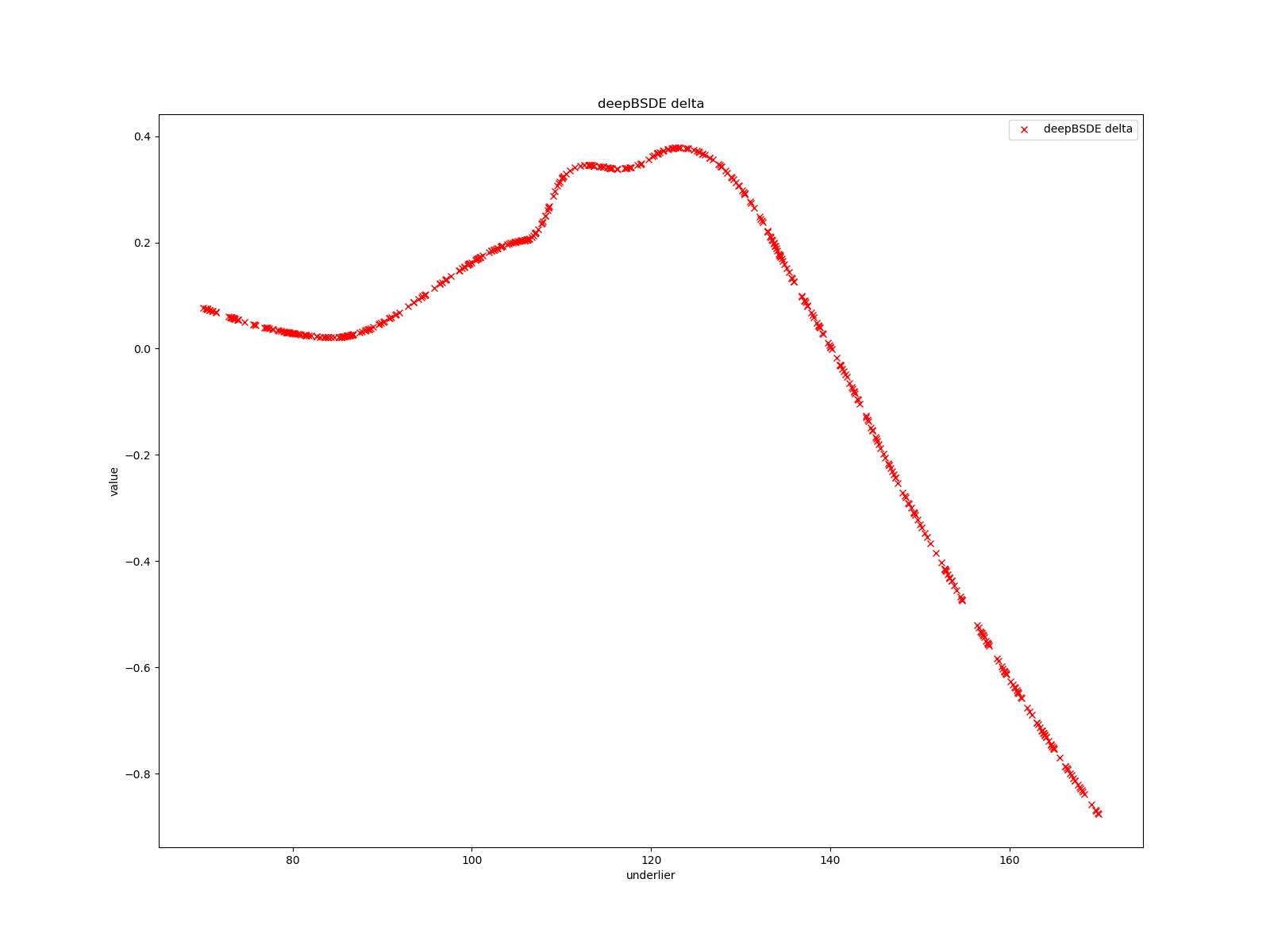}  
  \caption{$\pi_0$  network - Taylor}
\end{subfigure}
\caption{$Y_0$ initial network, roll-back, and $\pi_0$ initial network at 20000 mini-batches for 
different backward methods (random $X_0$) for the call combination example} \label{fig:CallComboRandomX0YCompareZCompare}
\end{figure}

\begin{figure}[p]
\begin{subfigure}{.5\textwidth}
  \centering
  \includegraphics[width=.9\linewidth]{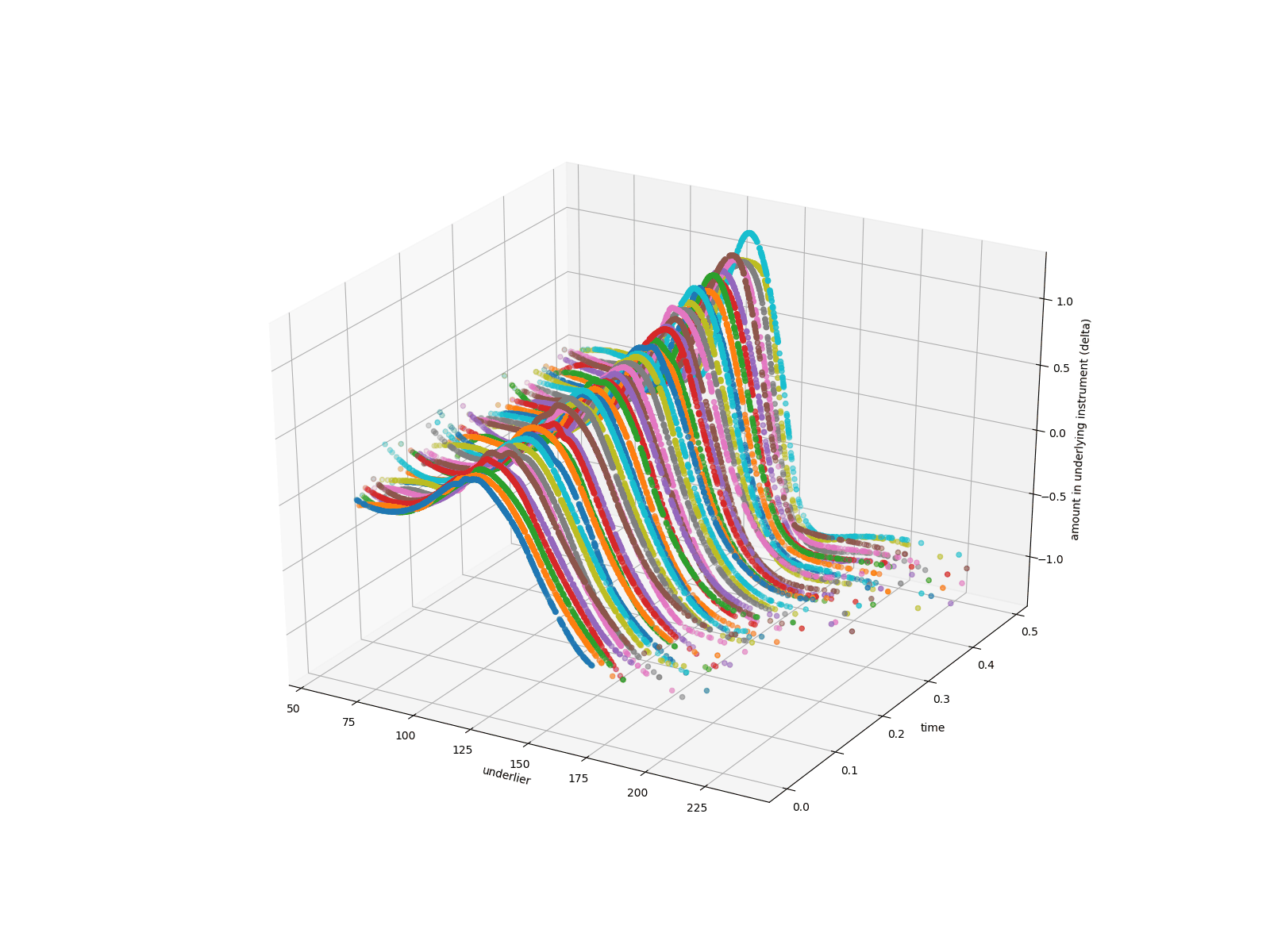}  
  \caption{Strategy - exact}
\end{subfigure}
\begin{subfigure}{.5\textwidth}
  \centering
  \includegraphics[width=.9\linewidth]{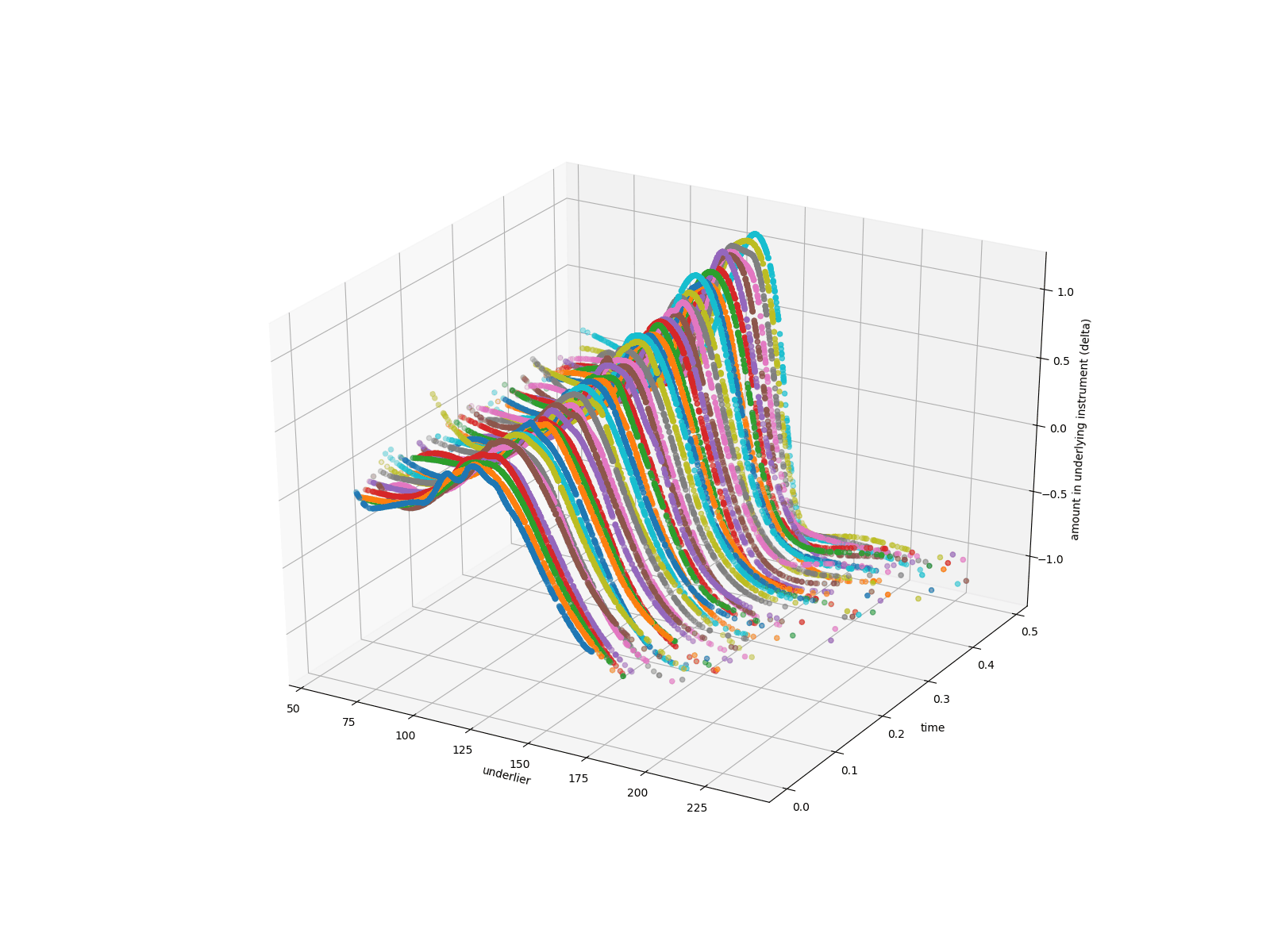}  
  \caption{Strategy - Taylor}
\end{subfigure}
\newline
\begin{subfigure}{.5\textwidth}
  \centering
  \includegraphics[width=.9\linewidth]{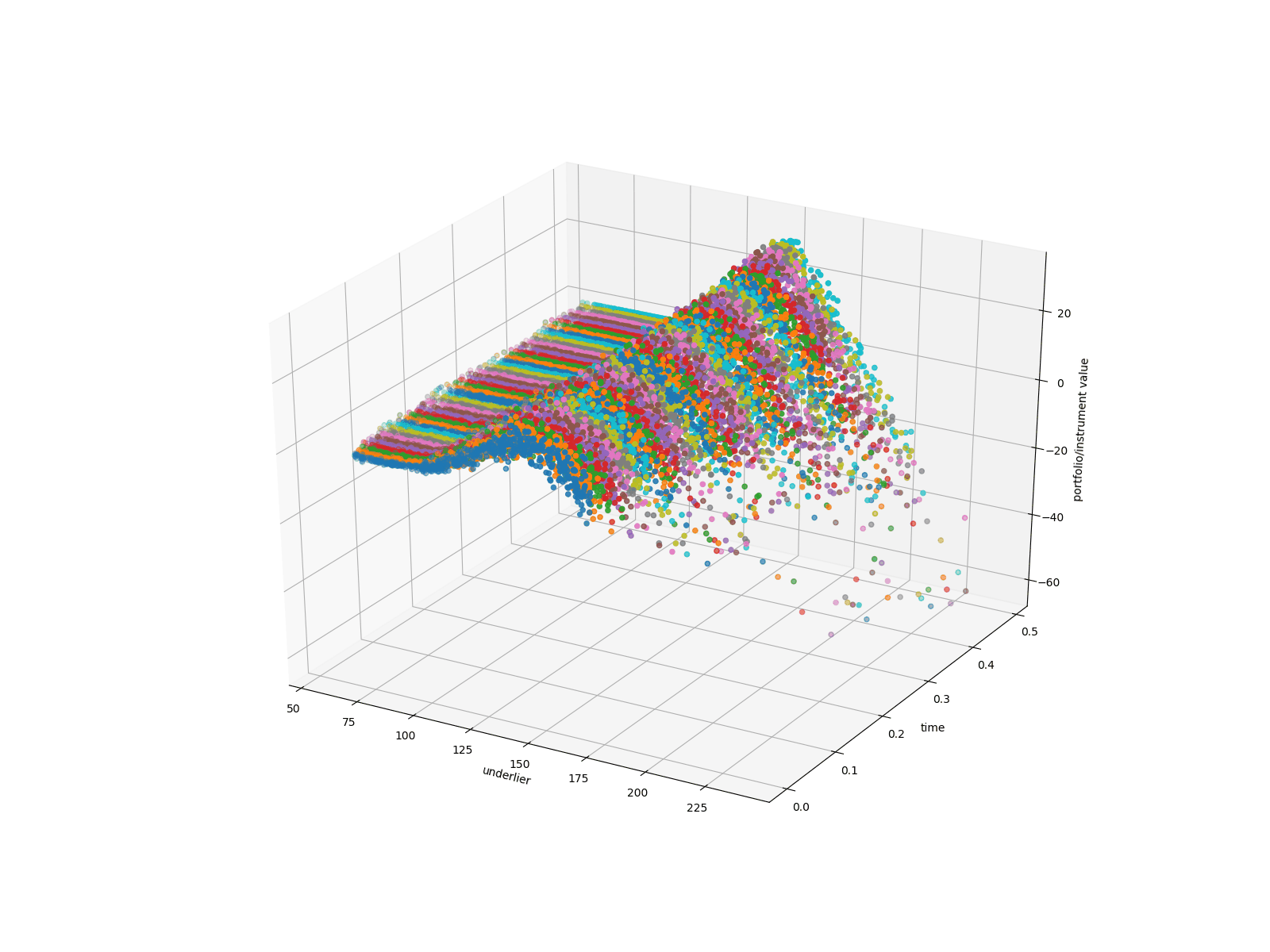}  
  \caption{$Y$ path values - exact}
\end{subfigure}
\begin{subfigure}{.5\textwidth}
  \centering
  \includegraphics[width=.9\linewidth]{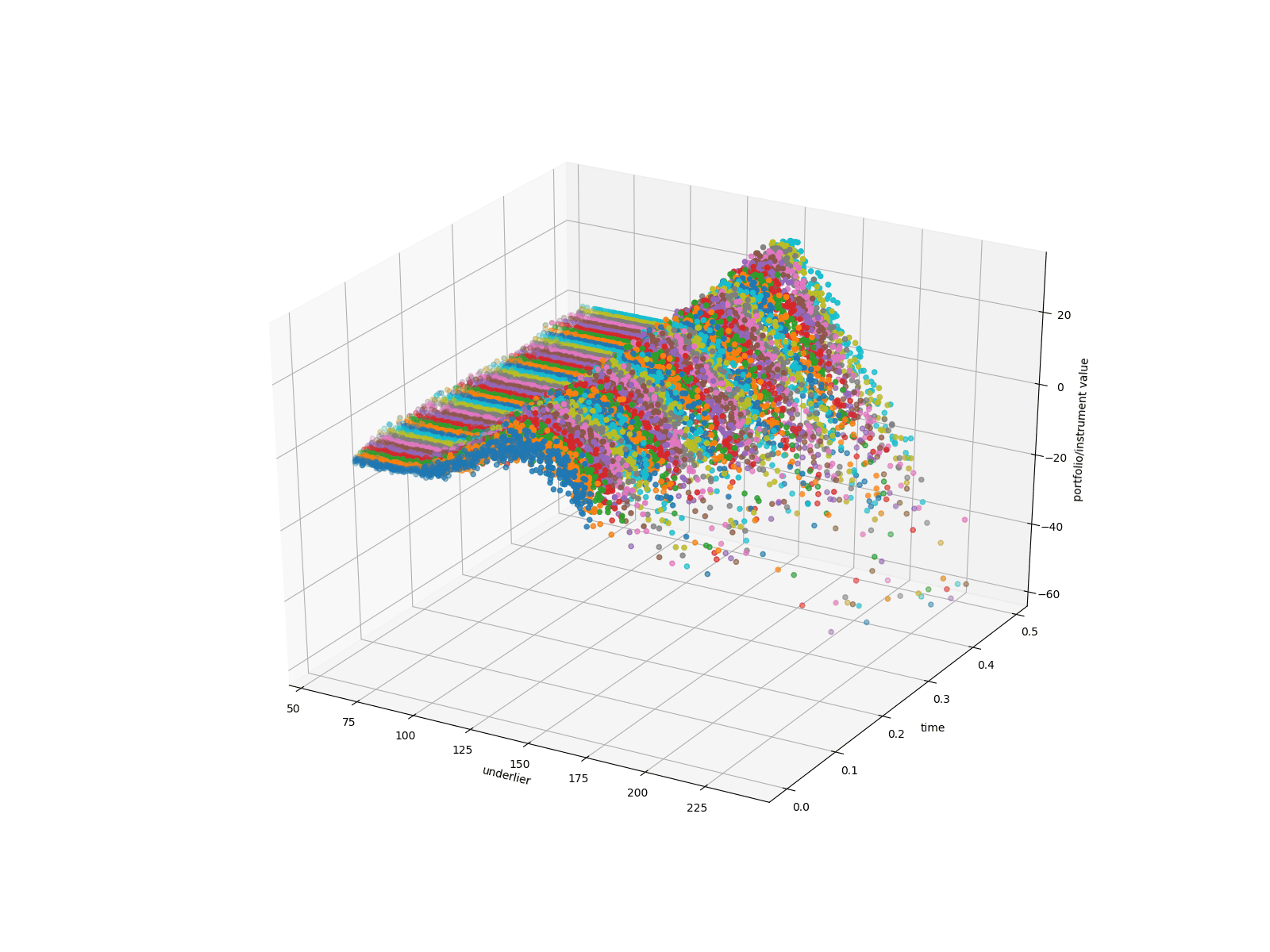}  
  \caption{$Y$ path values - Taylor}
\end{subfigure}
\caption{Strategies deltas and $Y$ path values at 20000 mini-batches for 
different backward methods (random $X_0$) for the call combination example} \label{fig:CallComoRandomX0StrategyYValues}
\end{figure}

We next show results for the random $X_0$ case for the call combination. Figure
\ref{fig:CallComboRandomX0losses} shows the loss functions. Figure
\ref{fig:CallComboRandomX0YCompareZCompare} shows initial $\mathsf{Yinit}$
network and rollback and the $Z_0$ network. Figure
\ref{fig:CallComoRandomX0StrategyYValues} shows the strategy DNN outputs and the
path $Y$ values. This example uses separate DNNs.

\subsection{Straddle}

Forsyth and Labahn picked the settings $\sigma=0.3$, $\mu=r_b=0.05$, and
$r_l=0.03$ \cite[Table 1 on page 28 in hjb.pdf]{forsyth2007numerical}. We used 100 time
steps (one of the numbers of time steps for which results are given in tables in
Forsyth and Labahn \cite{forsyth2007numerical}). The strike for the straddle is 100.0.

We first consider the fixed $X_0$ case. Like Forsyth and Labahn
\cite{forsyth2007numerical}, we pick initial spot $X_0$ to be 100.0 .

\begin{figure}[p]
\begin{subfigure}{\textwidth}
  \centering
  \includegraphics[width=.9\linewidth]{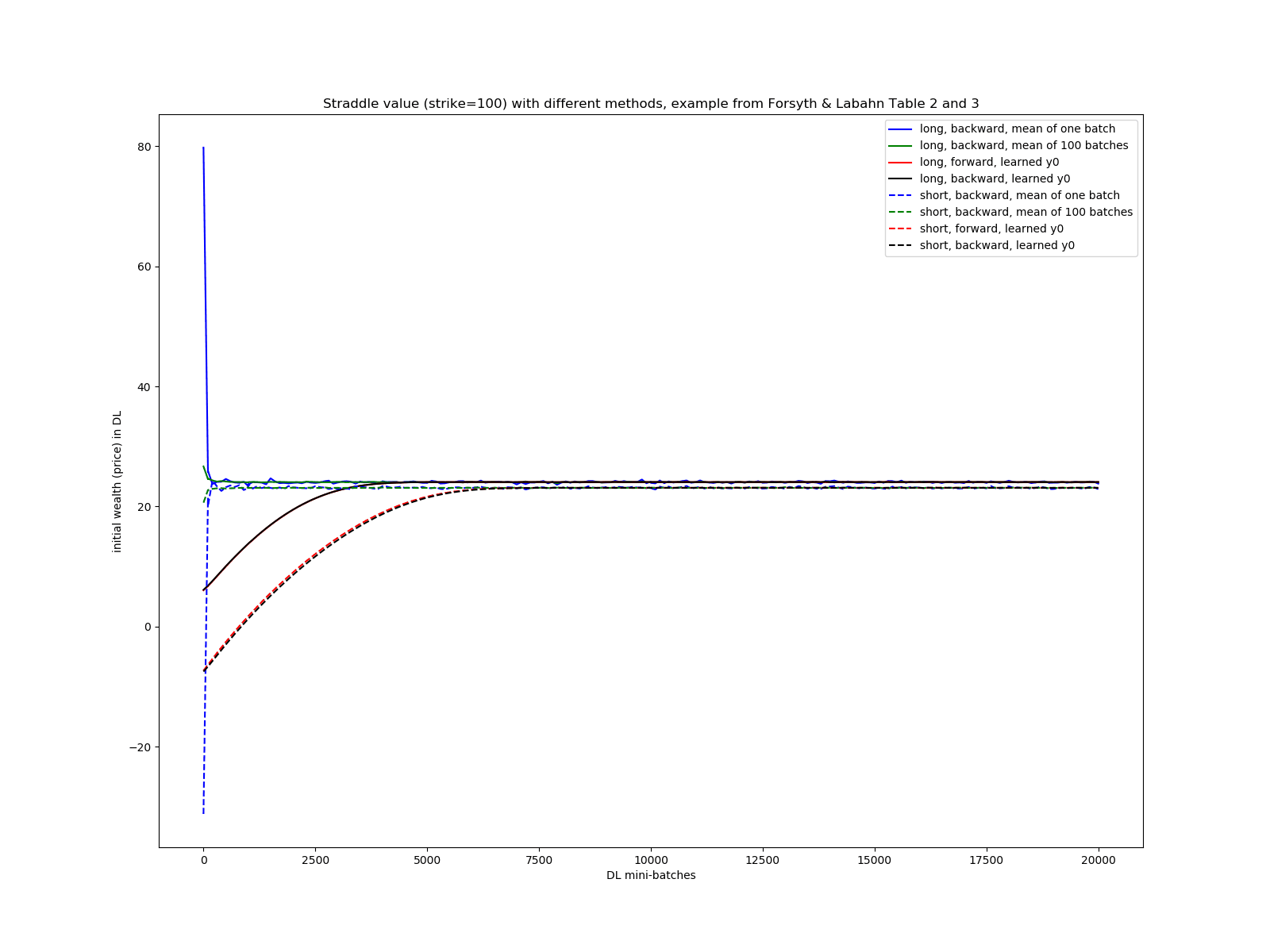}  
\end{subfigure}
\newline
\begin{subfigure}{\textwidth}
  \centering
  \includegraphics[width=.9\linewidth]{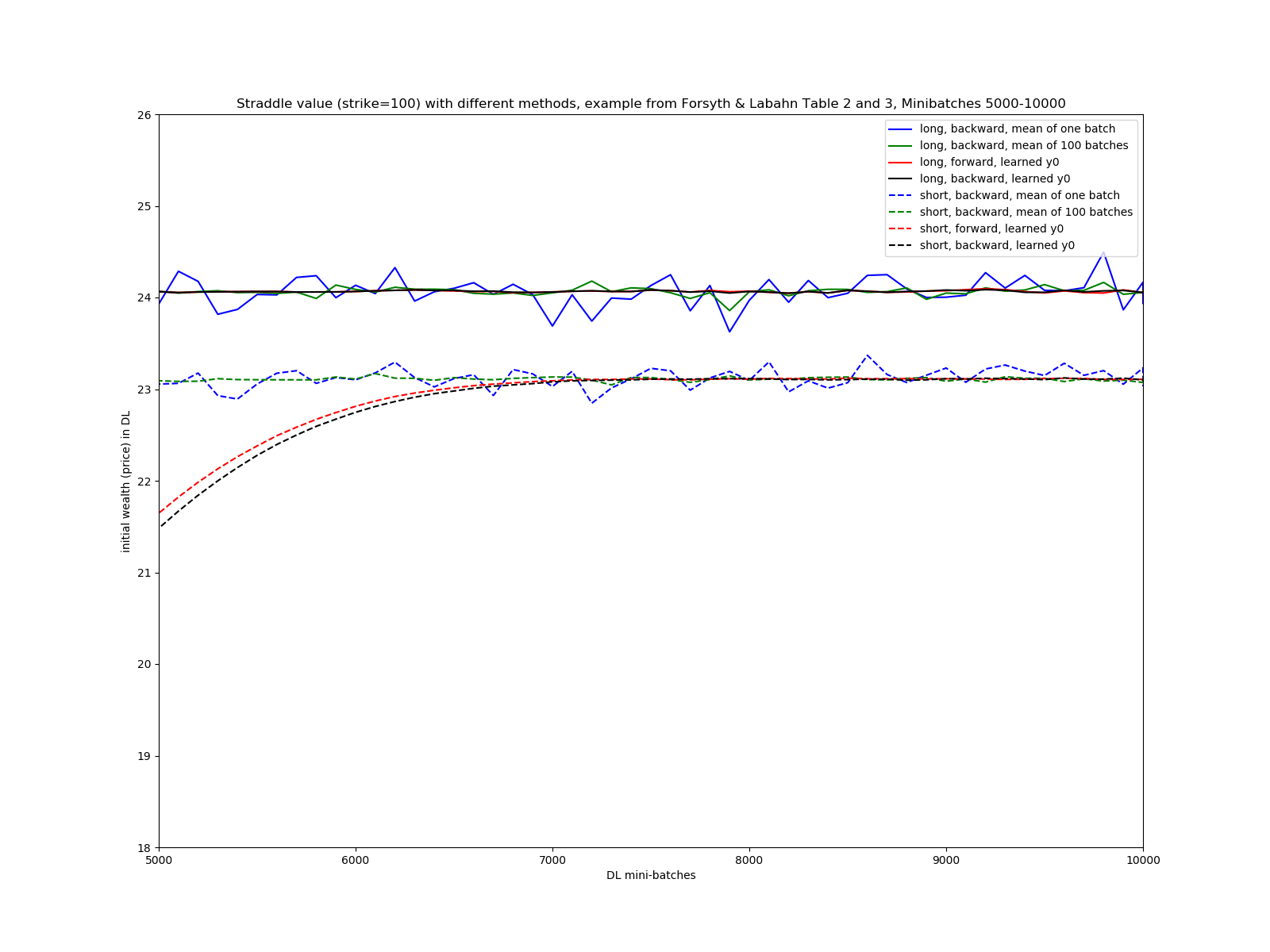}  
\end{subfigure}

\caption{$Y_0$ estimates or parameters for the straddle case - over 20000 mini-batches and 
detail for 5000-1000 minibatches - exact backward step - batch size 512} 
\label{fig:forsythlabahntableexampleexact}
\end{figure}

\begin{figure}[p]
\begin{subfigure}{\textwidth}
  \centering
  \includegraphics[width=.9\linewidth]{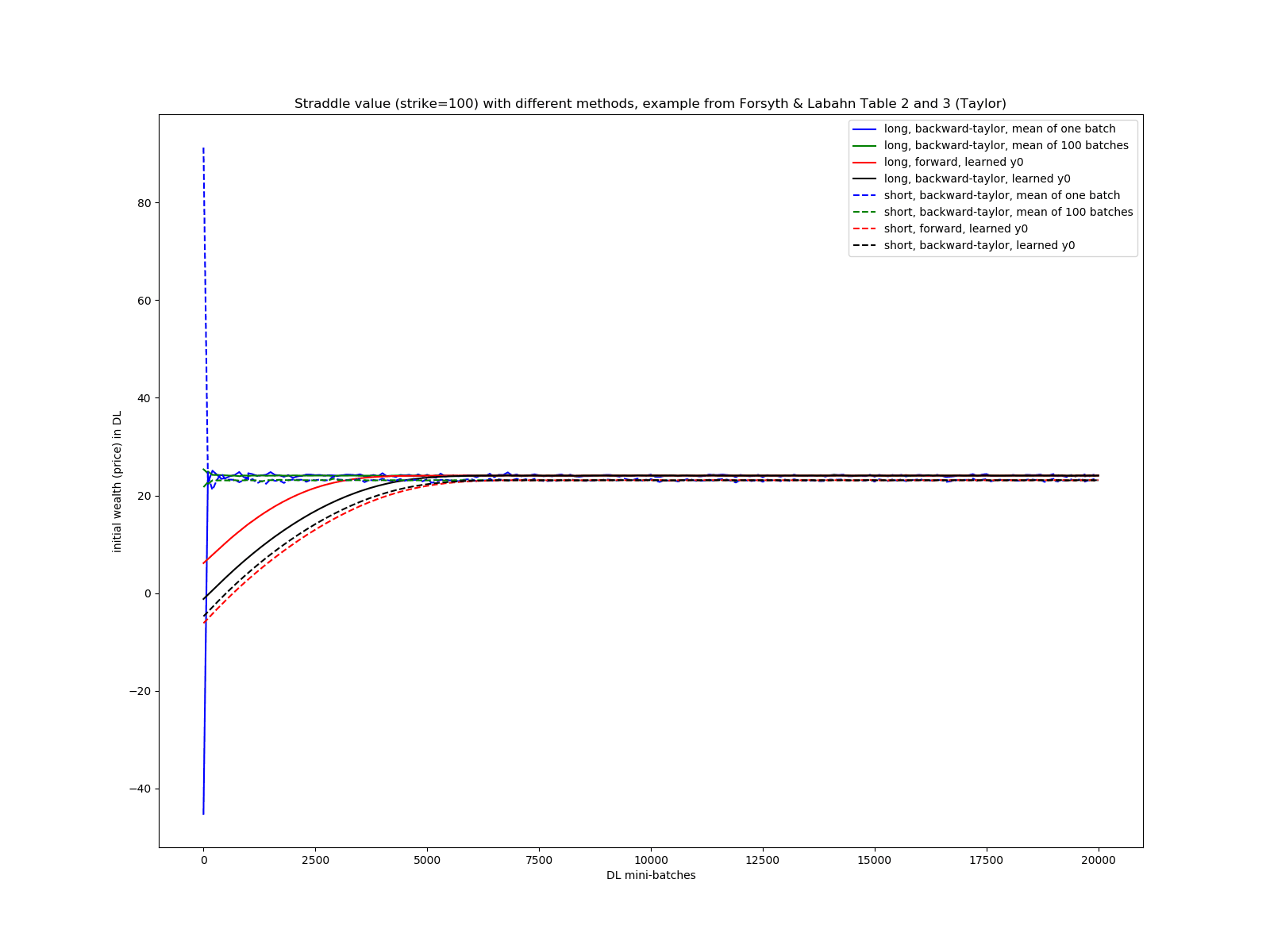}  
\end{subfigure}
\newline
\begin{subfigure}{\textwidth}
  \centering
  \includegraphics[width=.9\linewidth]{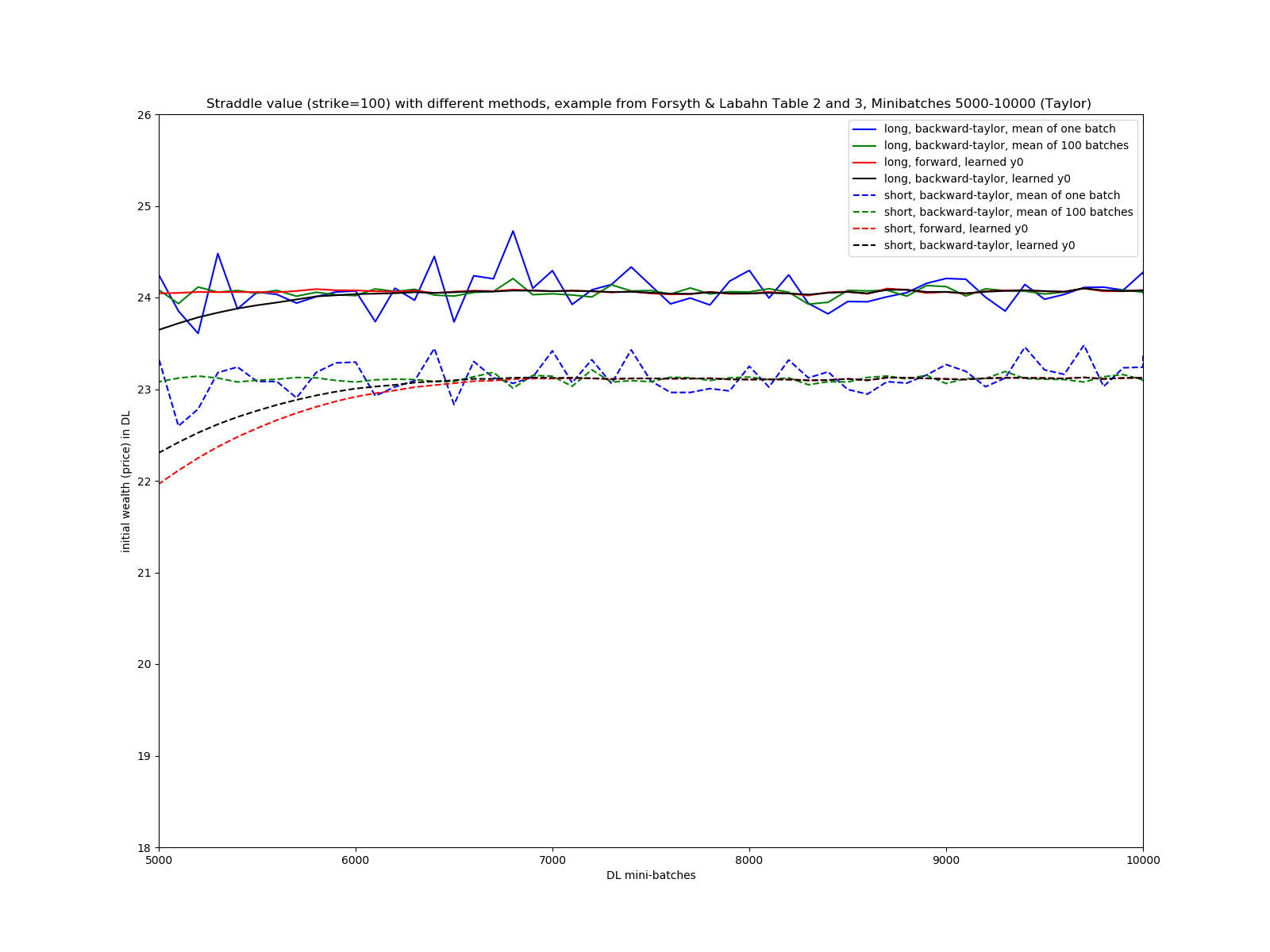}  
\end{subfigure}

\caption{$Y_0$ estimates or parameters for the straddle case - over 20000 mini-batches and 
detail for 5000-1000 minibatches - Taylor backward step - batch size 256}
\label{fig:forsythlabahntableexampletaylor}
\end{figure}

We plot the $Y_0$ estimates or parameters for different backward and forward
methods in figure \ref{fig:forsythlabahntableexampleexact} for exact and figure 
\ref{fig:forsythlabahntableexampletaylor} for Taylor backward step, for both long
and short straddles, together with a more detail view. We see that the method
that learns the $y0$ parameter initially converges more slowly than the
batch variance methods and that computing the mean over 100 mini-batches rather than one leads
to a faster and more smooth convergence. (And seemingly the learning $y0$
parameter method converges faster for upper price than for lower price).
However, once the learning $y0$ method gets close, its convergence is smoother
and better than the batch-variance methods. 

It would be interesting to see whether a method that updates the $y0$ parameter
based on a weighted average of the batch mean and the DL update would combine
the advantages of both methods. 

We compare our results against the results given in Forsyth and Labahn \cite[Tables 2 and 3]{forsyth2007numerical} in 
tables \ref{tab:straddlelowerprice} and \ref{tab:straddleupperprice} for the
lower and upper price, respectively. 

\begin{table}[p]

\begin{tabular}{|c|c|}
\hline
Method & Result \\
\hline
\multicolumn{2}{|c|}{Results from Forsyth and Labahn - 101 nodes} \\
\hline
Fully Implicit HJB PDE (implicit control) & 24.02047 \\
Crank-Nicolson HJB PDE (implicit control) & 24.0512  \\
Fully Implicit HJB PDE (pwc policy) & 24.01163 \\
Crank-Nicolson HJB PDE (pwc policy) & 24.0652 \\
\hline
\multicolumn{2}{|c|}{Forward deep BSDE - 20000 batches, size 256} \\
\hline
Learned $y0$ &  24.078833 (24.044819-24.098291) \\
\hline 
\multicolumn{2}{|c|}{Backward deep BSDE (exact) - 20000 batches, size 256} \\
\hline
Batch variance, 1 mini-batch mean &  24.119225 (23.912037-24.306263) \\
Batch variance, 100 mini-batch mean & 24.061472 (24.022112-24.132414) \\
Learned $y0$ & 24.072815 (24.043901-24.095972) \\
\hline
\multicolumn{2}{|c|}{Backward deep BSDE (Taylor) - 20000 batches, size 256} \\
\hline
Batch variance, 1 mini-batch mean &  24.119202 (23.911976-24.30629) \\
Batch variance, 100 mini-batch mean & 24.061443 (24.022062-24.132645) \\
Learned $y0$ & 24.072783 (24.043854-24.09594) \\
\hline
\multicolumn{2}{|c|}{Difference over all batches, exact-Taylor} \\
\hline
Batch variance, 1 mini-batch mean &  7.0242124e-05 (0.0040798187-0.0010948181)
\\
Batch variance, 100 mini-batch mean & 7.2354465e-05
(-0.00023078918-0.0051631927) \\ 
Learned $y0$ & 2.8257615e-05 ( -0.00012207031- 0.00032234192 )
\\
\hline
\end{tabular}
\caption{Upper Price}\label{tab:straddleupperprice}
\end{table}

\begin{table}[p]

\begin{tabular}{|c|c|}
\hline
Method  & Result \\
\hline
\multicolumn{2}{|c|}{Results from Forsyth and Labahn - 101 nodes} \\
Fully Implicit HJB PDE (implicit control) & 23.05854 \\
Crank-Nicolson HJB PDE (implicit control) & 23.08893 \\
Fully Implicit HJB PDE (pwc policy) & 23.06752 \\
Crank-Nicolson HJB PDE (pwc policy) & 23.09371 \\
\hline
\multicolumn{2}{|c|}{Forward deep BSDE - 20000 batches, size 256} \\
\hline
Learned $y0$ &  23.127728 (23.09221-23.139273) \\
\hline 
\multicolumn{2}{|c|}{Backward deep BSDE (exact) - 20000 batches, size 256} \\
\hline
Batch variance, 1 mini-batch mean &  23.1515 (22.834194-23.39135) \\
Batch variance, 100 mini-batch mean & 23.100569 (23.083847-23.133955) \\
Learned $y0$ & 23.126017 (23.091948-23.138735) \\
\hline
\multicolumn{2}{|c|}{Backward deep BSDE (Taylor) - 20000 batches, size 256} \\
\hline
Batch variance, 1 mini-batch mean &  23.151543 (22.834248-23.391354) \\
Batch variance, 100 mini-batch mean & 23.100618 (23.083895-23.133991) \\
Learned $y0$ & 23.12606 (23.091982-23.138788) \\
\hline
\multicolumn{2}{|c|}{Difference over all batches, exact-Taylor} \\
\hline
Batch variance, 1 mini-batch mean &  -3.9644332e-05  (-0.00045776367
-0.00018501282)
\\ 
 Batch variance, 100 mini-batch mean & -3.406489e-05 
 (-0.00035476685-9.536743e-05) \\
Learned $y0$ & -2.3744791e-05 (-0.00030136108-0.00034713745) 
\\
\hline
\end{tabular}
\caption{Lower Price}\label{tab:straddlelowerprice}
\end{table}

We see that we are close to the values given in  Forsyth and Labahn \cite[Tables 2 and 3]{forsyth2007numerical}.
 Forsyth and Labahn \cite[Tables 2 and 3]{forsyth2007numerical} have
 numbers for higher number of time steps and space steps as well which are even closer to our
results so it could be that if the  Forsyth and Labahn method would be run and
reported for more space steps but same number of time steps, it would give
values even closer to ours.

For random $X_0$ case, we pick $X_0$ uniformly within $[50,150]$ but plot results within $[80,120]$.

\begin{figure}[p]
\begin{subfigure}{0.5\textwidth}
  \centering
  \includegraphics[width=.9\linewidth]{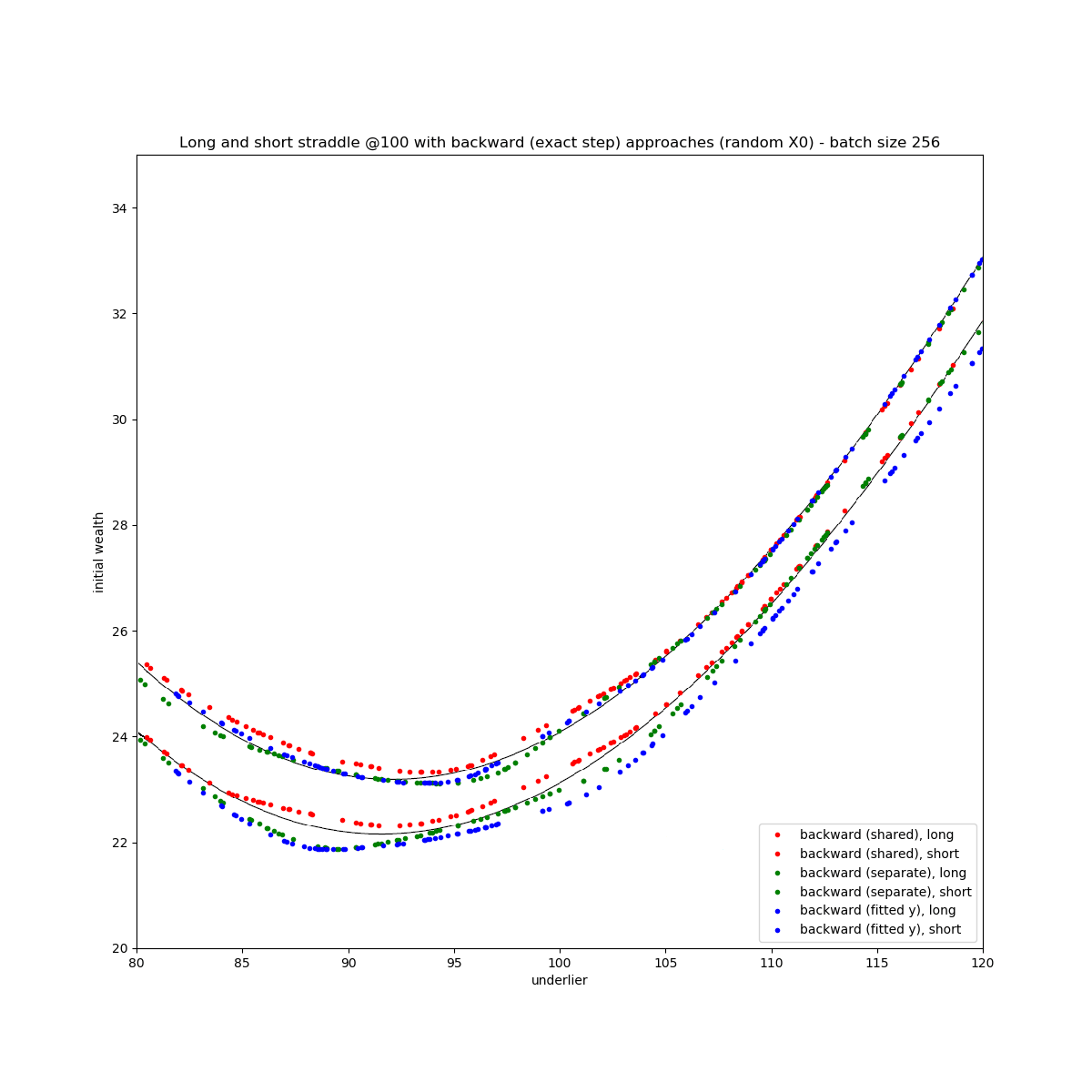}  
  \caption{Batch size 256}
\end{subfigure}
\begin{subfigure}{0.5\textwidth}
  \centering
  \includegraphics[width=.9\linewidth]{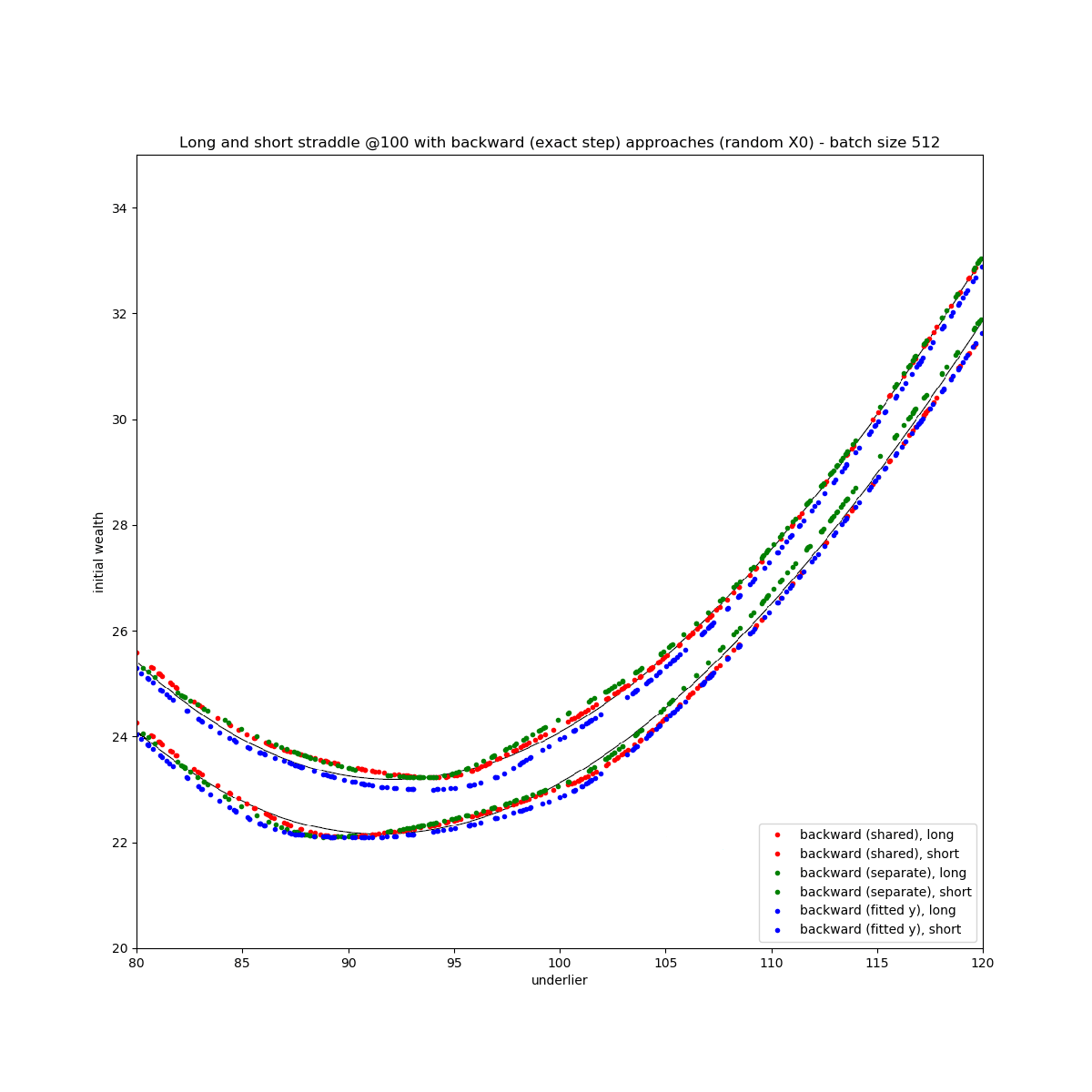}  
  \caption{Batch size 512}
\end{subfigure}
\newline
\begin{center}
\begin{subfigure}{0.5\textwidth}
  \centering
  \includegraphics[width=.9\linewidth]{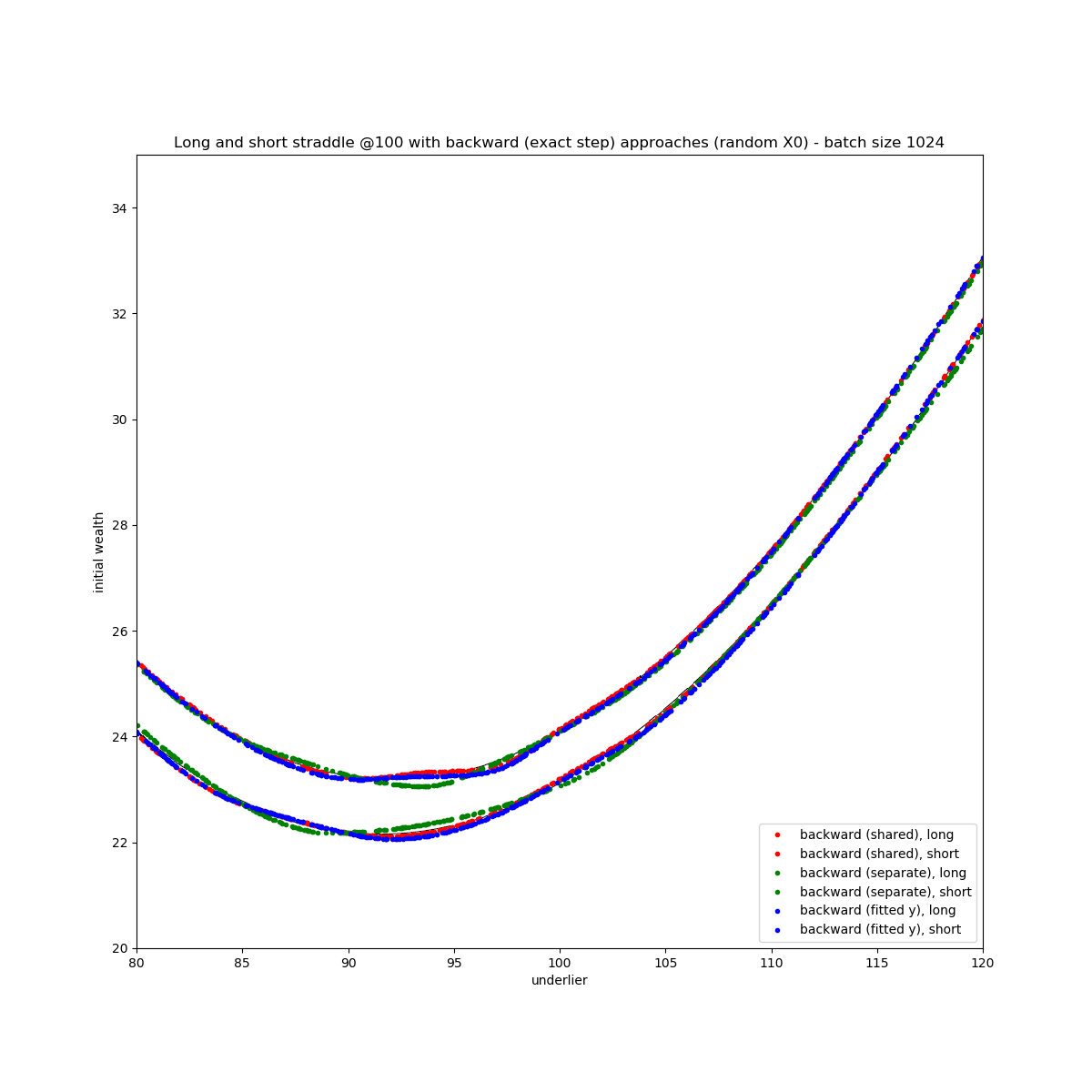}  
  \caption{Batch size 1024}
\end{subfigure}
\end{center}
\caption{$\mathsf{Yinit}(X_0)$ for various backward methods with exact backward step plotted over Forsyth and Labahn curves}
\label{fig:forsythplotexampleexactbackwardbatchsizes}
\end{figure}

\begin{figure}[p]
\begin{subfigure}{0.5\textwidth}
  \centering
  \includegraphics[width=.9\linewidth]{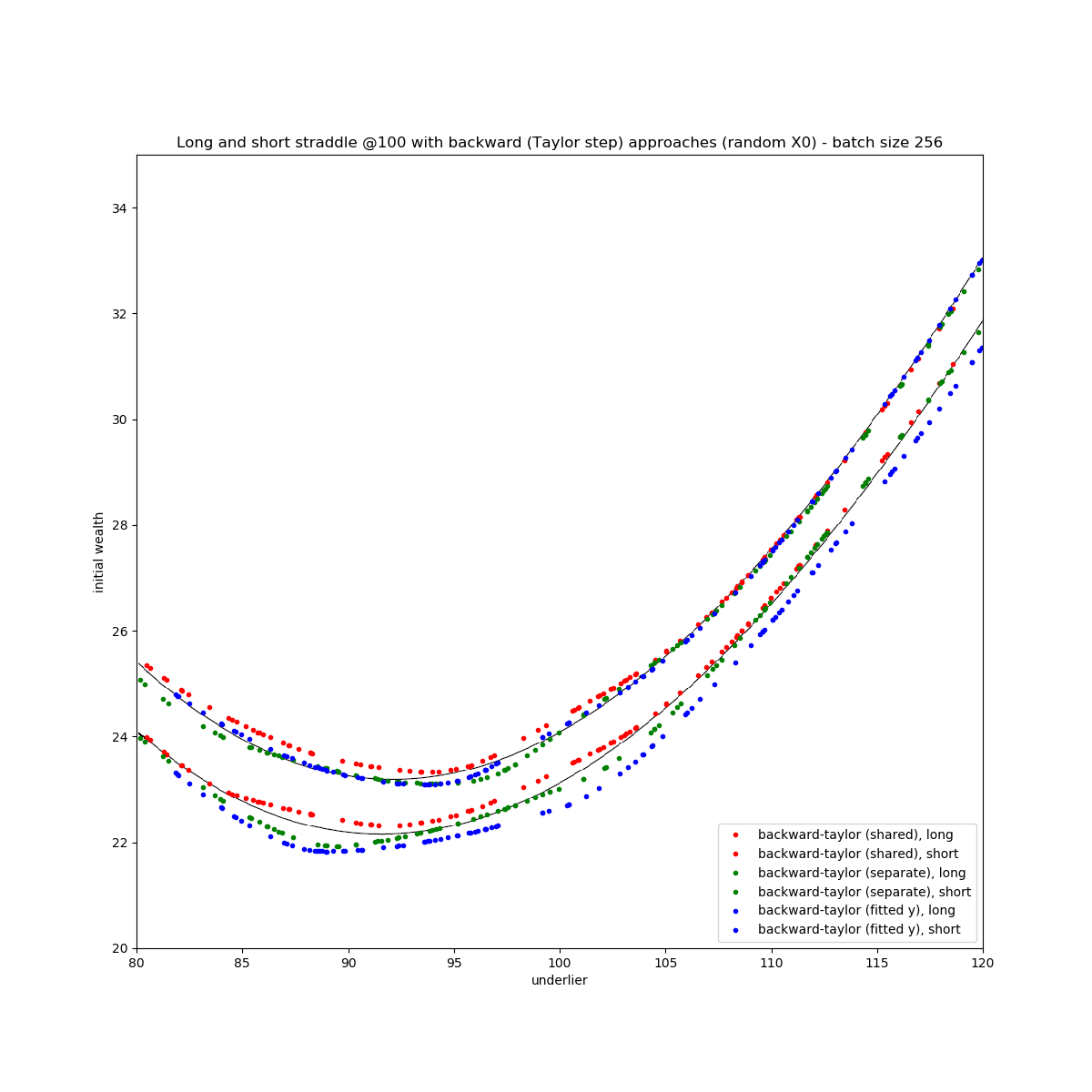}  
  \caption{Batch size 256}
\end{subfigure}
\begin{subfigure}{0.5\textwidth}
  \centering
  \includegraphics[width=.9\linewidth]{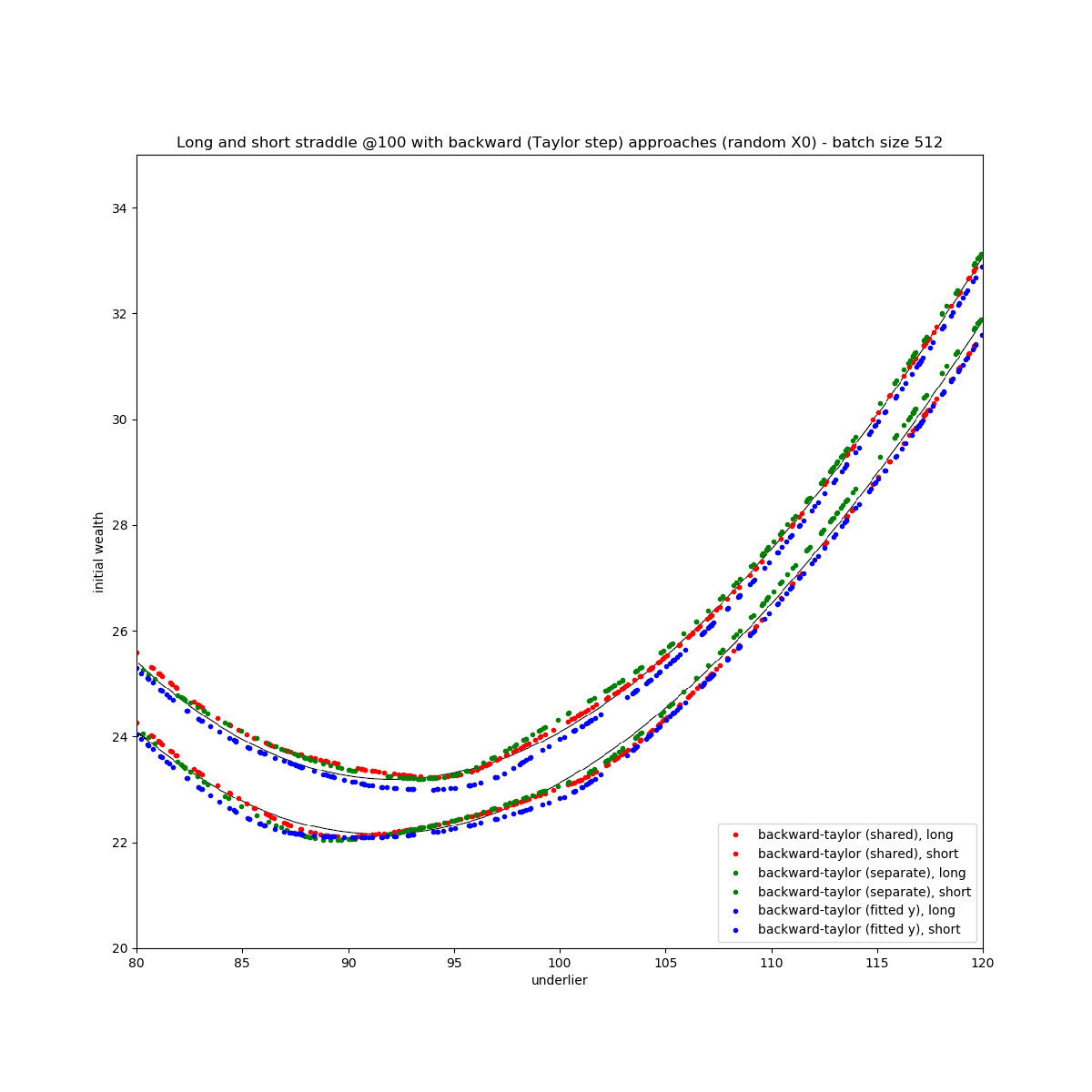}  
  \caption{Batch size 512}
\end{subfigure}
\newline
\begin{center}
\begin{subfigure}{0.5\textwidth}
  \centering
  \includegraphics[width=.9\linewidth]{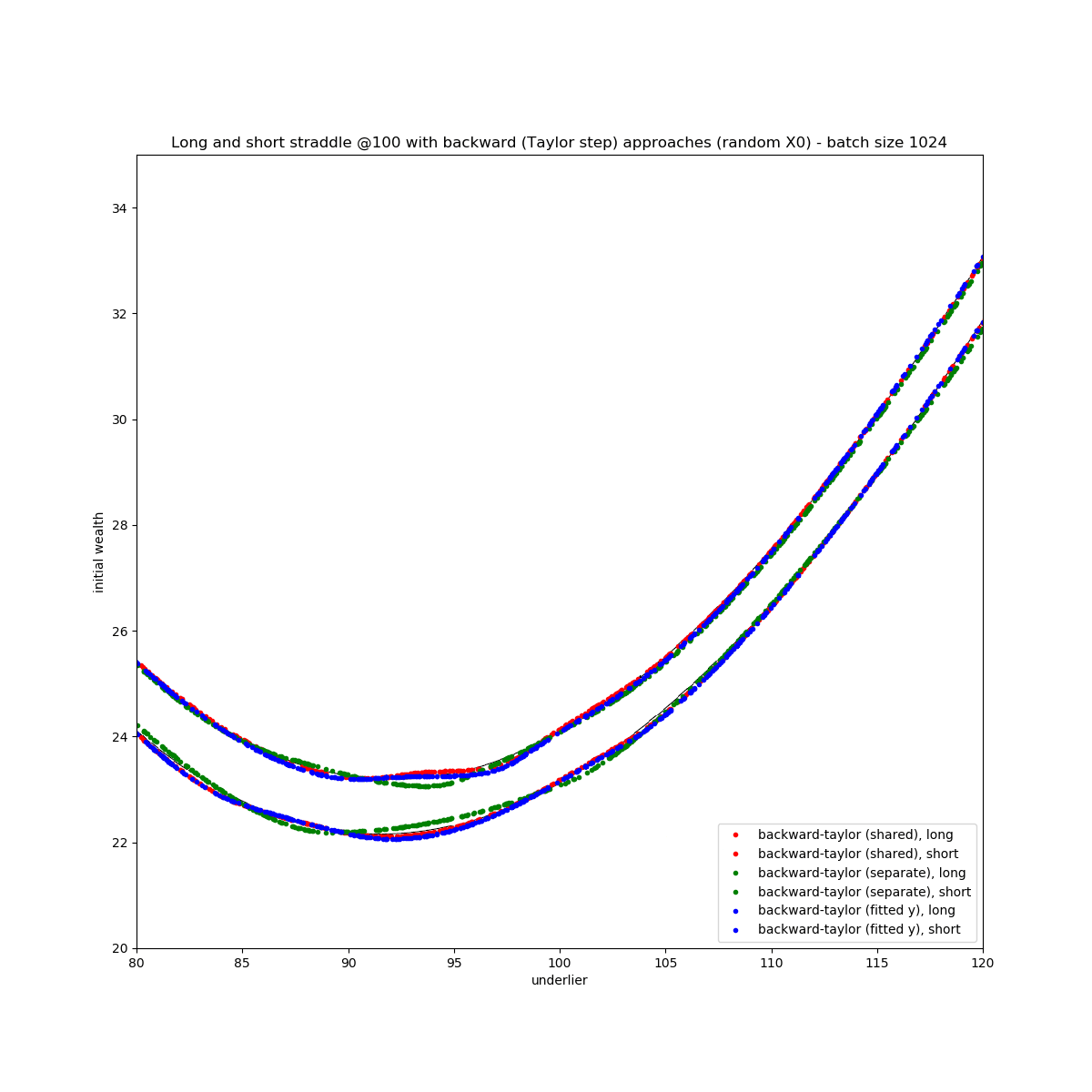}  
  \caption{Batch size 1024}
\end{subfigure}
\end{center}
\caption{$\mathsf{Yinit}(X_0)$ for various backward methods with Taylor backward step plotted over Forsyth and Labahn curves}
\label{fig:forsythplotexampletaylorbackwardbatchsizes}
\end{figure}

\begin{figure}[p]
\begin{subfigure}{0.5\textwidth}
  \centering
  \includegraphics[width=.9\linewidth]{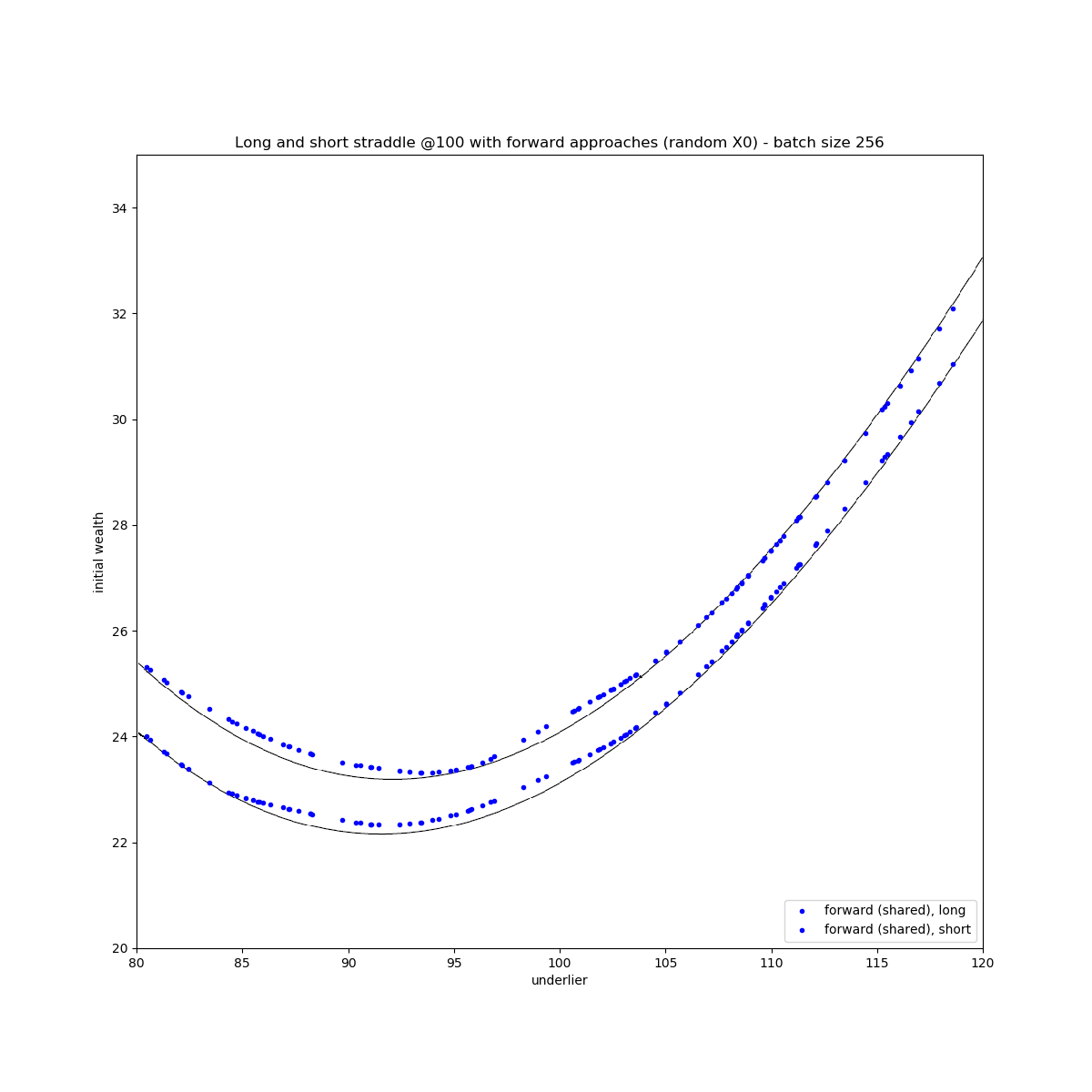}  
  \caption{Batch size 256}
\end{subfigure}
\begin{subfigure}{0.5\textwidth}
  \centering
  \includegraphics[width=.9\linewidth]{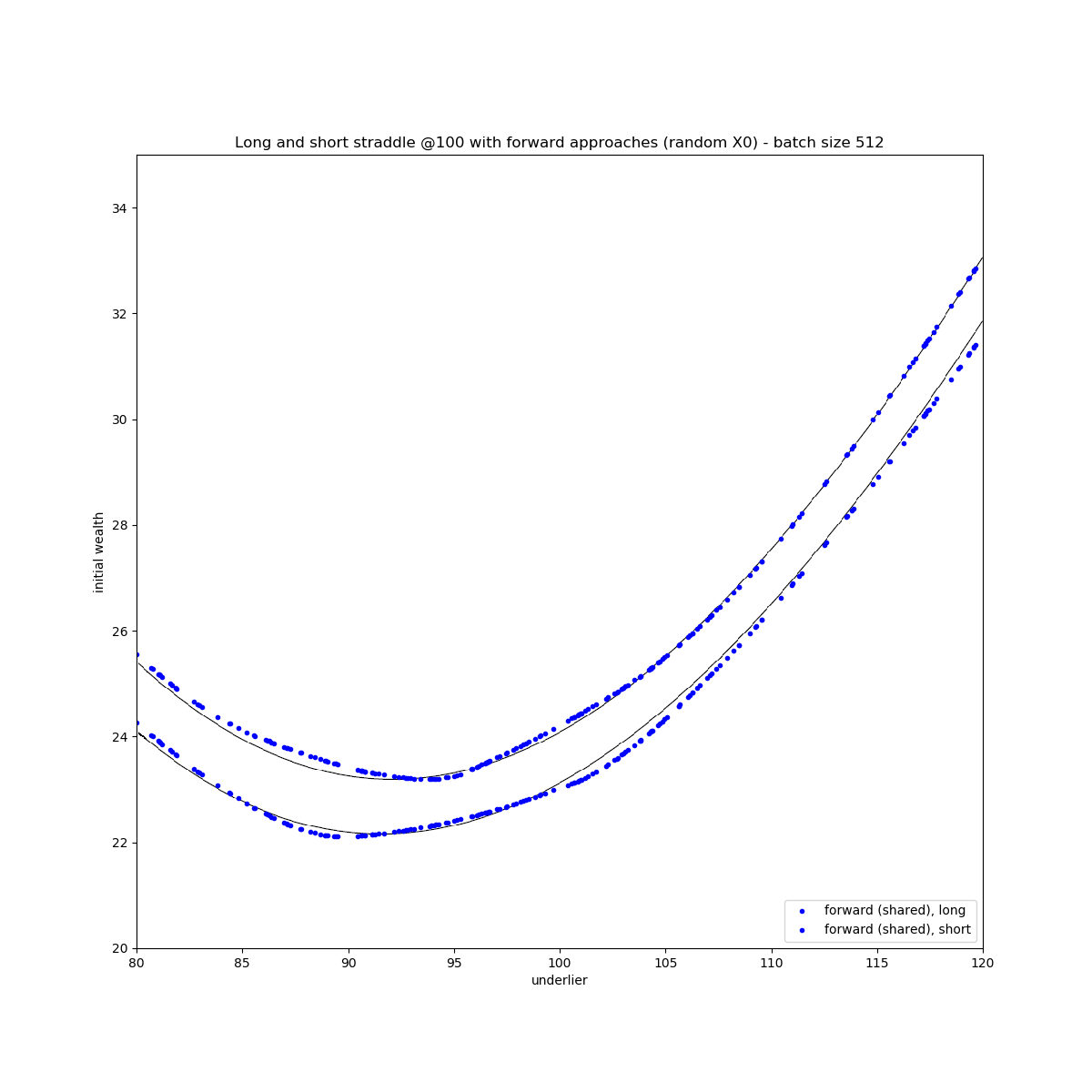}  
  \caption{Batch size 512}
\end{subfigure}
\newline
\begin{center}
\begin{subfigure}{0.5\textwidth}
  \centering
  \includegraphics[width=.9\linewidth]{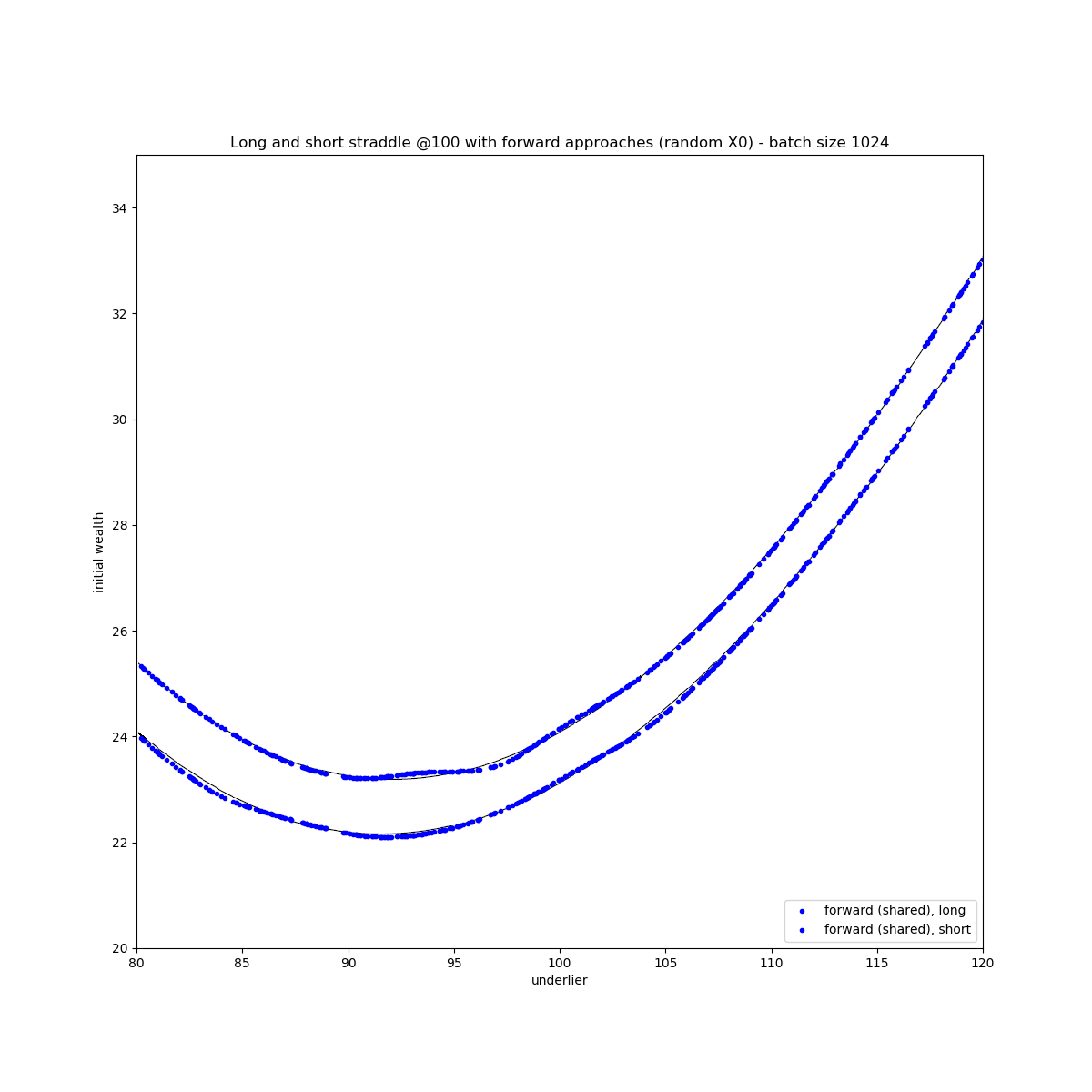}  
  \caption{Batch size 1024}
\end{subfigure}
\end{center}
\caption{$\mathsf{Yinit}(X_0)$ for various forward methods plotted over Forsyth and Labahn curves}
\label{fig:forsythplotexampleforwardbatchsizes}
\end{figure}

We extracted the curves from Figure 1 from Forsyth and Labahn \cite[Figure
1]{forsyth2007numerical} (the hjb PDF version) and plotted them as background in
all the figures (curves shown in black). Figure
\ref{fig:forsythplotexampleexactbackwardbatchsizes} shows different backward
methods with exact backward step for different batch sizes, figure
\ref{fig:forsythplotexampletaylorbackwardbatchsizes} shows different backward
methods with Taylor backward step  for different batch sizes, and figure
\ref{fig:forsythplotexampleforwardbatchsizes} shows forward method with random
$X_0$ for different batch sizes, all for long and short positions. We can see
that for increasing batch sizes, the agreement is improving.\footnote{Notice
that \cite[Figure
1]{forsyth2007numerical} do not give the number of space or time steps used for the plot.}

It can be seen that the results for exact backward step and Taylor backward step
are very close. 

\begin{figure}[p]
\begin{subfigure}{.5\textwidth}
  \centering
  \includegraphics[width=.9\linewidth]{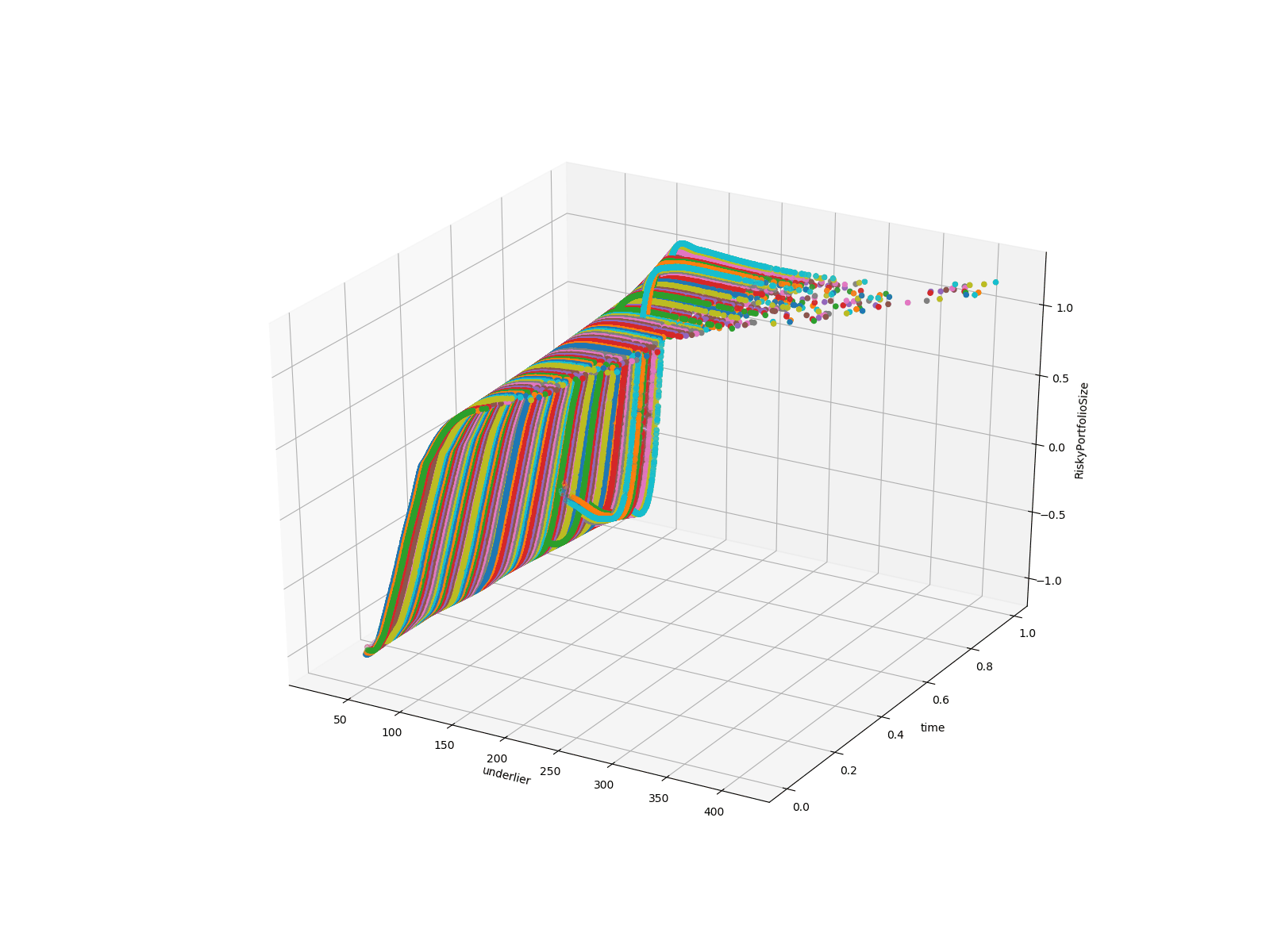}  
  \caption{Risky portfolio size (delta)}
\end{subfigure}
\begin{subfigure}{.5\textwidth}
  \centering
  \includegraphics[width=.9\linewidth]{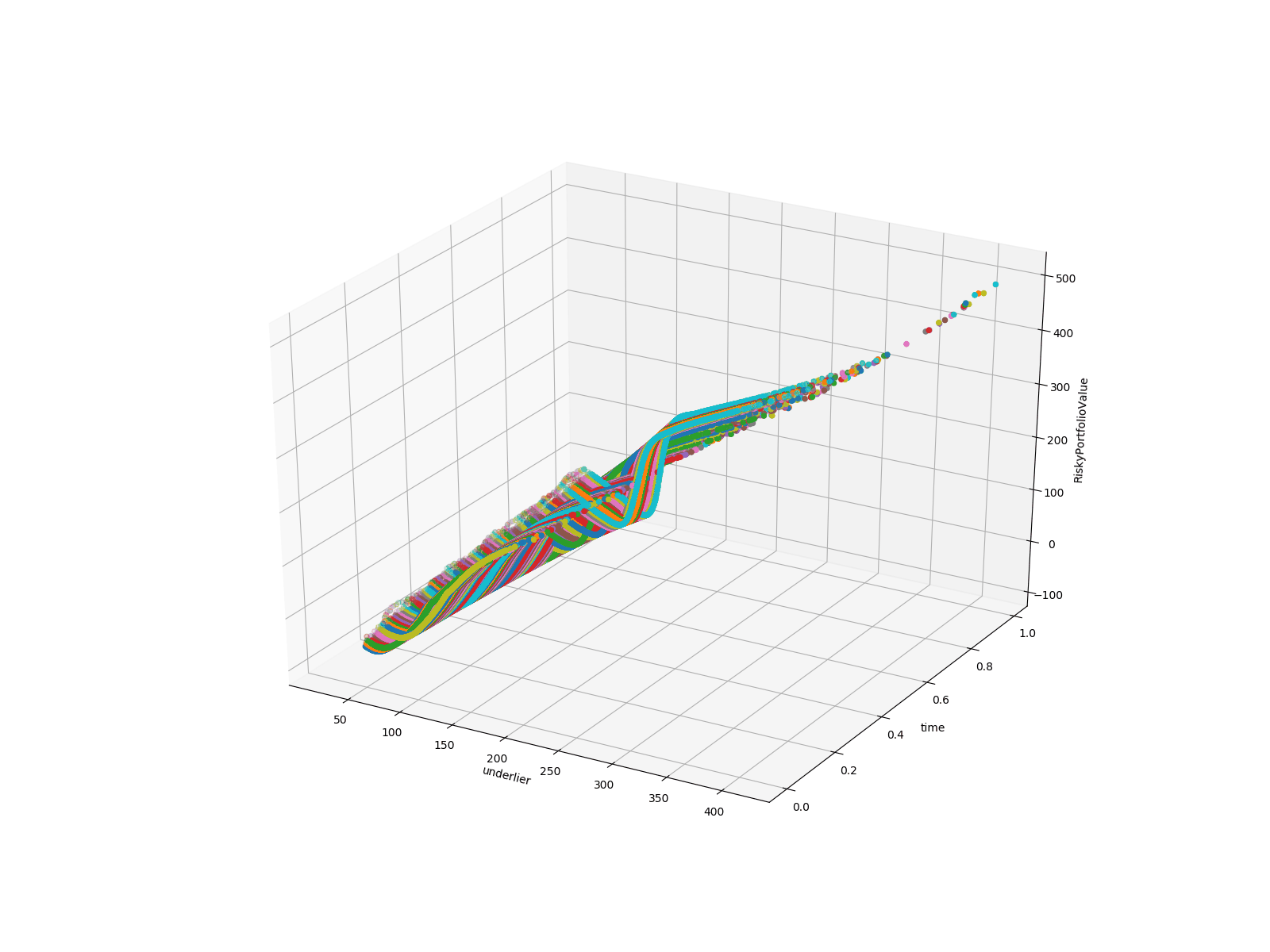}  
  \caption{Risky portfolio value (delta times stock price)}
\end{subfigure}
\newline
\begin{subfigure}{.5\textwidth}
  \centering
  \includegraphics[width=.9\linewidth]{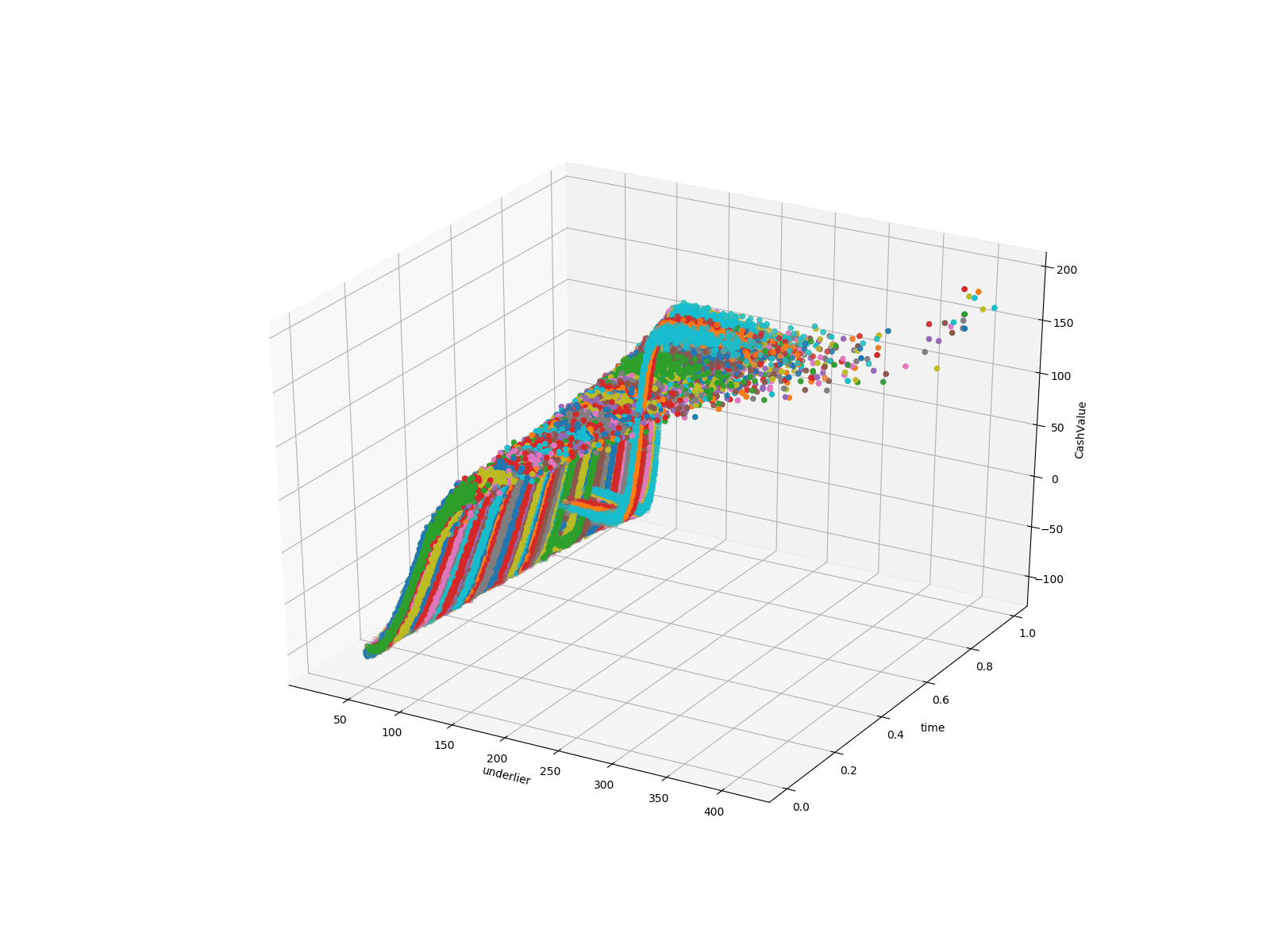}  
  \caption{Cash position value}
\end{subfigure}
\begin{subfigure}{.5\textwidth}
  \centering
  \includegraphics[width=.9\linewidth]{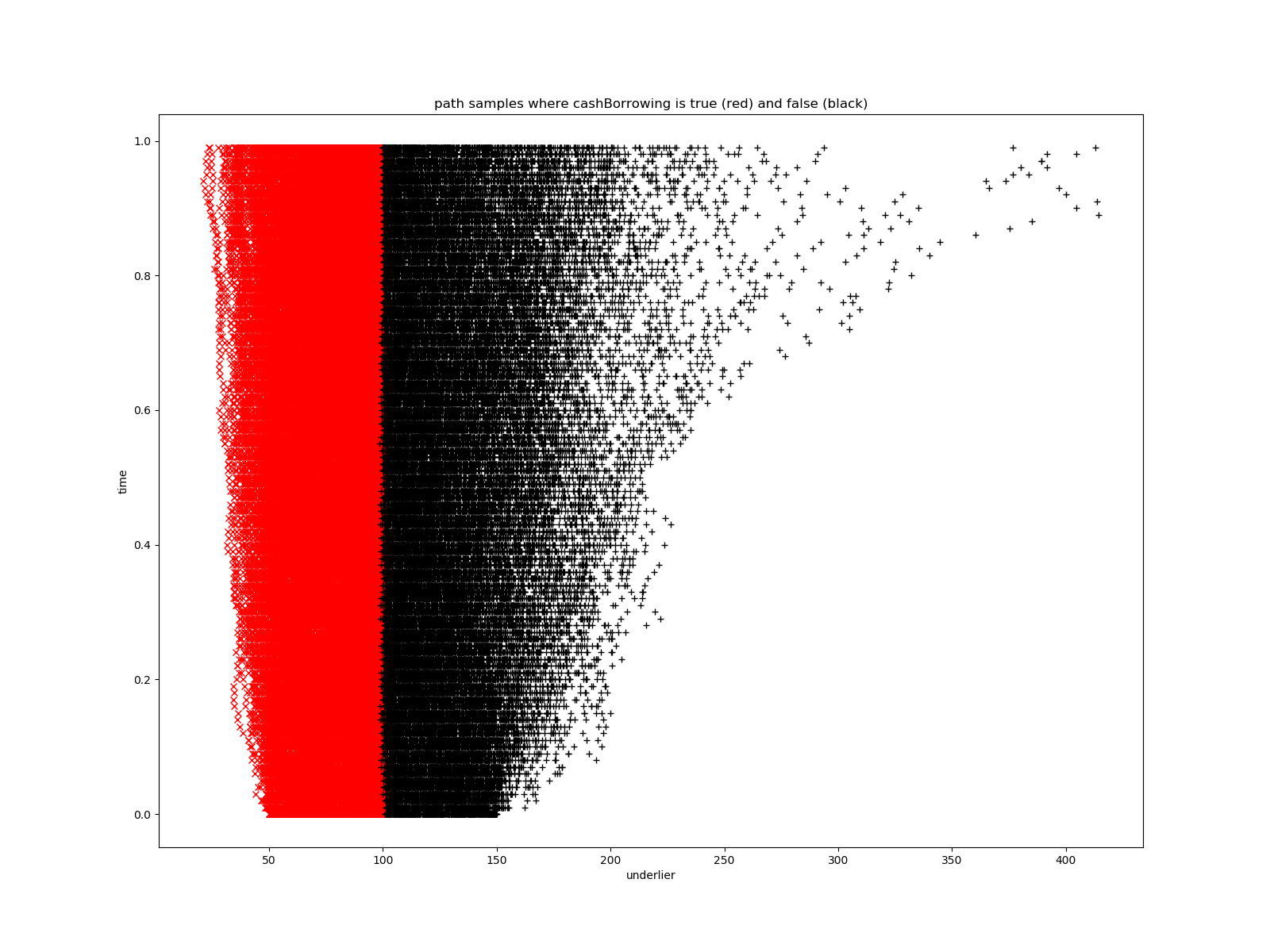}  
  \caption{Where strategy borrows/lends}
\end{subfigure}
\caption{Risky portfolio size, risky portfolio value, cash position value, and
locations where strategy borrows (red) or lends (black) for random $X_0$, exact
backward step, batch size 1024, long position/upper price}
\label{fig:forsythplotexamplestrategyvisualizationlong}
\end{figure}

\begin{figure}[p]
\begin{subfigure}{.5\textwidth}
  \centering
  \includegraphics[width=.9\linewidth]{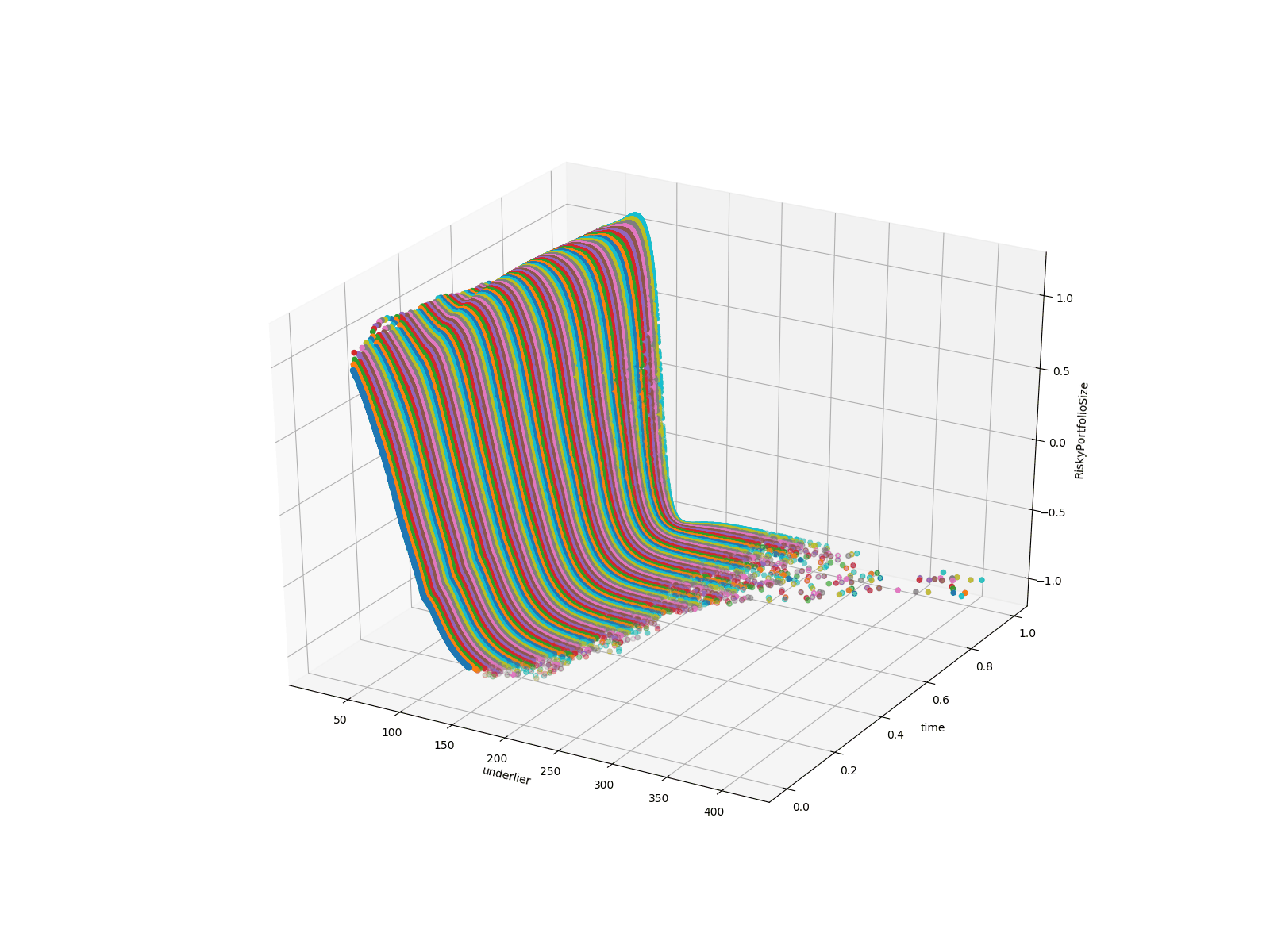}  
  \caption{Risky portfolio size (delta)}
\end{subfigure}
\begin{subfigure}{.5\textwidth}
  \centering
  \includegraphics[width=.9\linewidth]{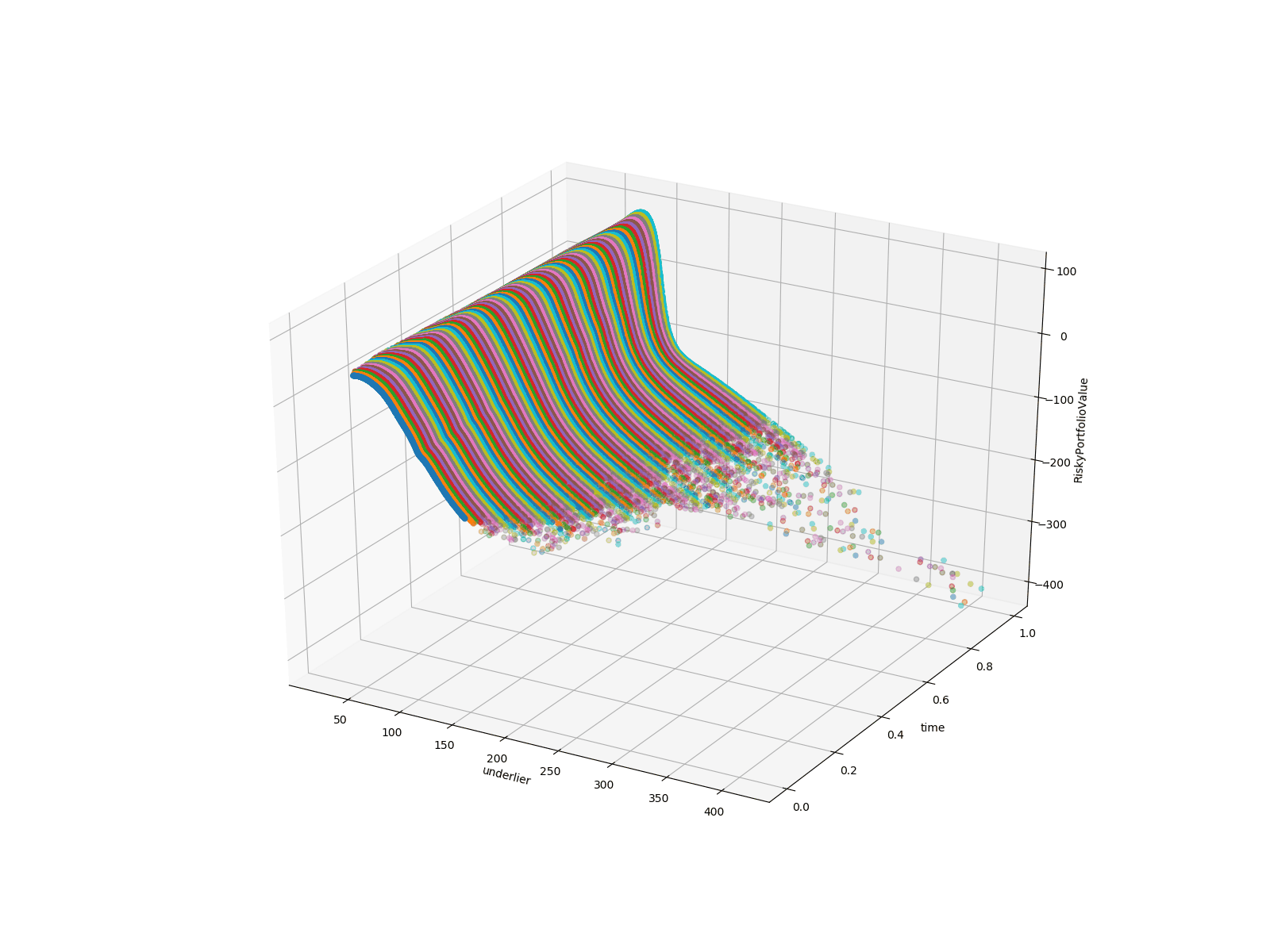}  
  \caption{Risky portfolio value (delta times stock price)}
\end{subfigure}
\newline
\begin{subfigure}{.5\textwidth}
  \centering
  \includegraphics[width=.9\linewidth]{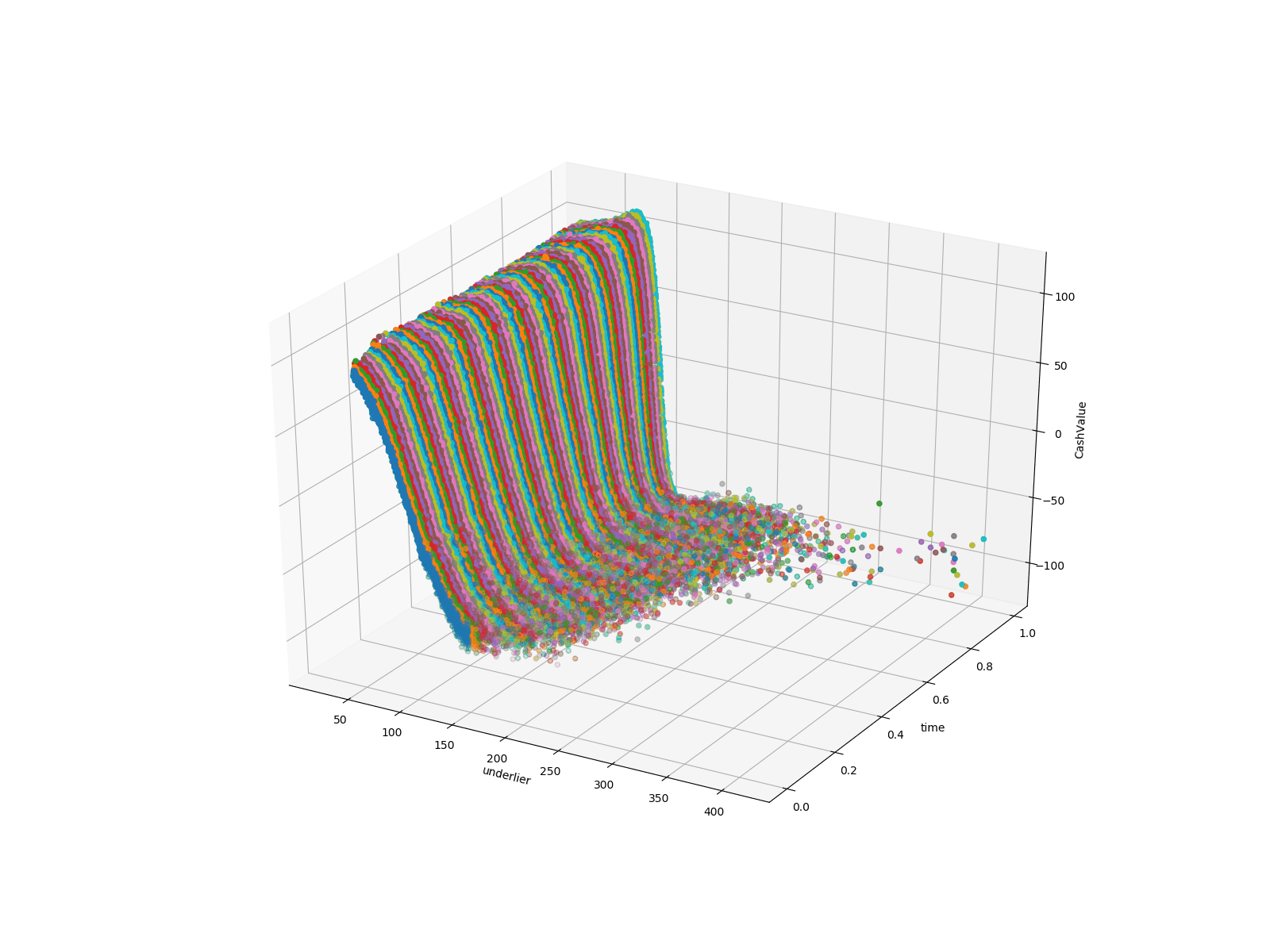}  
  \caption{Cash position value}
\end{subfigure}
\begin{subfigure}{.5\textwidth}
  \centering
  \includegraphics[width=.9\linewidth]{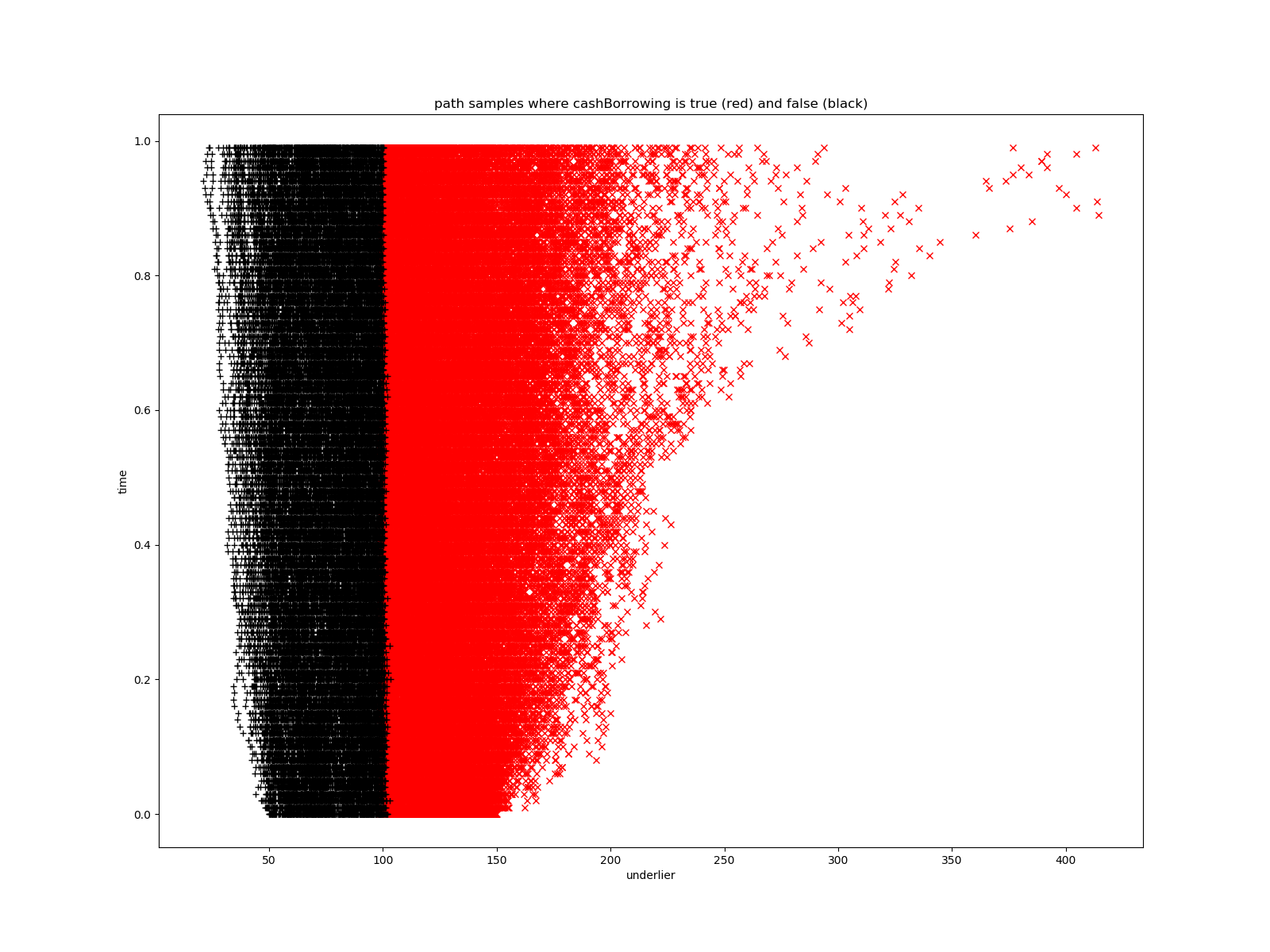}  
  \caption{Where strategy borrows/lends}
\end{subfigure}
\caption{Risky portfolio size, risky portfolio value, cash position value, and
locations where strategy borrows (red) or lends (black) for random $X_0$, exact
backward step, batch size 1024, short position/lower price}
\label{fig:forsythplotexamplestrategyvisualizationshort}
\end{figure}

Lastly, for the exact backward step for batch size 1024, we show RiskyPortfolio
size (delta), RiskyPortfolio value (delta times stock price), cash position
value for long and short straddle, and the regions with borrowing (red) and
lending (black) in figures \ref{fig:forsythplotexamplestrategyvisualizationlong}
and \ref{fig:forsythplotexamplestrategyvisualizationshort} respectively.

\section{Acknowledgements}

The authors would like to thank Orcan Ogetbil, Daniel Weingard and Xin Wang for
proof reading the draft and giving helpful feedback; Vijayan Nair for discussion
regarding methods, presentation, and results and for reviewing the paper; and
Agus Sudjianto for supporting this research.

\section{Conclusion}

In this paper, we first introduced FBSDE for general nonlinear problems, 
with particular details for the differential rates problem, time-discretize them, and
then derived exact and Taylor approximations for the backward step. We then
quickly described the forward and backward deep BSDE approaches that we consider
- both the batch-variance variant already described in the literature but also
the novel initial variable and network versions, the last one for random $X_0$. 
Then we applied
these methods for the differential rates problem for the call combination case
from Han, Jentzen and E \cite{han2018solving} and for the straddle case from
Forsyth and Labahn \cite{forsyth2007numerical}. We compare the results for a
case with fixed $X_0$ and for a case with varying $X_0$ with the results from
Forsyth and Labahn \cite{forsyth2007numerical} and see that they agree well. We
also compare methods for the exact backward step and the Taylor backward step
and in the straddle and call combination examples that we ran, the results seem
to be very close. We also visualized some of the results to show what they mean
in terms of trading strategy and borrowing and lending. 

The deepBSDE methods described in this paper are using a very different approach
from the PDE methods by Forsyth and Labahn \cite{forsyth2007numerical}, but they
give results very close to those published there. That makes us confident that
these methods can be used to generically and efficiently approximate solutions
to such nonlinear pricing problems, even with relatively small batch-size such
as 512 or 1024.

\section{Disclaimer}

Any opinions, findings and conclusions or recommendations expressed in this material are
those of the author and do not necessarily reflect the views of Wells Fargo Bank, N.A., its
parent company, affiliates and subsidiaries.

\bibliographystyle{alpha}
\bibliography{../FirstOverviewPaperDraft/reviewRefs}

\newcommand{\etalchar}[1]{$^{#1}$}
\begin{thebibliography}{WCS{\etalchar{+}}18}

\bibitem[EHJ17]{weinan2017deep}
Weinan E, Jiequn Han, and Arnulf Jentzen.
\newblock Deep learning-based numerical methods for high-dimensional parabolic
  partial differential equations and backward stochastic differential
  equations.
\newblock {\em Communications in Mathematics and Statistics}, 5(4):349--380,
  2017.
\newblock arXiv:1706.04702.

\bibitem[FL07]{forsyth2007numerical}
Peter~A Forsyth and George Labahn.
\newblock Numerical methods for controlled hamilton-jacobi-bellman pdes in
  finance.
\newblock {\em Journal of Computational Finance}, 11(2):1--44, 2007.
\newblock preprint version available online for instance at:
  \url{https://cs.uwaterloo.ca/~paforsyt/hjb.pdf}.

\bibitem[GYH20]{ganesan2019pricingbarriers}
Narayan Ganesan, Yajie Yu, and Bernhard Hientzsch.
\newblock Pricing barrier options with deep{B}{S}{D}{E}s.
\newblock {\em arXiv preprint arXiv:2005.10966}, May 2020.

\bibitem[Hie19]{hientzsch2019intro}
Bernhard Hientzsch.
\newblock Introduction to solving quant finance problems with time-stepped
  {F}{B}{S}{D}{E} and deep learning.
\newblock {\em arXiv preprint arXiv:1911.12231}, Nov 2019.
\newblock Also available at SSRN: https://ssrn.com/abstract=3494359 or
  http://dx.doi.org/10.2139/ssrn.3494359.

\bibitem[HJE18]{han2018solving}
Jiequn Han, Arnulf Jentzen, and Weinan E.
\newblock Solving high-dimensional partial differential equations using deep
  learning.
\newblock {\em Proceedings of the National Academy of Sciences},
  115(34):8505--8510, 2018.

\bibitem[HPW19]{hure2019some}
C{\^o}me Hur{\'e}, Huy{\^e}n Pham, and Xavier Warin.
\newblock Some machine learning schemes for high-dimensional nonlinear
  {P}{D}{E}s.
\newblock {\em arXiv preprint arXiv:1902.01599}, 2019.

\bibitem[LXL19]{liang2019deep}
Jian Liang, Zhe Xu, and Peter Li.
\newblock Deep learning-based least square forward-backward stochastic
  differential equation solver for high-dimensional derivative pricing.
\newblock {\em arXiv preprint arXiv:1907.10578}, 2019.
\newblock Also available at SSRN: https://ssrn.com/abstract=3381794 or
  http://dx.doi.org/10.2139/ssrn.3381794.

\bibitem[Mer15]{mercurio2015bergman}
Fabio Mercurio.
\newblock Bergman, {P}iterbarg, and beyond: pricing derivatives under
  collateralization and differential rates.
\newblock In {\em Actuarial Sciences and Quantitative Finance}, pages 65--95.
  Springer, 2015.
\newblock Also available at SSRN: https://ssrn.com/abstract=2326581 or
  http://dx.doi.org/10.2139/ssrn.2326581.

\bibitem[War18]{warin2018nesting}
Xavier Warin.
\newblock Nesting monte carlo for high-dimensional non linear {P}{D}{E}s.
\newblock {\em arXiv preprint arXiv:1804.08432}, 2018.

\bibitem[WCS{\etalchar{+}}18]{wang2018deep}
Haojie Wang, Han Chen, Agus Sudjianto, Richard Liu, and Qi~Shen.
\newblock Deep learning-based {B}{S}{D}{E} solver for {L}{I}{B}{O}{R} market
  model with application to bermudan swaption pricing and hedging.
\newblock {\em arXiv preprint arXiv:1807.06622}, 2018.
\newblock Also available at SSRN: https://ssrn.com/abstract=3214596 or
  http://dx.doi.org/10.2139/ssrn.3214596.

\end{thebibliography}

\end{document}